\documentclass[prb,aps,twocolumn,eqsecnum,showpacs,preprintnumbers,amsmath,amssymb]{revtex4}

\usepackage{graphicx}
\usepackage{dcolumn}
\usepackage{bm}
\newcommand{\bea}{\begin{eqnarray}}
\newcommand{\eea}{\end{eqnarray}}
\newcommand{\be}{\begin{equation}}
\newcommand{\ee}{\end{equation}}
\newcommand{\bse}{\begin{subequations}}
\newcommand{\ese}{\end{subequations}}
\newcommand{\nn}{\nonumber}
\newcommand{\ket}[1]{|#1\rangle}

\newcommand{\av}[1]{\langle #1\rangle}

\newcommand{\s}{{\sigma}}
\newcommand{\A}{\text{ASF}}
\newcommand{\M}{\text{MSF}}
\newcommand{\wt}{{\omega_0}}
\newcommand{\rt}{{r_0}}
\newcommand{\rts}{{r_{\s0}}}
\newcommand{\cH}{{\cal H}}
\newcommand{\E}{{\cal E}}
\newcommand{\rv}{{\bf r}}

\newcommand{\xv}{{\bf r}}

\newcommand{\kv}{{\bf k}}
\newcommand{\eps}{{\varepsilon}}

\newcommand{\vv}{{\bf v}}

\newcommand{\nv}{{\bf n}}

\newcommand{\Tr}{{\rm Tr}}

\newcommand{\T}{{\tilde T}}

\begin{document}
\title{Superfluidity and phase transitions in a resonant Bose gas}
\author{Leo Radzihovsky$^1$}
\author{Peter B. Weichman$^2$}
\author{Jae I. Park$^{1,3}$}

\affiliation{$^1$Department of Physics, University of Colorado,
Boulder, CO 80309}

\affiliation{$^2$BAE Systems, Advanced Information Technologies, 6
New England Executive Park, Burlington, MA 01803}

\affiliation{$^3$National Institute of Standards and Technology, 325
Broadway, Boulder, Colorado 80305-3328}

\date{\today}

\begin{abstract}

The atomic Bose gas is studied across a Feshbach resonance, mapping
out its phase diagram, and computing its thermodynamics and
excitation spectra. It is shown that such a degenerate gas admits
two distinct atomic and molecular superfluid phases, with the latter
distinguished by the absence of atomic off-diagonal long-range
order, gapped atomic excitations, and deconfined atomic
$\pi$-vortices. The properties of the molecular superfluid are
explored, and it is shown that across a Feshbach resonance it
undergoes a quantum Ising transition to the atomic superfluid, where
both atoms and molecules are condensed. In addition to its distinct
thermodynamic signatures and deconfined half-vortices, in a trap a
molecular superfluid should be identifiable by the absence of an
atomic condensate peak and the presence of a molecular one.

\end{abstract}

\pacs{Valid PACS appear here}
\maketitle

\section{Introduction}
\label{sec:intro}

\subsection{Background}
\label{sec:background}

Remarkable experimental advances in manipulating degenerate atomic
gases have opened a new era in studies of highly coherent,
interacting quantum many-body systems. One of the most striking
advances is the ability to finely control atomic two-body
interactions by tuning with a magnetic field the energy (detuning)
of the molecular Feshbach resonance (FR) through the atomic
continuum.\cite{F62,TTHK99} This technique has led to a realization
of a long-sought-after s-wave paired superfluidity in
bosonic\cite{WFHRH00,DCTW02} and fermionic atomic
gases.\cite{J03,GRJ04,Z04} For fermionic atoms, it also allowed the
system to be tuned between the BCS\cite{BCS57} regime of
weakly-paired, strongly overlapping Cooper pairs (familiar from
solid-state superconductors), and the BEC regime of tightly bound,
weakly-interacting Bose-condensed diatomic molecules.

Although this crossover has received considerable
attention,\cite{MRE93,TFMK01,OG02,MKH02,AGR04,CSL05,GR07} because of
the absence of {\em qualitative} differences between the BCS and BEC
s-wave paired {\em fermionic} superfluids, their equilibrium
properties are already {\em qualitatively} well described by early
seminal works.\cite{E69,L80,NS85}  In fact for a narrow FR
(unfortunately not realized by most current experimental systems),
the crossover can even be computed quantitatively, as a perturbation
series in the ratio of the FR width to the Fermi
energy.\cite{AGR04,GR07} In such narrow FR systems the crossover to
BEC takes place when the FR detuning $\nu$ (quasi-molecule's rest
energy) ranges from twice the Fermi energy $2\epsilon_F$ (when it
first becomes favorable to convert a finite fraction of the
Fermi-sea into molecules stabilized by Pauli-blocking) down to zero
energy, where all the fermions have become bound into Bose-condensed
diatomic molecules. The complementary broad resonance regime of most
experiments,\cite{GRJ05} particularly near a universal unitary
point\cite{H04} has been successfully studied using quantum Monte
Carlo\cite{CR05,BDM07,BPST06} and field theoretic
$\epsilon$-expansion\cite{Son06,NS06} and
$1/N$-expansion\cite{NS06,VSR06} methods borrowed from critical
phenomena.

As was recently pointed out\cite{RPW04,RDSS04} and is the subject of
this paper, the phenomenology of resonantly interacting degenerate
{\em bosonic} atoms contrasts strongly and qualitatively with this
picture.\cite{Nozieres} For a large {\em positive} detuning,
molecules are strongly energetically suppressed and unpaired atoms
(as in any bosonic system at zero temperature) form an {\em atomic}
superfluid (ASF), exhibiting atomic off-diagonal long-range order
(ODLRO).\cite{P51} In the opposite extreme of a large {\em negative}
detuning, free atoms are strongly disfavored (gapped), pairing up
into stable bosonic molecules, that then, at $T=0$, form a diatomic
{\em molecular} superfluid characterized by a molecular ODLRO. The
MSF does {\em not} exhibit {\em atomic} ODLRO, nor the associated
atomic superfluidity.  Together with a gapped atomic excitation
spectrum and correlation functions (characteristics that extend to
finite temperature), these features qualitatively distinguish it
from the ASF.

\begin{figure}[bth]

\includegraphics[width=\columnwidth]{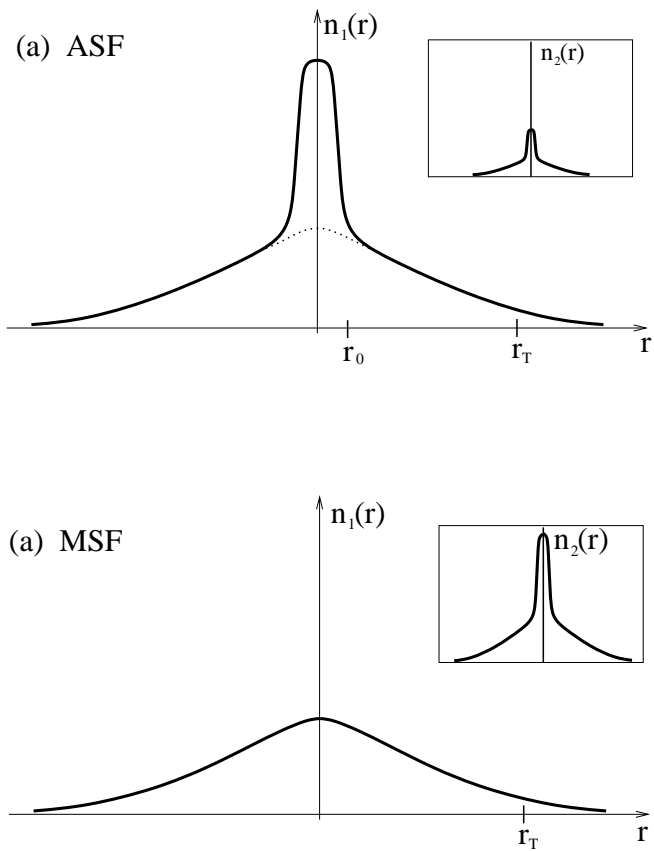}

\caption{Atomic density profiles, $n_1(r)$ in (a) the ASF and (b)
the MSF phases.  These are distinguished by the presence and absence
of atomic BEC peak, respectively. Each of these superfluid phases is
distinguished from the ``normal'' (thermal) state by the BEC peak in
the molecular density profile, $n_2(r)$, illustrated in insets. In
the dilute limit, the width $r_{0\sigma}$ ($\sigma = 1,2$) of the
BEC peak (set by the single-particle Gaussian ground state
wavefunction), and the extent $r_{T\sigma}$ of the thermal part of
the atomic cloud, are given by Eqs.\ (\ref{rs0}) and (\ref{rsT}),
respectively.}

\label{densityProfiles}
\end{figure}

In a trapped, dilute atomic gas the existence of these two
qualitatively distinct superfluid phases should be most directly
detectable through independent images of atomic and molecular
density profiles. As illustrated in Fig.\ \ref{densityProfiles}(a),
the {\em atomic} component should exhibit a BEC peak in the ASF
phase, that is absent in the MSF phase, shown in Fig.\
\ref{densityProfiles}(b). Both superfluid phases are distinguished
from the normal state by the BEC peak in the {\em molecular} density
profile, as illustrated in the insets to these figures.

\begin{figure}[bth]

\includegraphics[width=\columnwidth]{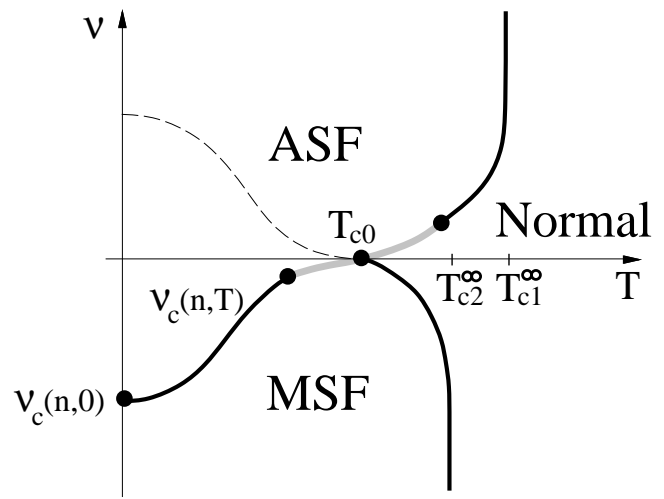}

\caption{The phase diagram (at fixed total density, $n$) for a
uniform condensate as a function of Feshbach resonance detuning
$\nu$ and temperature $T$. A curve of critical detuning $\nu_c(n,T)$
separates the atomic (ASF) and molecular (MSF) superfluid phases by
a phase transition, which is continuous between the ($T=0$) quantum
critical point $\nu_c(n,0)$ and a tricritical point $TC_1$. The
section of the critical curve (gray) between the two tricritical
points $TC_1$ and $TC_2$ denotes a first order transition boundary,
that terminates the continuous MSF--N phase boundary at a critical
end point $T_{c0}$, where three phases meet. The critical
temperatures $T^\infty_{c\s}$, $\s=1,2$, correspond to the far
detuned limits, $\nu/k_B T\to \pm \infty$.  The dashed curve inside
the ASF phase corresponds to a crossover line, $\nu_\times(T)$, at
which the molecules would condense on their own if there were no
Feshbach resonance coupling them to the atoms.}

\label{phasediagramNuT}
\end{figure}

Because of its paired nature, a complementary distinguishing
characteristic of a MSF are deconfined $\pi-$ (half-) vortices,
topological defects that, in contrast, are linearly confined in the
ASF state.  Consequently, as illustrated in Fig.\
\ref{phasediagramNuT}, a thermodynamically sharp quantum phase
transition, at an intermediate critical Feshbach resonance detuning
$\nu_c$, must separate the MSF and ASF phases. Each in turn is also
separated by a finite-temperature transition from the ``normal'' (N)
state lacking any order (i.e., breaking no symmetries).

Experimental observations of these and associated predictions have
so far been precluded by a short lifetime of the vibrationally hot
molecular state.\cite{vibration} The latter is believed to be
limited by 3-body recombination and strongly enhanced atom-molecule
scattering near the resonance. In contrast to Fermi
systems,\cite{J03,GRJ04,Z04} where Pauli exclusion greatly extends
the molecular lifetime for a positive scattering length and
stabilizes the Fermi-sea for negative scattering lengths by
suppressing multi-body collisions,\cite{PSS04} the resonantly
interacting bosonic atomic gas is observed to be highly unstable in
the negative two-body scattering length regime.\cite{D01,BM05}
Viable proposals for surmounting these problems are currently being
investigated. These include use of an adiabatic ramp of the detuning
through resonance,\cite{G03} or a two-photon Raman transition to
transfer the Feshbach molecular states to a lower lying vibrational
state.\cite{SM04}

Although no direct evidence for an equilibrium Bose molecular
condensate exists, observed resonant atomic loss in a stimulated
Raman transition in $^{87}$Rb (Ref.\ \onlinecite{Heinzen00}) and
time domain density oscillations in $^{85}$Rb (Ref.\
\onlinecite{DCTW02}) are consistent with a coherent transfer of
population from free bosonic atoms to diatomic
molecules.\cite{KH02,R98} It is not at the moment clear (at least to
the present authors) whether current experimental difficulties of
stabilizing bosonic atom-molecule mixtures near a FR are fundamental
or technical and system specific. One possible fundamental source of
instability in bosonic atom systems is the existence of Efimov bound
states of bosonic atom triplets.\cite{E71,GurarieDiscuss} At least
at a theoretical model level these can be suppressed by a
sufficiently strong three-body repulsion. Even in the unfavorable
scenario, where such phases of bosons near a FR are indeed
metastable (as most states of degenerate atoms ultimately are) one
expects that ideas discussed here should be important on
sufficiently short time scales and for understanding of the
associated nonequilibrium dynamics.

The rest of the paper is organized as follows. The Introduction is
concluded with a summary of the main results and their experimental
implications. In Section \ref{sec:2channelmodel} a microscopic
two-channel model, that is believe to accurately describe
resonantly-interacting atomic bose gas is introduced. The model is
first used to compute the two-body s-wave scattering, showing that
it correctly captures the Feshbach resonance phenomenology. Matching
the computed scattering amplitude to its measured counterpart allows
one to relate parameters appearing in the Hamiltonian to
experimental observables. In Section \ref{sec:symmetries} a general
symmetry-based discussion of the expected phases and associated
phase transitions in this system is presented. In Section
\ref{sec:mft}, by minimizing the corresponding imaginary-time
coherent state action, the generic mean field phase diagram for the
system is mapped out. In Sections \ref{diluteBEClimit} and
\ref{sec:excite}, this Landau analysis is supplemented by detailed
microscopic calculations of phase boundaries, spectra, condensate
depletion and superfluid density for a dilute, weakly-interacting
gas. The asymptotic nature of the ASF--MSF phase transition is
discussed in Section \ref{ASF_MSFtransition}. In Section
\ref{sec:BoseBCS} the mean field and perturbative analyses,
performed within a two-channel model, are supplemented with a
variational theory of a one-channel model. The latter is a better
description of bosons in which the paired state is absent (i.e.,
there is no long lived metastable paired state with distinct
internal quantum numbers) once the two-body attraction becomes too
weak to bind atom pairs (which includes, of course, the more
familiar regime of two-body repulsion). In Section \ref{sec:vortex}
topological defects, vortices and domain walls, in the ASF are
studied, and the ASF--MSF and SF-to-normal fluid transitions are
characterized in terms of a proliferation of these topological
defects. The paper is concluded in Section \ref{sec:summary}.

\subsection{Summary of results}
\label{sec:results}

In this paper a considerable elaboration and extension of
predictions reported in a recent Letter\cite{RPW04} are presented.
The primary results are summarized by the density profiles in Fig.\
\ref{densityProfiles} and the phase diagrams in Figs.\
\ref{phasediagramNuT} and \ref{phase_diagramAlphaMuA},
characterizing the phases and phase transitions of a resonant Bose
gas. As illustrated there, it is found that Feshbach-resonantly
interacting atomic Bose gas, in addition to the normal state
exhibits {\em two distinct} low-temperature superfluid states. The
first, appearing at positive detuning, is the more conventional
atomic superfluid, characterized by coexisting atomic and molecular
BEC and their associated ODLROs, with finite order parameters
$\Psi_{10}$ and $\Psi_{20}$, respectively. The other, more exotic,
MSF state, appearing at low temperature and negative detuning, is
characterized by superfluidity of diatomic molecules, with a finite
molecular condensate order parameter $\Psi_{20}$. It is
distinguished from the ASF by the absence of atomic ODLRO, i.e.,
inside the MSF phase $\Psi_{10} = 0$.

As illustrated in detail in Sec.\ \ref{sec:alphaneq0}, a finite
$\Psi_{10}$ always implies a finite $\Psi_{20}$. In the presence of an
atomic condensate, $\Psi_{10}$, the Feshbach resonance coupling allows
a scattering of two Bose-condensed atoms out of the atomic BEC into
the molecular BEC (i.e., ASF is really a superposition of
Bose-condensed open-channel atoms and Bose-condensed closed-channel
molecules) and therefore acts like an ordering ``field'' on the
molecular order parameter. This implies that a state in which atoms
are condensed but molecules are not is forbidden by general symmetry
principles.\cite{decoupledBEC}

As noted above, a vivid signature of two distinct superfluid orders
should be detectable via time-of-flight shadow images. In the dilute
regime (described by a BEC approximation), the resulting images are
schematically illustrated in Fig.\ \ref{densityProfiles}. At higher
densities, where a local density approximation is more appropriate,
it is expected that for a range of atom number and detuning, phase
boundaries as a function of chemical potential in the bulk system
(see e.g., Fig.\ \ref{phase_diagramAlphaMuA}) will translate into
shell structure\cite{shellsMott1,shellsMott2,shellsMott3} which
should also be observable experimentally in time-of-flight shadow
images.

\begin{figure}[tbp]

\centerline{\includegraphics[width=\columnwidth]{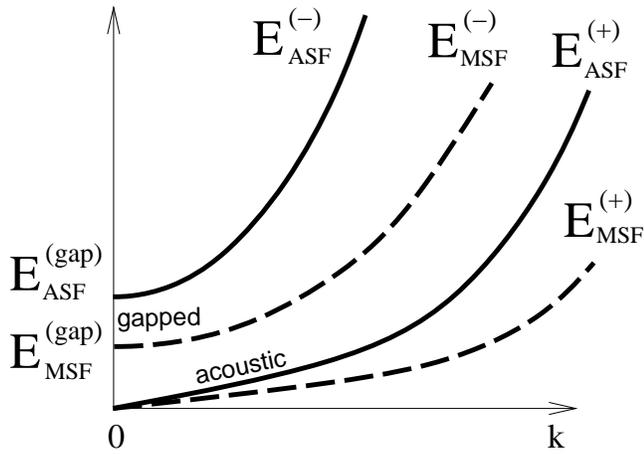}}

\caption{Schematic low-energy excitation spectra characterizing ASF
and MSF phases. In the ASF the acoustic and gapped branches of
excitations correspond to in-phase and out-of-phase fluctuations of
atomic and molecular condensates, respectively. In the MSF state the
acoustic branch is the standard gapless Bogoliubov mode. The gapped
branch corresponds to atom-like quasiparticle excitations that are
squeezed by the presence of the molecular condensate. At the
critical detuning $\nu_c$, the gap closes, signaling a quantum
MSF--ASF phase transition.}

\label{fig:spectrum}
\end{figure}

As for any neutral superfluid, ASF and MSF are each characterized by
an acoustic (Bogoliubov) ``sound'' mode, illustrated in Fig.\
\ref{fig:spectrum}, corresponding to long wavelength condensate
phase fluctuations, with long wavelength dispersions
\begin{equation}
E^+_\sigma({\bf k}) \approx c_\sigma \hbar k,
\end{equation}
where $c_\sigma$ (with $\sigma =$ ASF or MSF, or equivalently 1 or 2)
are the associated sound speeds with $c_{\rm MSF}$ given by
(\ref{4.27}) and $c_{\rm ASF}$ given by (\ref{4.44}) in terms of the
interaction parameters of the model (see Sec.\ \ref{subsec:modes}).

In the ASF state the gapless mode corresponds to in-phase
fluctuations of the atomic and molecular condensates. However, as
illustrated in Fig.\ \ref{fig:spectrum}, in contrast to ordinary
superfluids, ASF and MSF also each exhibit a gapped branch of
excitations,
\begin{equation}
E^-_\sigma(k) \approx E^{\rm gap}_\sigma + b_\sigma k^2,
\end{equation}
with gaps $E^{\rm gap}_\sigma$ given explicitly by (\ref{msfgap})
and (\ref{gapsASF}), while the quadratic corrections $b_{\sigma}$
may be inferred from the general forms (\ref{EkMSF}) and
(\ref{EkASF}) of the spectra. In the ASF the gap is controlled by
out-of-phase fluctuations of the atomic and molecular condensates,
and is set by the Feshbach coupling $\alpha$.

In the MSF, gapped excitations are single atom-like quasiparticles
akin to Bogoliubov excitations in the paired BCS state, that however
do not carry a definite atom number. These single-particle excitations
are ``squeezed'' by the presence of the molecular condensate, offering
a mechanism to realize atomic squeezed states.\cite{laser_analogy} We
expect that these gapped, atomic quantum fluctuations associated with
the presence of the molecular condensate can be measured by
interference experiments, similar to those reported in Ref.\
\onlinecite{Kasevich01}. As detailed below, the low-energy nature of
these excitations is guaranteed by the vanishing of the gap at the
MSF--ASF transition, $\nu_c$, with $E^{\rm gap}_{\rm MSF}(\nu_c) = 0$.

In the dilute weakly interacting limit appropriate to atomic gases,
the ASF--N and MSF--N transition temperatures, $T_{c1}(\nu)$ and
$T_{c2}(\nu)$, respectively, for a three-dimensional (3d) bulk uniform
system are well approximated by
\begin{equation}
T_{c\sigma}(\nu)\approx
\begin{cases}
T_{c0} \left[1 + a_\sigma \left(\frac{|\nu|}{k_B T_{c0}}
\right)^{\frac{1}{2}} \right],
& |\nu| \ll k_B T_{c0} \\
T_{c\sigma}^\infty = b_\sigma T_{c0},
& |\nu| \gg k_B T_{c0},
\end{cases}
\label{Tc}
\end{equation}
with $a_1 = 2^{9/2}\pi^{1/2}/3 c\zeta(3/2)$, $a_2 = a_1/8$, $b_1 =
c^{2/3}$, $b_2 = 2^{-5/3} c^{2/3}$, and $c = 1 + 2^{5/2}$. One sees
that $T_{c1} > T_{c2}$, with the asymptotic ratio $b_1/b_2 =
2^{5/3}$ set by the mass and boson number that both differ by a
factor of two between the two phases. When interactions are
included, for $\nu \neq 0$ the asymptotic nature of these thermal
transitions is in the well-studied classical 3d XY universality
class.

As illustrated in Fig.\ \ref{phasediagramNuT} (see also Figs.\
\ref{phase_diagramAlphaMuA} and \ref{phase_diagramAlphaMuB}), in the
vicinity of the critical endpoint $\nu=0$, $T=T_{c0}$, where the
three phases, N, MSF, ASF meet, a coupling of the molecular and
atomic superfluid order parameters converts a section (between the
tricritical points TC$_1$ and TC$_2$) of the (otherwise) continuous
N--ASF and MSF--ASF transitions to first order.\cite{BM05}  The
resulting crossing point at $T_{c0}$, that terminates a continuous
N--MSF transition is a critical endpoint (CEP).  In the dilute
limit, the CEP temperature is given by
\begin{equation}
T_{c0} \approx \frac{h^2}{2\pi m_1 k_B}
\left[\frac{n}{c\zeta(3/2)}\right]^{2/3},
\end{equation}
As illustrated in Secs.\ \ref{sec:mft} and \ref{ASF_MSFtransition},
the appearance of a first order transition (even in mean field
theory) near N--MSF continuous boundary, is a generic feature
resulting from the coupling of the ASF order parameter to the
critical MSF order parameter.\cite{RPW04}

The corresponding transition temperatures in a {\em trap} (here
distinguished from bulk quantities by a tilde) are also easily
computed and in 3d are given by
\begin{equation}
\tilde{T}_{c\sigma}(\nu)\approx
\begin{cases}
\T_{c0} \left[1 + a_\sigma \frac{|\nu|}{k_B \T_{c0}}\right],
& |\nu| \ll k_B \T_{c0} \\
\T_{c\sigma}^\infty\left[1-b_\sigma
e^{-|\nu|/\sigma\T_{c\sigma}^\infty}\right],
& |\nu| \gg k_B \T_{c0},
\end{cases}
\label{eq:trapTc}
\end{equation}
with $a_\sigma = 2\zeta(2)/9\sigma^2\zeta(3)$, and $b_\sigma =
2/3\sigma^2\zeta(3)$.  The transition temperatures in the limit of
asymptotically large positive ($\sigma=1$) and negative ($\sigma=2$)
detuning ($|\nu|/k_BT_{c\sigma}\gg 1$), and at the tricritical point
($\nu=0$), are given, respectively, by
\begin{eqnarray}
\tilde{T}_{c\sigma}^\infty
&=&\hbar\wt\left[\frac{N}{\sigma\zeta(3)}\right]^{1/3},
\label{eq:trapTcoo} \\
\tilde{T}_{c0}
&=&\hbar\wt\left[\frac{N}{3\zeta(3)}\right]^{1/3},
\label{eq:trapTc0}
\end{eqnarray}
where $\omega_0$ is the trap frequency. Comparing the first lines of
(\ref{Tc}) and (\ref{eq:trapTc}), note that the latter is now
approached {\em linearly} with, rather than as the square-root of the
reduced detuning from either side.

In the dilute limit the thermodynamics is also easily worked out. In
the 3d bulk system, the condensate densities for the atomic and
molecular BEC are given, respectively, by
\begin{widetext}
\begin{eqnarray}
n_{10}(T,\nu) &=& n
\left[1-\left(\frac{T}{T_{c1}}\right)^{3/2}
\frac{\zeta(3/2)+ 2^{5/2} g_{3/2}(e^{-\nu/k_B T})}
{\zeta(3/2)+2^{5/2}g_{3/2}(e^{-\nu/k_B T_{c1}})}\right],
\ \ \nu > 0,\ \ T < T_{c1}(\nu),
\label{n10result} \\
n_{20}(T,\nu) &=& \frac{1}{2}n
\left[1-\left(\frac{T}{T_{c2}}\right)^{3/2}
\frac{2^{5/2}\zeta(3/2)+g_{3/2}(e^{\nu/2k_B T})}
{2^{5/2}\zeta(3/2)+g_{3/2}(e^{\nu/2k_B T_{c2}})}\right],
\ \ \nu < 0,\ \ T < T_{c2}(\nu),
\label{n20result}
\end{eqnarray}
\end{widetext}
where $\zeta(3/2) \simeq 2.612$ and $g_{\alpha}(x) =
\sum_{n=1}^\infty x^n/n^\alpha$ is the extended zeta
function.\cite{GR}

As illustrated in Fig.\ \ref{phasediagramNuT} the MSF--ASF
transition takes place at a critical value of detuning $\nu_c(T,n)$
determined by the strength of atomic and molecular interactions,
shifting it away from its noninteracting value of $0$.  At zero
temperature this is a continuous quantum phase transition that for a
$d$-dimensional system is in the $(d+1)$-dimensional classical Ising
universality class\cite{commentFirstOrder,FB97,LL04} with
\begin{equation}
\nu_c(0,n) \approx -(g_2/2-g_{12})n - 2\alpha\sqrt{2n},
\label{nuc}
\end{equation}
where $g_1$, $g_{12}$, and $g_2$ are, respectively, the atom-atom,
atom-molecule and molecule-molecule interaction strengths, related
in the standard way to the corresponding scattering
lengths,\cite{GR07} and $\alpha$ is the Feshbach resonance coupling.
The transition at $\nu_c$ is characterized, upon approach from the
MSF side, by the vanishing of the single-atom excitation gap $E_{\rm
MSF}^{\rm gap}(\nu)$, and, upon approach from the ASF side, by the
disappearance of the atomic condensate $n_{10}(\nu)$. At zero
temperature, in the critical region these are predicted to vanish
according to
\begin{equation}
n_{10}(0,\nu) \sim |\nu - \nu_c|^{2\beta_I},\ \
E^{\rm gap}_{\rm MSF}(0,\nu) \sim |\nu - \nu_c|^{\nu_I},
\end{equation}
where $\beta_I$ and $\nu_I$ are, respectively, the order parameter and
correlation length exponents for the $(d+1)$-dimensional Ising
model. One may hope that when long-lived molecular condensates are
produced, nontrivial behavior of $E^{\rm gap}_{\rm MSF}(\nu)$ and the
full excitation spectra, $E^\pm_\sigma(k)$ may be observed in Ramsey
fringes\cite{DCTW02} and in Bragg and RF spectroscopy experiments
\cite{BraggKetterle,RFChin,RFKetterle,BraggBruunBaym}.

At finite temperature, away from the critical endpoint $T_{c0}$ the
transition is in the classical $d$-dimensional Ising universality
class. Scaling, together with the relevance (in the renormalization
group sense) of $T$ at the quantum critical point, also implies a {\em
universal} shape of the low $T$ part of the MSF--ASF phase boundary
\begin{equation}
\nu_c(n,T) \sim \nu_c(n,0) + a\, T^{1/\nu_I},
\end{equation}
illustrated in Fig.\ \ref{phasediagramNuT}.

The dashed curve, $\nu_\times(T)$ inside the ASF phase of the phase
diagram (Figs.\ \ref{phasediagramNuT} and \ref{phase_diagramAlphaMuA})
denotes a crossover (that becomes sharp with a vanishing Feshbach
resonance coupling $\alpha$) between ASF regimes with low and high
values of the molecular condensate $n_{20}$.  In the absence of the
coupling, the molecules would condense on their own for $\nu <
\nu_c(T)$.  For small $\alpha$ the weak symmetry breaking field
generated by the atomic condensate smears this transition into a
ASF-AMSF crossover, and leads to small, but finite, $n_{20}$ even for
$\nu > \nu_c(T)$.

\begin{figure}[tbp]

\includegraphics[angle=0,width=\columnwidth]{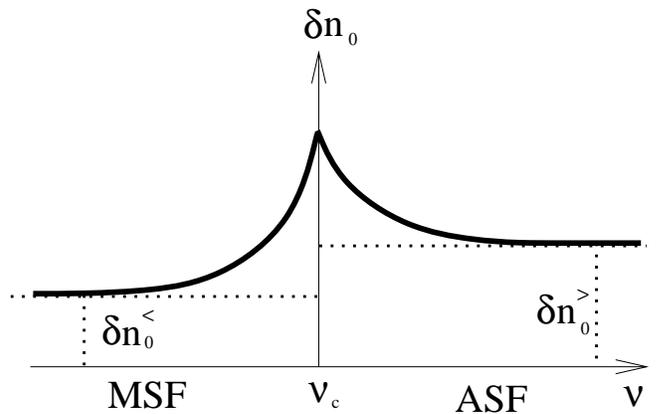}

\caption{Schematic of the zero-temperature depletion $\delta
n_0(0,\nu)$ as a function of detuning. The far detuned limits are
given by $\delta n_0^{>,<} = \s (8/3\sqrt{3})(a_\s n_{\s0})^{3/2}$,
where $>,<$ correspond to $\s=1,2$ respectively. The cusp peak is
given by (\ref{max_depletion}), displaying the power-law form
$|\nu-\nu_c|^p$, with $p=1$ in $d=3$.}

\label{fig:depletion}
\end{figure}

As for any superfluid, the ASF and MSF phases also exhibit
interaction-driven condensate depletion $\delta n_0^\sigma \equiv
n-n_{10}-2n_{20}$, quantifying the fact that, even at $T = 0$, not
all atoms are in the condensate. At $T = 0$ these are computed in
Sec.\ \ref{sec:td}. An interesting feature, illustrated in
Fig.\ \ref{fig:depletion}, is that $\delta n_0^\sigma(\nu)$ exhibit a
cusp maximum at $\nu_c$,
\begin{equation}
\delta n^\sigma_0(\nu) = \delta n_0(\nu_c)-c_\sigma|\nu-\nu_c|^p,
\label{cusp_depletion}
\end{equation}
associated with enhanced role of quantum fluctuations at the MSF--ASF
transition. The maximum depletion is given by
\begin{equation}
\delta n_0(\nu_c) \approx \frac{16}{3\sqrt{\pi}} \left(n_{2 0} a_2
\right)^{3/2} + \frac{1}{3 \pi^2} \left(\frac{m_1 \alpha \sqrt{n_{2
0}}}{\hbar^2} \right)^{3/2}
\label{max_depletion}
\end{equation}
where $a_2$ is the molecule-molecule s-wave scattering length.
Outside the critical region one expects $p=1$, crossing over to
$p=1-\alpha_I$ inside it, where $\alpha_I$ is the
$(d+1)$-dimensional Ising specific heat exponent.

\begin{figure}[bth]

\includegraphics[width=\columnwidth]{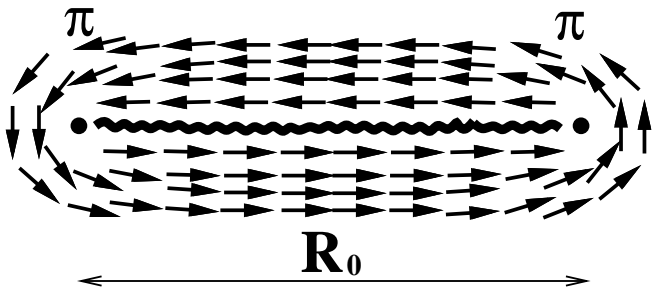}

\caption{$2\pi$ atomic condensate vortex in the ASF splits into a
$\pi+\pi$ vortex pair connected by a ``normal'' domain wall, whose
length $R_0$ increases as the FR coupling $\alpha$ becomes weaker.}

\label{vortex_pi}
\end{figure}

Another important qualitative distinction between the ASF and MSF
phases is the nature of their topological excitations, namely
vortices. The paired nature of the MSF allows for $\pi-$ (half-)
vortices (as in a BCS superconductor), while the ASF, being (from a
symmetry point of view) a standard ``charge-one'' superfluid, admits
only standard $2\pi$-vortices. However, pairing correlations present
in the ASF lead to an interesting $\pi$-vortex experimental
signature even in the atomic superfluid. Thus, in the ASF phase a
seemingly standard $2\pi$-vortex in an atomic condensate, $\Psi_{10}
> 0$, will generically split into two $\pi$-vortices (see Fig.\
\ref{vortex_pi}) confined by a domain wall of length
\begin{equation}
R_0 \approx \frac{\pi\hbar}{\sqrt{m\alpha n_{20}^{1/2}}}
\sqrt{1+2\frac{n_{20}}{n_{10}}},
\label{R0result}
\end{equation}
that diverges as the ASF--MSF phase boundary is approached from the
ASF side. Sufficiently close to the transition, it is expected to
track the associated correlation length.

This confinement arises because in the large Feshbach coupling limit a
$2\pi$-vortex in the atomic condensate induces a $4\pi$-vortex in the
molecular condensate. Such a double molecular vortex is unstable to
two fundamental $2\pi$ molecular vortices that, in 2d, repel
logarithmically, but are confined linearly inside the ASF phase.  This
provides a complementary formulation of the ASF--MSF transition as a
confinement-deconfinement transition of $\pi$ (half) atomic vortices.


\section{Two-channel Feshbach resonant model}
\label{sec:2channelmodel}

\begin{figure}[bth]

\includegraphics[width=\columnwidth]{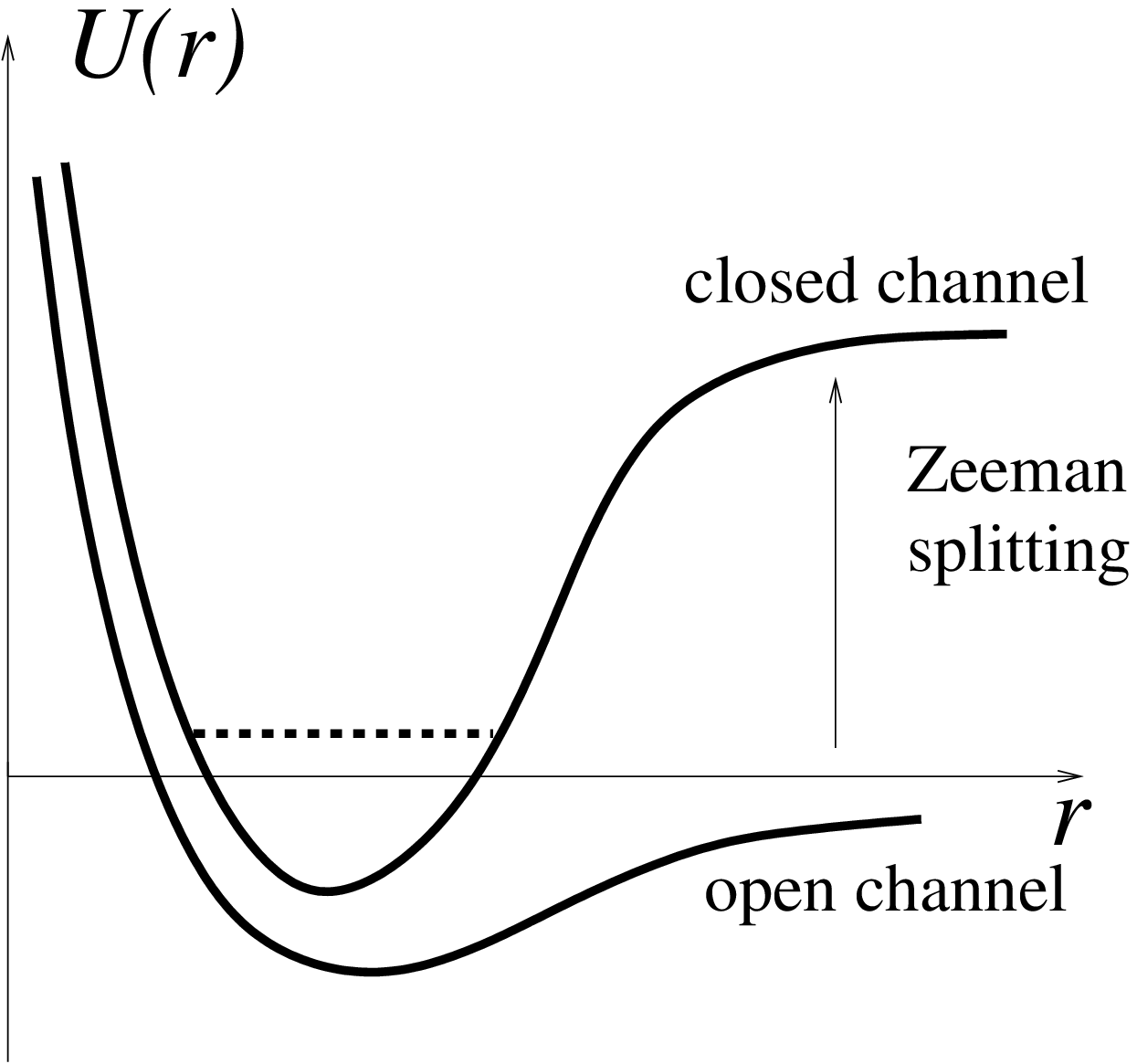}
\includegraphics[width=\columnwidth]{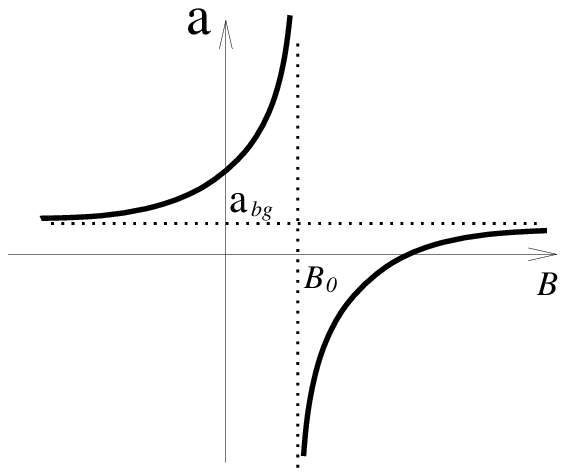}

\caption{\textbf{Above:} Schematic illustration of a Feshbach
resonance, modeled by two coupled channel interaction potentials
(distinguished by two-atom electronic spin states) as a function of
inter-particle separation.  The (so-called) ``open'' channel is too
shallow to support a bound state, while the other, ``closed''
channel supports a bound state or a resonance (indicated by a dashed
line) that is tuned with a magnetic field via the Zeeman splitting
between the two channels. \textbf{Below:} At resonance $\nu=0$, when
a bound state first appears, the s-wave atomic scattering length
diverges according to (\ref{eq:as_exp}) or (\ref{eq:as}).}

\label{fig:feshbach}
\end{figure}

The goal here is to model a resonantly interacting atomic Bose gas of
the type realized in recent experiments\cite{DCTW02}. A resonant
interaction is the key feature special to a select class of atomic
(fermionic\cite{GRJ04,Z04} and bosonic\cite{DCTW02}) systems.  A fully
microscopic description of such resonant interactions is quite
complex, involving a full set of internal nuclear and electronic spin
degrees of freedom characterized by hyperfine states, mixed upon
scattering by the interatomic exchange interaction.  However, in the
vicinity of a resonance, two-atom scattering in the ``open'' channel
is dominated by hybridization with a two-atom molecular bound state in
the ``closed'' channel, thereby allowing one to neglect all other
off-resonant channels. As illustrated in Fig.\ \ref{fig:feshbach}, the
two channels are distinguished by the two-atom {\em electron} spins,
with the open-channel an approximate spin-triplet and closed-channel
an approximate singlet.\cite{nuclearspin} Consequently they have
different Zeeman energies, allowing the center of mass rest energy
$\nu$ of the closed-channel molecule (bound state) to be tuned,
relative to the open-channel two-atom continuum, via an external
magnetic field.  This yields an unprecedented tunability of the
effective atomic interaction strength by varying a magnetic field.

The two-channel model describing the resonant atom-molecule system
is characterized by the following grand-canonical Hamiltonian:
\begin{widetext}
\begin{equation}
\hat{H} = \int d\xv \left\{\sum_{\s=1}^2 \left[
\hat{\psi}^\dagger_\s(\xv) \hat{h}_\sigma \hat{\psi}_\s(\xv)
+ \frac{1}{2} g_\s \hat{\psi}^\dagger_\s(\xv)^2 \hat{\psi}_\s(\xv)^2
\right]
+ g_{12}\hat{\psi}^\dagger_1(\xv) \hat{\psi}^\dagger_2(\xv)
\hat{\psi}_2(\xv) \hat{\psi}_1(\xv)
- \frac{1}{2} \alpha \left[\hat{\psi}^\dagger_1(\xv)
\hat{\psi}^\dagger_1(\xv) \hat{\psi}_2(\xv) + {\rm h.c.} \right]
\right\}
\label{H2channel}
\end{equation}
\end{widetext}
where $\hat{\psi}^\dagger_\s({\xv}),\hat{\psi}_\s({\xv})$ are
bosonic creation and annihilation field operators for atoms ($\s=1$)
and molecules ($\s=2$). They are described by respective
single-particle Hamiltonians
\begin{equation}
\hat{h}_{\s} = -\frac{\hbar^2}{2m_\s} \nabla^2 + \mu_\s + V_\s(\xv),
\label{h}
\end{equation}
with atomic and molecular masses $m_1 = m$ and $m_2 = 2m$ and
effective chemical potentials $\mu_1 = \mu$ and $\mu_2 = 2\mu -
\nu_0$.  The (bare) detuning parameter $\nu_0$ is related to the
energy of a (closed-channel) molecule at rest, that can be
experimentally controlled with an external magnetic field. In the
ensemble of fixed {\em total} number of atoms $N$ (free and bound
into molecules), relevant to trapped atomic gas experiments, the
chemical potential $\mu$ is determined by the total atom number
equation
\begin{equation}
\label{Neqn}
N=\int d\xv\left[\langle\hat{\psi}^\dagger_1(\xv)
\hat{\psi}_1(\xv)\rangle
+ 2\langle\hat{\psi}^\dagger_2(\xv)
\hat{\psi}_2(\xv)\rangle\right].
\end{equation}
The positive local pseudo-potential parameters $g_1, g_2, g_{12}$
measure background (nonresonant) repulsive atom-atom, atom-molecule
and molecule-molecule interactions, respectively, and in the dilute
limit are proportional to corresponding background 2-body s-wave
scattering lengths. The Feshbach resonance coupling $\alpha$
characterizes the \emph{coherent} atom-molecule interconversion rate
(hyperfine interaction driven hybridization between open and closed
channels), encoding the fact that a molecule can decay into two
open-channel atoms\cite{laser_analogy,nuclearspin}. The external
potentials $V_\sigma(\xv)$ describe the atomic and molecular traps,
which for most of the paper will be taken to be a ``box'', modeled
(for convenience) using periodic boundary conditions.

\begin{figure}[tbp]

\centerline{\includegraphics[width=\columnwidth]{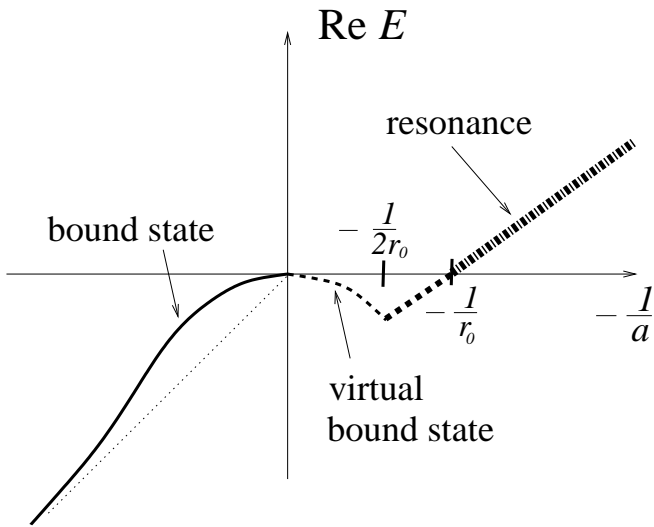}}

\caption{The real part of the pole of the scattering amplitude
$f_0(E)$, (\ref{Ep_neg}), as a function of detuning, parameterized
here by $-1/a_s = (\nu/\hbar) \sqrt{m/\Gamma_0}$, with $-1/r_0 =
\sqrt{m\Gamma_0}/(2\hbar)$. As discussed in the text, bound states
and resonances must correspond to physical solutions of the
Schr\"odinger equation with proper boundary conditions. The thin
dotted line indicates asymptotic linear behavior of the bound state
for small positive $a_s$.}

\label{fig:poleplot}
\end{figure}

In principle it is possible to obtain the above Hamiltonian
$\hat{H}$ from a microscopic analysis of atoms interacting via a
Feshbach resonance.\cite{TTHK99,GR07} However, its validity is
ultimately justified by the fact that for two atoms in a vacuum (for
which one takes $\mu=0$) it reproduces the experimentally observed
Feshbach-resonance phenomenology. Namely, it predicts an atomic
scattering resonance for positive detuning, a true molecular bound
state for negative detuning (illustrated in Fig.\
\ref{fig:poleplot}), and an s-wave scattering length of the
experimentally observed form
\begin{equation}
a_s = a_{\rm bg} \left(1-\frac{B_w}{B-B_0}\right).
\label{eq:as_exp}
\end{equation}
Here, $a_{\rm bg}$ is the background (nonresonant) scattering length,
$B_w$ is the experimental width (not to be confused with the width of
the Feshbach resonance\cite{GR07}), and $B_0$ is the value of the
magnetic field at which the Feshbach resonance is tuned to zero
energy.

These properties follow directly from the s-wave atomic scattering
amplitude $f_0(E)$ that for two atoms in a vacuum can be computed
exactly.\cite{AGR04,GR07,SRaop} Focusing for simplicity on the resonant
part of the interaction, (i.e., taking $g_{\s} = g_{12} = 0$) the
scattering amplitude is given by
\begin{equation}
f_0(E) = -\frac{\hbar}{\sqrt{m}}\frac{\sqrt{\Gamma_0}}
{E - \nu + i \sqrt{\Gamma_0 E}},
\label{eq:f0}
\end{equation}
with $\nu$ the renormalized (physical) detuning and $\Gamma_0$ a
parameter measuring the width of the resonance.  These are given by
\begin{eqnarray}
\nu &=& \nu_0 - \frac{\alpha^2 m}{\hbar^2}
\int\frac{d^3 p}{(2\pi)^3} \frac{1}{p^2},
\label{eq:nu} \\
\Gamma_0 &\equiv& \frac{\alpha^4 m^3}{16\pi^2\hbar^6}.
\label{Eq:width}
\end{eqnarray}
The latter is related to an effective range parameter
\begin{equation}
r_0 = -2\hbar/\sqrt{m\Gamma_0}.
\label{eq:r0}
\end{equation}
The integral in (\ref{eq:nu}) is implicitly cut off by the
ultraviolet (uv) scale  $\Lambda \approx 2\pi/d_0$, set by the
inverse of the size $d_0$ of the closed-channel (molecular) bound
state [below which the point interaction approximation inherent in
the Hamiltonian (\ref{H2channel}) breaks down], so that
\begin{equation}
\nu = \nu_0 - \frac{\alpha^2 m \Lambda}{2\pi^2\hbar^2}
\label{eq:nu2}
\end{equation}
relates the bare and physical detuning.

The s-wave scattering length $a_s = -f_0(0)$ is then given by
\begin{equation}
a_s = -\sqrt{\frac{\Gamma_0}{m}}\frac{\hbar}{\nu}.
\label{eq:as}
\end{equation}
Thus, to reproduce the experimentally observed scattering length
variation with magnetic field (\ref{eq:as_exp}), one fixes the
detuning to be $\nu \approx 2\mu_B(B-B_0)$, with the approximate
Bohr magneton proportionality constant set by the Zeeman energy
difference between approximate electronic spin-triplet (open) and
spin-singlet (closed) channels. Matching (\ref{eq:as}) to
(\ref{eq:as_exp}) also allows one to relate the Feshbach resonance
coupling to the ``width'' $B_w$, giving
\begin{equation}
\Gamma_0 \approx 4m \mu_B^2 a_{\rm bg}^2 B_w^2/\hbar^2.
\end{equation}

Interpretation of the scattering physics in terms of an intermediate
molecular bound or quasi-bound state follows from the poles of
$f_0(E)$, together with appropriate constraints arising from
boundary conditions on the molecular wavefunction.  From
(\ref{eq:f0}) the physical pole is given by
\begin{equation}
E_p = E_r - i\Gamma/2,
\label{Ep_neg}
\end{equation}
where $E_r = \nu-\Gamma_0/2$ and $\Gamma = \Gamma_0 (4\nu/\Gamma_0 -
1)^{1/2}$. For negative detuning, $\nu < 0$, the pole is purely real
and negative, corresponding to a bound state with energy
\begin{equation}
E_p^{-} \approx
\begin{cases}
-\nu^2/\Gamma_0, & |\nu| \ll \Gamma_0,\\
\nu, & |\nu| \gg \Gamma_0.
\end{cases}
\label{Ep_pos}
\end{equation}
and infinite lifetime. The corresponding wavevector $k_p^- = i
\sqrt{2m |E_p|}$ lies on the positive imaginary axis, and
corresponds, as required, to a wavefunction $\sim e^{-|k_p^-| r}$
that decays exponentially at infinity. As $\nu \to 0^+$ one has $E_p
\to 0^-$, and the bound state coincides with the bottom of the
continuum.

For $\nu > 0$, one would like to be able to interpret the state
created by $\hat \psi_2$ as a metastable molecule with a finite
decay time into two atoms. Such an interpretation makes sense only
if the real and imaginary parts of $E_p$ are positive and negative,
respectively, and $\mathrm{Re} E_p > |\mathrm{Im} E_p|$---$E_p$ is
called a resonance in this case, with the inequality being the
condition that a well defined resonance have a width that is
narrower than its energy.  However, as seen in Fig.\
\ref{fig:poleplot}, for a range of small positive $\nu$, rather than
moving to positive values, $E_p$ remains real and moves back to
negative values.  But this does not indicate a restored bound state
because $k_p^+ = -i \sqrt{2m |E_p|}$ now lies on the negative
imaginary axis and corresponds to a wavefunction $\sim e^{+|k_p^-|
r}$ that grows at infinity.  Thus, in this range of detuning the
pole no longer corresponds to a true bound state and is often
referred to as a virtual bound state.\cite{LandauQM}

Although for $\Gamma_0/4 < \nu < \Gamma_0/2$, one has $\mathrm{Im}
E_p < 0$, indicating a finite decay time, the real part of $E_p$
remains negative. Only for $\nu > \Gamma_0/2$ is $\mathrm{Re} E_p >
0$, and only for $\nu$ significantly larger than $\Gamma_0/2$ does
one obtain a true resonance, with a restored molecular
interpretation for the $\hat \psi_2$ field.

This behavior is summarized in Fig.\ \ref{fig:poleplot}, where
$\mathrm{Re} E_p$ is plotted as a function of detuning.  It should be
emphasized that the issue here is strictly one of physical
interpretation of the microscopic scattering states.  The model
remains well defined, and is a valid description of experiments in the
FR regime, over the full parameter range.\cite{nonmonopotential}. The
thermodynamic phases that will be derived in later sections are also
well defined for all parameters.

In addition to a gas parameters $n^{1/3} a_{\rm bg}^{(i)}$,
corresponding to {\em background} scattering lengths associated with
couplings $g_1$, $g_2$, and $g_{12}$ (that are constant in the
neighborhood of the Feshbach resonance), the two-channel model
(\ref{H2channel}) is characterized by a dimensionless parameter
\begin{equation}
\gamma \equiv \sqrt{\frac{\Gamma_0}{k_B T_{\rm BEC}}}
= \sqrt{\frac{2}{\pi}} \frac{1}{n^{1/3} |r_0|},
\label{eq:gamma}
\end{equation}
that measures the effective strength of the Feshbach-resonant
interaction relative to the kinetic energy $k_B T_\mathrm{BEC} =
(2\pi \hbar^2/m) n^{2/3}$ set by the total atomic density $n$. As
long as these are all small, i.e., the gas is dilute with respect to
background scattering lengths and the Feshbach resonance is narrow,
the description of {\em phases} (i.e., properties away from any
phase transitions) can be accurately given by a perturbative
expansion in these dimensionless interaction
parameters.\cite{AGR04,GR07,SRaop} This will be verified through
explicit calculations in Sec.\ \ref{sec:td}. Physically, this narrow
resonance limit, $\gamma \ll 1$ corresponds to molecules that are
predominantly in the closed-channel, i.e., have a long lifetime
before decaying into two free, open-channel atoms.

In the opposite, broad resonance ($\gamma \gg 1$) limit the
molecular wave function is strongly hybridized with the
open-channel, and is characterized by a high density of continuum
states above threshold. For such a system, although for negative
detuning a bound molecular state still exists, no resonance remains
for positive detuning---the physics of this regime can no longer be
interpreted in terms of populations of coexisting atoms and
(metastable) molecules. In this limit the dispersion of the bare
molecular field can be neglected and $\hat{\psi}_2$ can be
adiabatically eliminated (integrated out, ignoring the subdominant
atom-molecule and molecule-molecule density interactions) in favor
of two open-channel bosons.\cite{GR07,SRaop} The resulting
single-channel broad resonance model Hamiltonian is then given by
\bea
\hat{H}_{1-ch} &=& \int d\xv \left\{\hat \psi_1^\dagger(\xv)
\hat{h}_1\hat \psi_1(\xv)
+ \frac{1}{2} g \hat\psi_1^\dagger(\xv)^2 \hat \psi_1(\xv)^2 \right.
\nn \\
&&\left. +\ \frac{1}{6} w \hat \psi_1^\dagger(\xv)^3 \hat \psi_1(\xv)^3
\right\},
\label{Ha}
\eea
where $g$ is the effective atom-atom interaction coupling,
approximately given by $g = g_1-\alpha^2/\nu$, and a stabilizing
(against collapse) three-body interaction has been added, with a
coupling $w > 0$. In contrast to the $\gamma \ll 1$ two-channel
model, due to the divergence of $g$, the one-channel model is
strongly interacting when the resonance is tuned to low energy, $\nu
\approx 0$, and the s-wave scattering length exceeds the
inter-particle spacing, i.e., $n a^3_s \gg 1$. Consequently, in this
regime predictions derived from a perturbative analysis of the
one-channel model can only be qualitatively trustworthy. Moreover,
as $|g|$ and $w$ increase, quantitative predictive power would
require one to include higher than three-body interactions, so
(\ref{Ha}) really only makes sense when $|g|$ is not too large.

Considered as a function of $\mu_1$ and $g$, the single channel
model (\ref{Ha}) also exhibits the same three N, MSF and ASF phases
(with MSF requiring $g < 0$). So long as $|g|$ is not too large
(i.e., so long as the associated scattering length obeys $n |a_s|^3
\ll 1$), a variational BCS approach can be used to accurately
compute the thermodynamics. This approach is discussed in Sec.\
\ref{sec:BoseBCS}.

With the model defined by $\hat H$,\cite{RGpart} Eq.\
(\ref{H2channel}), the thermodynamics as a function of a chemical
potential $\mu$ (or equivalently atom density, $n$), detuning $\nu$
and temperature $T$ can be worked out in a standard way by computing
the partition function $Z = \Tr[e^{-\beta\hat{H}}]$ ($\beta \equiv
1/k_B T$) and the corresponding free energy $F = -k_B T\ln Z$.  The
trace over quantum mechanical states can be conveniently
reformulated in terms of an imaginary-time ($\tau$) functional
integral over coherent-state atomic ($\s=1$) and molecular ($\s=2$)
fields $\psi_\s(\xv,\tau)$,
\begin{equation}
Z = \int  D\bar\psi_\s D\psi_\s \, e^{-S/\hbar},
\label{Z}
\end{equation}
where the imaginary-time action is given by\cite{NegOrl}
\begin{equation}
S = \int_0^{\beta\hbar} d\tau \int d{\xv} \left[\sum_{\s=1}^2
\psi^*_\s\hbar \partial_\tau \psi_\s + H(\psi^*_\s,\psi_\s) \right].
\label{S}
\end{equation}
The total atom number constraint (\ref{Neqn}) that allows one to
eliminate $\mu$ in favor of $N$ is then simply given by
\begin{equation}
N = -\frac{\partial F}{\partial\mu}.
\end{equation}

\section{Symmetries, Phases and Phase Transitions}
\label{sec:symmetries}

Before delving into detailed calculations it is instructive to first
consider the general symmetry-based characterization of phases and
transitions between them. Since both atoms ($\hat{\psi}_1$) and
molecules ($\hat{\psi}_2$) are bosonic and can therefore Bose
condense, the system's thermodynamics is determined by two
Bose-condensate order parameters, $\Psi_{10}$ and $\Psi_{20}$,
respectively. As usual, microscopically, these label thermodynamic
averages of the corresponding field operators, or, for weakly
interacting system, equivalently, are single-particle wavefunctions
into which all bosons (atoms and molecules, respectively)
Bose-condense.

The condensate fields $\Psi_{10}$, $\Psi_{20}$ are legitimate order
parameters that uniquely characterize the nature of the possible
phases. Naively one would expect four phases: (i) Normal (N)
($\Psi_{10} = \Psi_{20} = 0$), (ii) ($\Psi_{10} \neq 0$, $\Psi_{20}
= 0$), (iii) MSF ($\Psi_{10} = 0$, $\Psi_{20} \neq 0$), and (iv)
AMSF ($\Psi_{10} \neq 0$, $\Psi_{20} \neq 0$), corresponding to four
different combinations of vanishing and finite order parameters.
However, a finite Feshbach resonance interaction {\em explicitly}
breaks $U(1) \times U(1)$ symmetry of the $\alpha=0$ Hamiltonian
down to $U(1) \times \mathbb{Z}_2$.  Physically, this corresponds to
the fact that only the {\em total} number of atoms
\bea
N &=& N_1 + 2N_2,
\label{Nconstraint}
\\
&=& (\langle\hat{\psi}_1^\dagger \hat{\psi}_1\rangle +
2\langle\hat{\psi}_2^\dagger \hat{\psi}_2\rangle) V
\nn
\eea
($V$ is the system volume) is conserved in the presence of Feshbach
resonant scattering (break up) of a molecule into two atoms, rather
than a separately conserved number of atoms, $N_1$ and a number of
molecules, $N_2$. This is why only a single chemical potential $\mu$
is introduced in $\hat{H}$, Eq.\ (\ref{H2channel}). Consequently, a
Feshbach resonant interaction requires a condensation of molecules
(ordering of $\hat{\psi}_2$) whenever atoms are Bose-condensed
($\hat{\psi}_1$ is ordered) and thereby forbids the existence of the
state ($\Psi_{10} \neq 0$, $\Psi_{20} = 0$). As a result, the system
of resonantly interacting bosonic atoms exhibits only {\em three}
distinct phases: N, MSF, and AMSF;\cite{commentOP} since the
atom-only condensate is impossible, for brevity of notation the AMSF
state will often be referred to as simply ASF, using the two names
interchangeably.

In addition to the fully ``disordered'' normal state that does not
break any symmetries, the above distinct thermodynamic phases are
associated with different ways that the $U(1) \times \mathbb{Z}_2$
symmetry is broken.\cite{commentOP}  Bose-condensation of molecules
breaks the $U(1)$ subgroup and corresponds to a N--MSF transition to
the MSF phase. Since it is characterized by ordering of a complex
scalar field, $\hat{\psi}_2$, this transition is in an
extensively-explored and well-understood XY-model universality
class\cite{ZinnJustin}. The low-energy excitations in the MSF phase
are gapless Goldstone-mode phase fluctuations of the condensate
$\Psi_{20}$, associated with the broken $U(1)$ symmetry. In addition
there are gapped excitations associated with the magnitude
fluctuations of $\Psi_{20}$.

The breaking of the remaining $\mathbb{Z}_2$ symmetry is associated
with Bose-condensation of atoms, $\hat\psi_1$, in the presence of a
molecular condensate.  As will be shown explicitly in Sec.\
\ref{ASF_MSFtransition}, the corresponding MSF--AMSF transition is
associated with the ordering of a real scalar field, and one would
therefore expect the MSF--AMSF transition to be in the well-explored
Ising universality class.  However, as will be seen, a coupling of
the scalar order parameter to the strongly-fluctuating Goldstone
mode of the MSF phase has a nontrivial effect on the Ising
transition, quite likely driving it first order sufficiently close
to the transition.\cite{FB97,LL04}

Bose-condensation of atoms (ordering of $\hat{\psi}_1$) directly
from the normal state breaks the full $U(1)\times \mathbb{Z}_2$
symmmetry and corresponds to a direct, continuous N--AMSF phase
transition. Since it is associated with the ordering of a complex
scalar field, one expects (and finds) it also to be in the
well-studied XY-model universality class, and to exhibit a single
Goldstone mode corresponding to common (locked) phase fluctuations
of the condensate fields $\Psi_{01}$ and $\Psi_{02}$.

There is an instructive isomorphism of this description, in terms of
two complex scalar order parameters $\Psi_{10}$ and $\Psi_{20}$, to
that in terms of a two-dimensional rank-1 vector (${\bf M}_0$),
together with a rank-2, traceless, symmetric tensor (${\cal Q}_0$)
order parameter. The latter description is well known in the studies
of ferroelectric nematic liquid crystals.\cite{deGennes} There,
ordering of ${\cal Q}_0$ describes the isotropic-nematic transition,
where the principal axes of mesogenic molecules align
macroscopically, breaking the 2d rotational symmetry modulo $\pi$
rotation.  The latter remains unbroken in the nematic phase. This is
isomorphic to the N--MSF transition discussed above. For polar
molecules, at lower temperature this isotropic-nematic transition
can be followed by vector ordering (of, e.g., the molecular electric
dipole moments) of ${\bf M}$.  In the non-rotationally invariant,
quadrupolar environment of the nematic phase, this corresponds to
spontaneous breaking of the remaining $\mathbb{Z}_2$ symmetry.
Therefore, the subsequent nematic-polar (ferroelectric) transition
corresponds to a spontaneous selection between two equivalent $0$
and $\pi$ orientations of ${\bf M}$ relative to the molecular
principal axes characterized by the nematic ${\bf Q}_0$ order.
Clearly, this latter transition can be identified with the MSF--AMSF
transition in the atomic system.  The mathematical mapping between
the two descriptions is elaborated on in more detail in Appendix
\ref{mapNematic}.


\section{Mean field theory}
\label{sec:mft}

The first goal is to determine the nature of the phases and
corresponding phase transitions exhibited by the resonant bosonic
atom-molecule model introduced above.  To this end one must evaluate
the free energy, which in the presence of interactions and
fluctuations can only be carried out perturbatively. However, away
from continuous phase transition boundaries, i.e., well within the
ordered phases, fluctuations are small. The functional integral in
$Z$ is then dominated by field configurations
$\psi_\s(\xv,\tau)\approx\Psi_{\s0}(\xv)$ that minimize the action
$S$, and therefore can be evaluated via a saddle-point
approximation. For time-independent solutions that characterize
thermodynamic phases, the order parameters $\Psi_{\s0}(\xv)$
equivalently minimize the variational energy functional
$H[\Psi_{\s0}]$, in which one substitutes the classical order
parameters for the field operators in (\ref{H2channel}). For the
case of a uniform bulk system with $V_\s(\xv) = 0$ (that is the
focus of this section) one expects that $H[\Psi_{\s0}]$ is minimized
by a spatially uniform solution (see however Ref.\
\onlinecite{solitonGurarie}) $\Psi_{\s0} = |\Psi_{\s0}|
e^{i\theta_\s}$. A mean field analysis then reduces to a
minimization of the energy density
\bea
\label{HmuMFT}
{\cal H}_\text{mf} &=& H[\Psi_{10},\Psi_{20}]/V
\\
&=& -\mu_1 |\Psi_{10}|^2 + \frac{g_1}{2} |\Psi_{10}|^4
-\mu_2 |\Psi_{20}|^2 + \frac{g_2}{2} |\Psi_{20}|^4
\nn \\
&& + g_{12} |\Psi_{10}|^2 |\Psi_{20}|^2 - \alpha
\mathrm{Re}[\Psi_{20}^* \Psi_{10}^2],
\nn
\eea
Total atom number conservation ensures a global $U(1)$ symmetry with
respect to uniform, $\sigma$-independent phase rotation. The
Feshbach resonance interaction
\begin{equation}
\alpha \mathrm{Re}[\Psi_{20}^* \Psi_{10}^2]
=\alpha |\Psi_{20}| |\Psi_{10}|^2 \cos(2\theta_1 - \theta_2)
\end{equation}
clearly locks atomic and molecular phases together, analogously to two
Josephson-coupled superconductors, with the energy minimized by
\begin{equation}
\theta_2 = 2 \theta_1 (\mathrm{mod}\, 2\pi).
\label{thetaLock}
\end{equation}
where without loss of generality $\alpha$ is taken to be positive; for
$\alpha < 0$, the molecular phase $\theta_2$ is simply shifted by
$\pi$.  The corresponding saddle point equations are given by
\bse
\bea
0 &=& \frac{\partial{\cal H}_\text{mf}}{\partial \Psi_{10}^*}
\label{saddlepointA} \\
&=& - \alpha \Psi_{10}^* \Psi_{20} + \Psi_{10}
\left(-\mu_1 + g_1 |\Psi_{10}|^2 + g_{12} |\Psi_{20}|^2 \right)
\nn \\
0 &=& \frac{\partial{\cal H}_\text{mf}}{\partial \Psi_{20}^*}
\label{saddlepointB} \\
&=& -\frac{1}{2} \alpha \Psi_{10}^2
+ \Psi_{20} \left(-\mu_2 + g_2 |\Psi_{20}|^2
+ g_{12} |\Psi_{10}|^2 \right).
\nn
\eea
\label{saddlepoint}
\ese
For a trapped system with a fixed total number of atoms appropriate
to atomic gas experiments, these equations must be supplemented by
the total atom number (\ref{Neqn})---given by $N/V = |\Psi_{10}|^2 +
2 |\Psi_{20}|^2$ within the mean field approximation---so as to map
out the phase diagram as a function of atom number $N$ and detuning
$\nu$. However, it is simpler to instead treat the atomic ($\mu_1$)
and molecular ($\mu_2$) chemical potentials as independent variables
and first map out the phase behavior as a function of $\mu_1$ and
$\mu_2$.

\subsection{Vanishing Feshbach resonance coupling, $\alpha=0$}
\label{sec:alphaeq0}

To complete a minimization of ${\cal H}_\text{mf}$ it is instructive
to first consider a special case of vanishing Feshbach resonance
coupling, $\alpha = 0$, for which saddle-point equations reduce to
\bse
\bea
\label{SPalpha=0a}
0 &=& \Psi_{10} \left(-\mu_1 + g_1 |\Psi_{10}|^2 + g_{12}
|\Psi_{20}|^2 \right)\\
\label{SPalpha=0b}
0 &=& \Psi_{20} \left(-\mu_2 + g_2 |\Psi_{20}|^2
+ g_{12} |\Psi_{10}|^2 \right).
\eea
\ese
In this special case the model corresponds to an easy-plane
(XY-model) limit of two energetically-coupled ferromagnets, a model
that has appeared in a broad variety of physical
contexts.\cite{ChaikinLubensky,RTtubules}

In the $\alpha = 0$ limit, the model exhibits four different phases,
corresponding to the four different combinations of zero or nonzero
order parameters $\Psi_{10}$, $\Psi_{20}$.

For $\mu_1 < 0$ and $\mu_2 < 0$, ${\cal H}_{\rm mf}$ is convex with a
unique minimum at
\begin{equation}
|\Psi_{10}|= |\Psi_{20}| = 0,\ \ \mbox{for}\ \mu_1, \mu_2 < 0,
\label{mus<0}
\end{equation}
corresponding to the normal state.

For $\mu_1 > 0$ and $\mu_2 < 0$, the minimum continuously shifts to
\begin{equation}
|\Psi_{10}| = \sqrt{\mu_1/g_1},\ |\Psi_{20}| = 0,\ \ \mbox{for}\
\mu_1 > 0,\ \mu_2 < \mu_1 \frac{g_{12}}{g_1},
\label{mu1>0}
\end{equation}
corresponding to a continuous transition at
\begin{equation}
\mu_1 = 0,\ \mu_2 < 0\  \mbox{(Normal--ASF transition line)}
\label{N-ASF}
\end{equation}
from the normal state to the atomic superfluid (ASF), where atoms are
Bose condensed but molecules are not. Substituting this solution into
the second saddle-point equation, (\ref{SPalpha=0b}), one confirms
that $\Psi_{20} = 0$ is indeed a minimum so long as $\mu_2 < \mu_1
g_{12}/g_1$.

In a complementary fashion, for $\mu_2 > 0$, $\Psi_{20}$ becomes
nonzero, while $\Psi_{10}$ continues to vanish so long as $\mu_1 <
\mu_2 g_{12}/g_2$.  Hence the normal state undergoes a transition to
the molecular superfluid (MSF) along the line
\begin{equation}
\mu_1 < 0,\ \mu_2 = 0\ \mbox{(N--MSF transition line)}.
\label{N-MSF}
\end{equation}
The MSF phase is characterized by order parameters
\begin{equation}
|\Psi_{20}| = \sqrt{\mu_2/g_2},\ |\Psi_{10}| = 0,\ \mbox{for}\
\mu_2 > 0,\ \mu_1 < \mu_2 \frac{g_{12}}{g_2},
\label{mu2>0}
\end{equation}
i.e., Bose condensed molecules but uncondensed atoms.

Along the transition line
\begin{equation}
\mu_2 = \mu_1 \frac{g_{12}}{g_1},\ \mu_1 > 0\
\mbox{(ASF--AMSF transition line)}
\label{ASF-AMSF}
\end{equation}
the ASF phase becomes unstable to development of a nonzero molecular
Bose condensate. Conversely along the line
\begin{equation}
\mu_1 = \mu_2 \frac{g_{12}}{g_2},\ \mu_2 > 0\
\mbox{(MSF--AMSF transition line)}
\label{MSF-AMSF}
\end{equation}
the MSF phase becomes unstable to a finite atomic Bose condensate. It
is clear that as long as $g_{12}^2 < g_1 g_2$ the region defined by
above two boundaries is finite and corresponds to atomic-molecular
superfluid (AMSF) in which both atoms and molecules are condensed,
with
\bea
&&|\Psi_{10}| = \sqrt{\frac{g_2 \mu_1 - g_{12} \mu_2}
{g_1g_2-g_{12}^2}},\
|\Psi_{20}| = \sqrt{\frac{g_1 \mu_2 - g_{12} \mu_1}
{g_1g_2-g_{12}^2}},
\nn \\
&&\mbox{for}\ \mu_1,\ \mu_2 > 0,\
\frac{g_{12}}{g_2} < \mu_1/\mu_2 < \frac{g_1}{g_{12}}.
\label{3.13}
\eea
Within this mean field analysis, for this range of parameters all
transitions above are second order.

\begin{figure}[tbp]

\includegraphics[width=0.9\columnwidth]{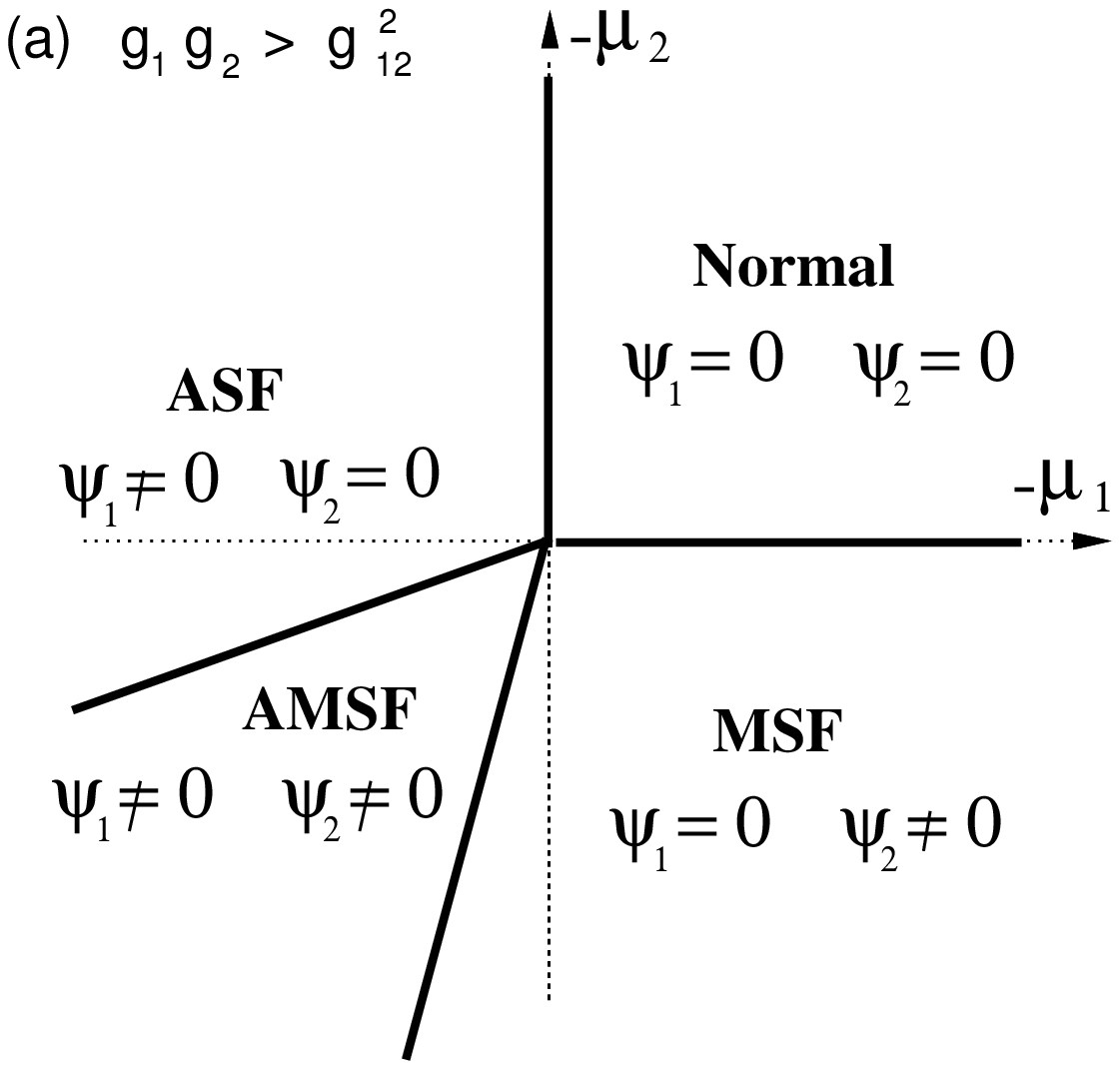}
\includegraphics[width=0.9\columnwidth]{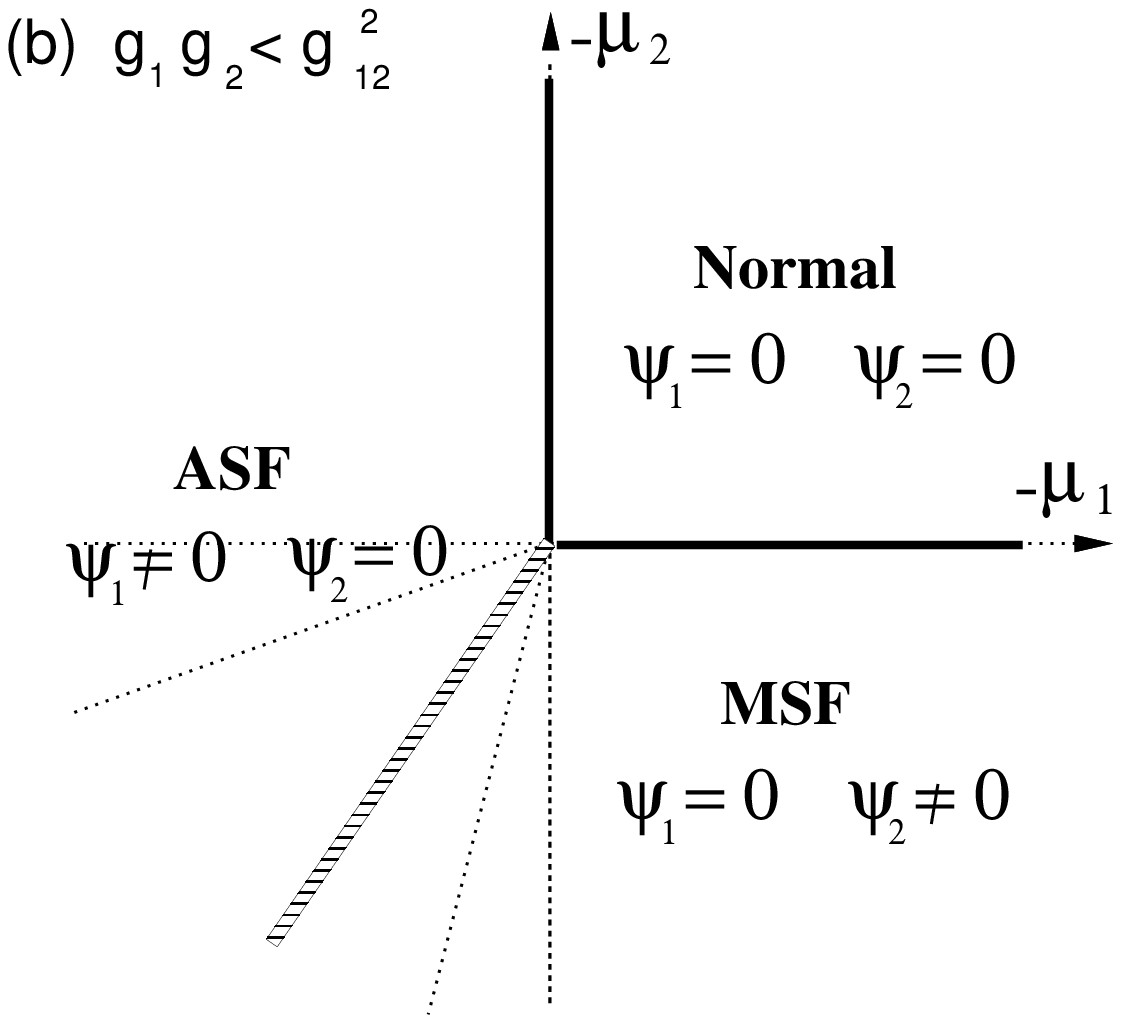}

\caption{The two possible $\mu_1$-$\mu_2$ mean field phase diagrams
for a special case of a vanishing Feshbach resonance coupling,
$\alpha = 0$. (a) For $g_1 g_2 > g_{12}^2$ there are four distinct
phases separated by four second order transition lines, meeting at a
tetracritical point. (b) For $g_1 g_2 < g_{12}^2$ there are only
three distinct phases, meeting at the bicritical point. The uniform
AMSF phase, exhibiting both atomic and molecular superfluidity, is
unstable and is replaced by a direct first order ASF--MSF transition
denoted by the hatched double line.  On this line the dominant
atom-molecule repulsion drives the system to phase separate into ASF
and MSF regions.  The other two transitions remain continuous. In
(b) the dashed lines $\mu_1 = \mu_2 g_{12}/g_2$ and $\mu_2 = \mu_1
g_{12}/g_1$ denote spinodals.}

\label{phase_diagramMu}
\end{figure}

For $g_{12}^2 > g_1 g_2$ this fourth AMSF phase is absent because
energies $-\mu_\s^2/2g_\s$ of the ASF and MSF minima cross {\em
before} either becomes locally unstable.  Consequently, instead of
continuous transition to AMSF, the system undergoes a first order
transition between the atomic and molecular superfluids at
\begin{equation}
\frac{\mu_1}{\mu_2} = \sqrt{\frac{g_1}{g_2}}, \ \
\mbox{for}\ g_{12}^2 > g_1 g_2,\ \mbox{(first order ASF--MSF line)},
\label{3.14}
\end{equation}
where the ASF and MSF minima become degenerate. In this case, the
lines $\mu_1 = \mu_2g_{12}/g_2$ and $\mu_2 = \mu_1g_{12}/g_1$ are
spinodals, beyond which the phase on the opposite side of the first
order line becomes locally, not just globally, unstable. As usual
with a first order transition, for a fixed total atom number
(relevant to trapped atomic gas experiments), along this first order
line corresponds to a coexistence region where the system phase
separates into coexisting atomic and molecular superfluids. The two
possible phase diagrams for $\alpha = 0$ are illustrated in Fig.\
\ref{phase_diagramMu}.

\subsection{Finite Feshbach resonance coupling, $\alpha \neq 0$}
\label{sec:alphaneq0}

As is clear from the Hamiltonian (\ref{H2channel}), and its mean
field form (\ref{HmuMFT}), the full system is characterized by a
finite Feshbach resonance coupling $\alpha > 0$. If $\mu_1,\mu_2 <
0$ it is easy to see from (\ref{saddlepoint}) that the normal phase
$\Psi_{10} = \Psi_{20} = 0$ remains a local minimum of
$\cal{H}_\text{mf}$, and a nonzero $\alpha$ does not affect the
boundaries of the normal phase, with (\ref{N-ASF}) and (\ref{N-MSF})
remaining valid.

\begin{figure}

\centerline{\includegraphics[width=\columnwidth]{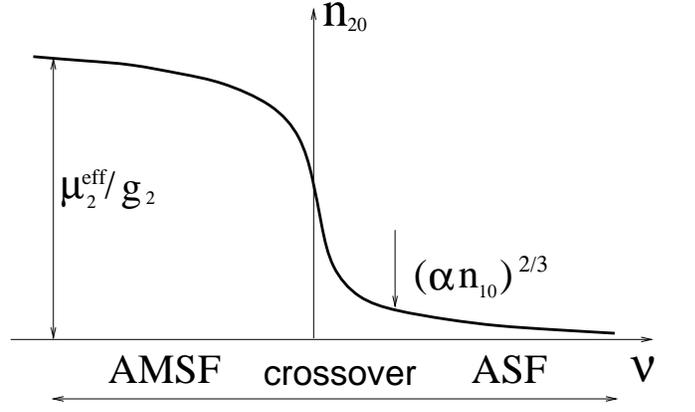}}

\caption{Schematic illustration of the crossover from small
molecular condensate fraction (in a region in which $n_{20} =
|\Psi_{20}|^2$ would vanish exactly for $\alpha = 0$) to a large
molecular BEC. The horizontal axis represents a path, parameterized
by $\mu_1$ and $\mu_2$, which intersects the crossover line in Fig.\
\ref{phase_diagramAlphaMuA}.  The estimates for the exhibited sizes
of $n_{20}$ deep in the AMSF phase, and in the vicinity of the
crossover line, follow from the discussion surrounding equation
(\ref{eq:xoverscaling}).}

\label{fig:crossover}
\end{figure}

A key physical consequence of a finite $\alpha$ is that, in a phase
where atoms are condensed, a finite Feshbach coupling [that
mathematically acts like an ordering field on the molecular
condensates; see (\ref{saddlepointB})] scatters pairs of condensed
atoms into a molecular BEC.  Equivalently, it hybridizes states of a
pair of atoms and a molecule, and as a result a finite molecular
condensate is {\em always} induced in a state where atoms are
condensed. Consequently (much like an external magnetic field
eliminates the distinction between the paramagnetic and
ferromagnetic states), as anticipated in Sec.\ \ref{sec:symmetries},
a finite $\alpha$ eliminates the ASF phase, replacing it by the
AMSF. A finite Feshbach coupling thereby converts the ASF--AMSF
transition into a {\em crossover},
\begin{equation}
\mu_2 = \mu_1 \frac{g_{12}}{g_1},\ \mu_1 > 0\
\mbox{(ASF--AMSF crossover)},
\label{ASF-AMSFcrossover}
\end{equation}
between two regimes of AMSF with low and high density of a molecular
condensate, with the quantitative distinction and crossover between
these becoming sharp in the small $\alpha$ limit. This is
illustrated in Fig.\ \ref{fig:crossover}. For simplicity of
notation, from here on, both of these regimes will be referred to as
simply ASF.

For small $\alpha$ the value of the atomic and molecular condensates
throughout the ASF phase can be estimated from the saddle point
equations (\ref{saddlepoint}). As for the case of $\alpha = 0$, for
$\mu_1 > 0$ one has $\Psi_{10} \approx \sqrt{|\mu_1|/g_1}$, which
(through the atom-molecule repulsion $g_{12}$) acts to shift the
effective molecular chemical potential to $\mu^{\rm eff}_2 = \mu_2 -
\mu_1 g_{12}/g_1$. Vanishing of $\mu^{\rm eff}_2$ defines the ASF
crossover line (\ref{ASF-AMSFcrossover}). To the left of and above
this crossover boundary, the effective molecular chemical potential
is negative and the molecular condensate is small.  It is induced to
be finite, via (\ref{saddlepointB}), only by virtue of a finite
Feshbach resonance coupling to the atomic condensate. This gives
\bea
\Psi_{20} &=& \frac{\alpha \Psi_{10}^2}
{2(-\mu_2 + g_{12} |\Psi_{10}|^2)} + O(\alpha^2)
\label{Psi2weak} \\
&\approx& \frac{\alpha \mu_1}{2(g_{12}\mu_1 - g_1 \mu_2)},
\ \ \mbox{for}\ \mu_1 > 0,\ \mu_2 < \mu_1 \frac{g_{12}}{g_1}.
\nn
\eea
On the other hand below the crossover line, $\mu_2^{\rm eff} > 0$, a
would-be spontaneous (for $\alpha=0$) molecular condensate is only
weakly modified from its $\alpha=0$ value
\bea
|\Psi_{20}|^2 &=& \frac{\mu_2^{\rm eff}}{g_2}
\label{Psi2strong} \\
&\approx& \frac{g_1\mu_2-g_{12}\mu_1}{g_1 g_2},
\ \ \mbox{for}\ \mu_1 > 0,\ \mu_2 > \mu_1 \frac{g_{12}}{g_1}.
\nn
\eea
In the intermediate regime, in the vicinity of the crossover line
itself, one expects $\Psi_{20}$ still to vanish with $\alpha$, but
more slowly than linearly.  Thus, if $|\Psi_{20}| \gg
\alpha/g_{12}$, then (\ref{saddlepointA}) becomes $\mu_1 -
g_1|\Psi_{10}|^2 \approx g_{12} |\Psi_{20}|^2$.  Substituting this
into the second term on the right hand side of (\ref{saddlepointB})
one obtains to leading order the relation
\be
\Delta = \tau |\Psi_{20}| + |\Psi_{20}|^3.
\label{eq:xovereq}
\ee
in which
\be
\Delta = \frac{\alpha \mu_1}{2(g_1 g_2-g_{12}^2)},\
\tau = \frac{g_{12}\mu_1  - g_1 \mu_2}{g_1g_2-g_{12}^2}
\label{eq:xovervars}
\ee
are the scaled Feshbach coupling and deviation from the crossover
line, respectively. The solution may be obtained in the scaling form
\begin{equation}
|\Psi_{20}| = \Delta^{1/3} X(\tau/\Delta^{2/3})
\label{eq:xoverscaling}
\end{equation}
in which the scaling function $X(x)$ is the solution to the cubic
equation
\be
1 = x X(x) + X(x)^3,
\label{eq:scalefn}
\ee
with $X(0)=1$.  Close to the crossover line, where
$|\tau|/\Delta^{2/3} = O(1)$, one sees that $\Psi_{20}$ is of order
$(\alpha |\Psi_{10}|^2)^{1/3}$.  For large positive
$\tau/\Delta^{2/3}$ one enters the linear scaling regime where
$|\Psi_{20}| \approx \Delta/\tau$, which is consistent with
(\ref{Psi2weak}) so long as the constraint $|\Psi_{20}| \gg
\alpha/g_{12}$ is obeyed. This leads to the condition $(g_{12} \mu_1
- g_1 \mu_2)/g_{12} \mu_1 \ll 1$ on the deviation from the crossover
line. The crossover behavior is sketched in Fig.\
\ref{fig:crossover}.

\begin{figure}[bth]

\includegraphics[width=\columnwidth]{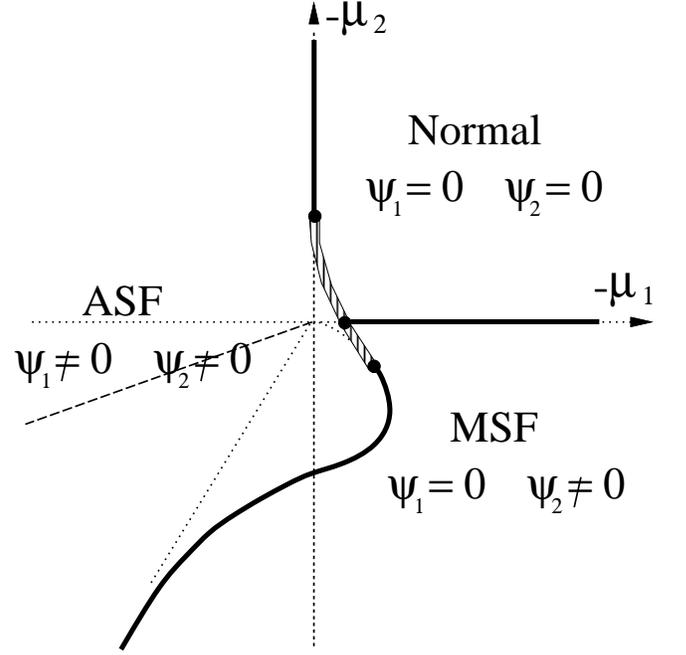}

\caption{Mean field phase diagram in the $\mu_1$-$\mu_2$ plane for a
finite Feshbach resonance coupling $\alpha \neq 0$ and $g_1 g_2 >
g_{12}^2$.  In contrast with the $\alpha = 0$ limit shown in Fig.\
\ref{phase_diagramMu}(a), a finite $\alpha$ eliminates the
distinction between the ASF and AMSF phases, converting the
ASF--AMSF transition, indicated by the dashed line, into a
crossover. Feshbach resonance scattering also strongly modifies the
MSF--ASF phase boundary, and for small chemical potentials drives
the N--ASF and MSF--ASF transitions first order (indicated by
hatched curves), with the first order section terminated by two
tricritical points. The point where the three phases meet, and the
continuous N--MSF phase boundary terminates at the first order
boundary, is a critical endpoint.}

\label{phase_diagramAlphaMuA}
\end{figure}

\begin{figure}[bth]

\includegraphics[width=\columnwidth]{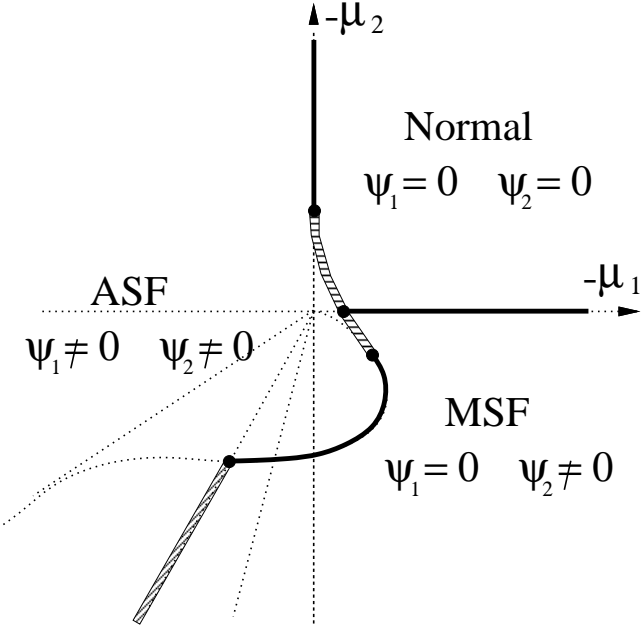}

\caption{Mean field phase diagram in the $\mu_1$-$\mu_2$ plane for a
finite Feshbach resonance coupling $\alpha \neq 0$ and $g_1 g_2 <
g_{12}^2$.  In contrast with the $\alpha = 0$ limit in Fig.\
\ref{phase_diagramMu}(b), finite $\alpha$ eliminates the distinction
between the ASF and AMSF phases. Feshbach resonance scattering also
strongly modifies the MSF--ASF phase boundary, and for small
chemical potentials drives a segment of the N--ASF and MSF--ASF
transitions first order (indicated by hatched curves), with the
first order section terminated by two tricritical points. The point
where the three phases meet, and the continuous N--MSF phase
boundary terminates at the first order boundary, is a critical
endpoint.  At large positive $\mu$'s the MSF--ASF transition retains
its $\alpha = 0$ first order character, separated from the
continuous section of this transition by another tricritical point.}

\label{phase_diagramAlphaMuB}
\end{figure}

It is clear from (\ref{saddlepoint}) that $\Psi_{10} = 0$ is still a
solution, so that the N--MSF transition line (\ref{N-MSF}) is not
modified by the Feshbach resonance coupling. Then equation
(\ref{saddlepointB}) still gives $|\Psi_{20}| = \sqrt{\mu_2/g_2}$ in
the MSF phase. However, the subsequent MSF--ASF transition boundary
is modified, as a finite $\alpha$ shifts the effective chemical
potential of the atomic condensate (in addition to the shift due to
the atom-molecule repulsion, $g_{12}$) to
\be
\mu_1^{\rm eff} = \mu_1-g_{12}|\Psi_{20}|^2+\alpha|\Psi_{20}|.
\ee
with the MSF--ASF transition located by $\mu_1^{\rm eff}=0$.  Using
$|\Psi_{20}|=\sqrt{\mu_2/g_2}$ of the MSF one obtains an estimate of
the MSF--ASF transition boundary
\begin{equation}
\mu_1 = \frac{g_{12}}{g_2} \mu_2 - \alpha
\sqrt{\frac{\mu_2}{g_2}}\ \
(\mbox{MSF--ASF transition line}).
\label{MSF-ASF}
\end{equation}
For small $\mu_2$ (specifically, $0 \leq \mu_2 \ll \alpha^2
g_2/g_{12}^2$), the second term on the right hand side dominates, and
the boundary displays a sharp square-root singularity into negative
values of $\mu_1$ (near the origin preempted by a first order
transition: see below) illustrated in the phase diagrams for $g_1 g_2
> g_{12}^2$ and $g_1 g_2 < g_{12}^2$ in Figs.\
\ref{phase_diagramAlphaMuA} and \ref{phase_diagramAlphaMuB},
respectively. In the opposite limit ($\mu_2 \gg \alpha^2
g_2/g_{12}^2$), the boundary asymptotes to the $\alpha = 0$ phase
boundary, (\ref{MSF-AMSF}).


Another important consequence of a finite Feshbach resonance
coupling is that for small chemical potentials it drives the N--ASF
and MSF--ASF transitions first order. The somewhat technical
calculation of the corresponding first order phase boundaries,
illustrated in Figs.\ \ref{phase_diagramAlphaMuA} and
\ref{phase_diagramAlphaMuB}, are relegated to Appendix
\ref{mftDetails}. Here a more approximate, but more transparent,
analysis is presented.  To this end, one can use an approximation to
(\ref{saddlepointB})
\be
\Psi_{20} \approx \frac{\alpha}{2|\mu_2|}\Psi_{10}^2,
\label{Psi12relationA}
\ee
valid for sufficiently negative $\mu_2$ and $\mu_1$, to eliminate
$\Psi_{20}$ from ${\cal H}_\text{mf}[\Psi_{10},\Psi_{20}]$ in favor of
$\Psi_{10}$. The resulting energy density in the normal state is well
approximated by
\be
{\cal H}_\text{mf}\approx -\mu_1 |\Psi_{10}|^2 +
\frac{1}{2}\left(g_1-\frac{\alpha^2}{2|\mu_2|} \right)|\Psi_{10}|^4
+\frac{\alpha^2 g_{12}}{4 \mu_2^2}|\Psi_{10}|^6.
\ee
Clearly, for $\mu_1 < 0$ and sufficiently large {\em negative}
effective quartic coupling, $u_4 = g_1-\alpha^2/2|\mu_2|$ a
secondary minimum at $\Psi_{10}\neq 0$ develops, that can compete
with the normal state $\Psi_{10} = \Psi_{20} = 0$ minimum. It is
easy to show that the corresponding ASF state minimum becomes
degenerate with the normal state at a critical value $\mu_1^c = -3
u_4^2/16 u_6$ ($u_6 = \alpha^2 g_{12}/4 \mu_2^2$ is the effective
$|\Psi_{10}|^6$ coupling), that translates into a first order N--ASF
boundary
\be
\mu_1^c \approx -\frac{3\alpha^2}{4g_{12}}
\left(\frac{|\mu_2|g_1}{\alpha^2}-\frac{1}{2}\right)^2,
\ee
as illustrated in Figs.\ \ref{phase_diagramAlphaMuA} and
\ref{phase_diagramAlphaMuB}. On the other hand, for sufficiently
small $\mu_2 > 0$, the relation between $\Psi_{20}$ and $\Psi_{10}$
following from (\ref{saddlepointB}) becomes (keeping only the $g_2
|\Psi_{20}|^2$ term inside the parentheses on the right hand side),
\be
\Psi_{20} \approx \left(\frac{\alpha}{2g_2}\right)^{1/3}
\Psi_{10}^{2/3},
\label{Psi12relationB}
\ee
which, when inserted into $\cal{H}_\text{mf}$, Eq.\ (\ref{HmuMFT}),
determines the first order MSF--ASF phase boundary, in a way
detailed in Appendix \ref{mftDetails}.


\section{Dilute BEC limit}
\label{diluteBEClimit}

Thus far, the phase diagram in the $\mu_1$-$\mu_2$ plane, has been
studied by treating the atomic and molecular chemical potentials as
independent tuning parameters. As seen, within a mean field
approximation, the temperature then plays no apparent role.

However, to make a direct contact with trapped degenerate atomic gas
experiments, where it is the total number of atoms $N$ and the
detuning $\nu$ that are varied, one needs to eliminate the chemical
potentials $\mu_\sigma$ in favor of the atom density $n$, detuning
$\nu = 2\mu_1-\mu_2$, and temperature $T$. For an interacting system,
this is a nontrivial change of variables that can usually only be
carried out perturbatively. However, in the dilute limit, appropriate
to atomic gas systems, the transition out of the normal state can be
treated by ignoring weak atomic interactions (with corrections in
powers of $n a_\s^3$ and $\gamma$), thereby reducing the problem to an
easily calculable BEC limit.\cite{WRFS} The system then reduces to two
independent ideal Bose gases, coupled only through the overall
constraint of fixed density $n = n_1 + 2n_2$, Eq.\
(\ref{Nconstraint}).

\subsection{Bulk N--ASF and N--MSF BEC transitions}

In the noninteracting limit, for a bulk (uniform) system the free
energy and atom density in $d$ spatial dimensions are easily
calculated and are given by\cite{ZUK77}
\bea
f_0 &=& \frac{1}{\beta V} \sum_{\kv,\s=1,2}
\ln\left[1-e^{-\beta(\eps_{\kv\s} - \mu_\s)}\right]
\nn \\
&=& -\frac{1}{\beta \Lambda_T^d} \sum_{\s=1,2}
\s^{d/2} g_{\frac{d+2}{2}}(z_\s)
\label{fBEC} \\
n &=& -\sum_{\s=1,2} \s \frac{\partial f}{\partial \mu_\s}
= \frac{1}{V} \sum_{\kv,\s=1,2} \frac{\s}
{e^{\beta(\eps_{\kv\s} - \mu_\s)} - 1}
\nn \\
&=& \frac{1}{\Lambda_T^d} \sum_{\s=1,2}
\s^{(d+2)/2} g_{\frac{d}{2}}(z_\s)
\label{nBEC}
\eea
where $\eps_\s = \hbar^2 k^2/2m_\s$ (with $m_1 = m$ and $m_2 = 2m$)
are the single particle energies, $z_\s = e^{\beta \mu_\s}$ are the
fugacities, $\Lambda_T = h/\sqrt{2\pi m k_B T}$ is the (atomic)
thermal de Broglie wavelength, and
\begin{equation}
g_\alpha(z) = \frac{1}{\Gamma(\alpha)}
\int_0^\infty dx \frac{x^{\alpha-1}}{z^{-1}e^x -1}
= \sum_{n=1}^\infty \frac{z^n}{n^\alpha}
\label{3.35}
\end{equation}
is the extended zeta function.

For {\em positive} detuning, $\nu > 0$, atoms (being less
energetically costly than molecules) condense first at a critical line
$\mu_1 = 0$, where $\mu_2 = 2\mu_1-\nu = -\nu < 0$. The corresponding
critical temperature $T_{c1}(\nu)$ for N--ASF transition is easily
determined by the fixed density condition
\begin{equation}
n = \Lambda_{T_{c1}}^{-d} \left[\zeta(d/2)
+ 2^{(d+2)/2} g_{d/2}\left(e^{-\beta_{c1} \nu} \right) \right],
\label{3.36}
\end{equation}
with $\zeta(\alpha) = g_\alpha(1)$. As usual, the transition exists
only for $d > 2$. In the far detuned limit, $\nu/k_B T_{c1} \gg 1$
($z_2 \ll 1$), the molecular population is exponentially suppressed,
the second term in (\ref{3.36}) may be neglected and one obtains the
standard BEC result
\begin{equation}
T_{c1}^\infty \approx \frac{h^2}{2\pi m k_B}
\left[\frac{n}{\zeta(d/2)}\right]^{2/d},
\label{3.37}
\end{equation}
that is approached exponentially as $e^{-\beta_{c1}\nu}$.  In the
opposite limit, $0 < \nu/k_B T \ll 1$, the expansion
\cite{ZUK77,Erdelyi}
\begin{equation}
g_\alpha(e^{-x}) = \Gamma(1-\alpha) x^{\alpha-1}
+ \sum_{n=0}^\infty  \frac{\zeta(\alpha-n)}{n!} (-x)^n
\label{gAsymptotics}
\end{equation}
may be used to obtain
\begin{equation}
\frac{T_{c1}}{T_{c0}}-1 \approx
\frac{2^{(d+6)/2} \Gamma\left(\frac{4-d}{2}\right)}
{d(d-2)\left[1+2^{(d+2)/2} \right] \zeta(d/2)}
\left(\frac{\nu}{k_B T_{c0}} \right)^{\frac{d-2}{2}}
\label{3.39}
\end{equation}
valid for the range $2 < d < 4$ of interest to us, where
\begin{equation}
T_{c0} = \frac{h^2}{2\pi m k_B}
\left\{\frac{n}{\left[1+2^{(d+2)/2} \right]
\zeta(d/2)}\right\}^{2/d}
\label{3.40}
\end{equation}
is the transition temperature at zero detuning, $\nu = 0$
corresponding to the critical endpoint in Fig.\
\ref{phasediagramNuT}.\cite{foot:1order}

Similarly, for {\em negative} detuning, $\nu < 0$, molecules are
energetically less costly and therefore condense first. The
corresponding N--MSF critical line, given by $\mu_2 = 0$, with
$\mu_1 = \nu/2 < 0$, using the fixed density condition translates
into a $T_{c2}(\nu)$, determined implicitly by
\begin{equation}
n = \Lambda_{T_{c2}}^{-d} \left[2^{(d+2)/2}\zeta(d/2)
+  g_{d/2}(e^{\beta_{c2} \nu/2}) \right].
\label{3.41}
\end{equation}
In the far detuned limit the N--MSF transition temperature
approaches
\begin{equation}
T_{c2}^\infty \approx \frac{h^2}{4\pi m k_B}
\left[\frac{n}{2 \zeta(d/2)}\right]^{2/d}
\label{3.42}
\end{equation}
exponentially as $e^{-\beta_{c2}|\nu|}$, while in the small detuning
limit $|\nu|/k_BT \ll 1$ it approaches $T_{c0}$ according
to\cite{foot:1order}
\begin{equation}
\frac{T_{c2}}{T_{c0}}-1 \approx
\frac{2^{(6-d)/2} \Gamma\left(\frac{4-d}{2}\right)}
{d(d-2)\left[1+2^{(d+2)/2} \right] \zeta(d/2)}
\left(\frac{|\nu|}{k_B T_{c0}} \right)^{\frac{d-2}{2}}.
\label{3.43}
\end{equation}

The resulting ratios
\bea
&&\frac{T_{c1}^\infty}{T_{c2}^\infty} = 2^{(d+2)/d}
\nn \\
&&\frac{T_{c1}^\infty}{T_{c0}} = \left[1 + 2^{(d+2)/2} \right]^{2/d}
\nn \\
&&\frac{T_{c2}^\infty}{T_{c0}} = \left[1 + 2^{-(d+2)/2} \right]^{2/d}
\nn \\
&&\frac{T_{c1}(\nu) - T_{c0}}{T_{c2}(-\nu) - T_{c0}} = 2^d,\
0 < \frac{\nu}{k_BT_{c0}} \ll 1.
\label{3.44}
\eea
are noteworthy.  The normalized transition temperatures,
$T_{c\sigma}(\nu)/T_{c0}$, give the corresponding phase boundaries
(a function of $\nu/ k_B T_{c0}$) displayed in the phase diagram in
Fig.\ \ref{phasediagramNuT}.

The thermodynamics of this dilute Bose gas mixture above the
transition temperature (i.e., inside the normal state) can be
obtained by using the atom number constraints (\ref{nBEC}) to
express the chemical potentials $\mu_1 \equiv \mu$, $\mu_2 = 2\mu -
\nu$ as functions of temperature and detuning. In the neighborhood
(above) the N--ASF and N--MSF transitions this can be done
analytically using (\ref{nBEC}) and (\ref{gAsymptotics}). To leading
order, for $2 < d < 4$ and $|\mu| \ll \nu$ one obtains near the
N--ASF line:
\begin{equation}
\left(\frac{T_{c1}}{T} \right)^{d/2} - 1
\approx \frac{\Gamma(\frac{2-d}{2}) (\beta|\mu|)^{(d-2)/2}}
{\zeta(d/2) + 2^{(d+2)/2} g_{d/2}(e^{-\beta \nu})},
\label{Tc1BEC}
\end{equation}
while in the neighborhood of the N--MSF line, defining $\delta \mu
\equiv \mu - \nu/2 \ll |\nu|$, one obtains
\begin{equation}
\left(\frac{T_{c2}}{T} \right)^{d/2} - 1
\approx \frac{2^d \Gamma(\frac{2-d}{2}) (\beta|\delta\mu|)^{(d-2)/2}}
{2^{(d+2)/2} \zeta(d/2) +  g_{d/2}(e^{\beta \nu/2})}.
\label{Tc2BEC}
\end{equation}
Finally, for $\nu = 0$ one finds
\begin{equation}
\left(\frac{T_{c0}}{T} \right)^{d/2} - 1
\approx \frac{\left(1 + 2^d \right) \Gamma(\frac{2-d}{2})
(\beta|\mu|)^{(d-2)/2}}
{\zeta(d/2) \left(1 + 2^{(d+2)/2} \right)},
\label{Tc0BEC}
\end{equation}
Therefore, for $\s = 0,1,2$, corresponding to $\nu = 0$, $\nu > 0$,
and $\nu < 0$, respectively, one obtains
\begin{equation}
\frac{|\delta\mu_\s|}{k_B T_{c\s}} \approx A_\s
\left[\frac{T}{T_{c\s}} - 1 \right]^{2/(d-2)},
\label{3.48}
\end{equation}
where $T_{c\s}$ are transition temperatures evaluated at the given
values of $\nu$, $n$, the chemical potential deviations are given by
$\delta \mu_1 = \delta \mu_0 = \mu$, $\delta \mu_2 = \delta \mu$, and
the amplitudes are
\begin{widetext}
\bea
A_1(\nu) &=& \left\{\frac{d(d-2)}
{4\Gamma\left(\frac{4-d}{2}\right)}
\left[\zeta(d/2) + 2^{(d+2)/2}
g_{d/2}(e^{-\nu/k_B T_{c1}}) \right] \right\}^{2/(d-2)}
\nn \\
A_2(\nu) &=& \left\{\frac{d(d-2)}{2^{d+2}
\Gamma\left(\frac{4-d}{2}\right)}
\left[2^{(d+2)/2}\zeta(d/2)
+ g_{d/2}(e^{\nu/2k_B T_{c2}}) \right] \right\}^{2/(d-2)}
\nn \\
A_0 &=& \left\{\frac{d(d-2) \zeta(d/2)}
{4\Gamma\left(\frac{4-d}{2}\right)}
\frac{1 + 2^{(d+2)/2}}{1 + 2^d} \right\}^{2/(d-2)}.
\label{3.49}
\eea
For $d > 4$ the exponent $2/(d-2)$ sticks at unity, so that
$\delta \mu$ varies linearly with the temperature deviation.
\cite{foot:fisher}

For $d = 3$, using $\Gamma(1/2) = \sqrt{\pi}$ and $\zeta(3/2) \simeq
2.612$, all of the above results reduce to those quoted in the
Introduction and summarized by the phase diagram, Fig.\
\ref{phasediagramNuT}.

The rest of the thermodynamics in this noninteracting limit now
follows in a standard fashion, leading to the familiar Gaussian
model critical behavior. For example, the atomic and molecular
condensate densities below their respective normal-superfluid
transition temperatures $T_{c\s}(\nu)$ are easily computed. For $\nu
> 0$ the atomic chemical potential $\mu_1 = \mu$ vanishes before the
molecular one $\mu_2 = 2\mu - \nu$ and for $T < T_{c1}(\nu)$ an
atomic condensate $n_{10}(T,\nu) = |\Psi_{10}|^2$ develops with
\begin{eqnarray}
n_{10} &=& n\left[1-\left(\frac{T}{T_{c1}}\right)^{d/2}
\frac{\zeta(d/2)+2^{\frac{d+2}{2}}g_{d/2}(e^{-\nu/k_B T})}
{\zeta(d/2)+2^{\frac{d+2}{2}}g_{d/2}(e^{-\nu/k_B T_{c1}})}\right],
\ \ \text{for}\ \ \nu > 0,\ \ T < T_{c1}(\nu),
\label{n10}
\end{eqnarray}
as the gas transitions to the ASF in the BEC limit. Close to
$T_{c1}(\nu)$, the atomic condensate (\ref{Tc1BEC}) grows linearly
with reduced temperature
\begin{equation}
n_{10}(T,\nu) \sim n\left(1-\frac{T}{T_{c1}(\nu)}\right),
\ \ \text{for}\ \ \nu > 0,\ \ T \rightarrow T_{c1}^{-}(\nu),
\end{equation}
consistent with the expected order parameter exponent $\beta = 1/2$.

Similarly, for $\nu < 0$ the molecular chemical potential $\mu_2 =
2\mu - \nu$ vanishes before the atomic one $\mu_1 = \mu$, and for $T
< T_{c2}(\nu)$ a molecular condensate $n_{20}(T,\nu) =
|\Psi_{20}|^2$ develops with
\begin{eqnarray}
n_{20} &=& \frac{1}{2} n\left[1-\left(\frac{T}{T_{c2}}\right)^{d/2}
\frac{2^{\frac{d+2}{2}} \zeta(d/2)+g_{d/2}(e^{\nu/2k_B T})}
{2^{\frac{d+2}{2}} \zeta(d/2)+g_{d/2}(e^{\nu/2k_B T_{c2}})}\right],
\ \ \text{for}\ \ \nu < 0,\ \ T < T_{c2}(\nu),
\label{n20}
\end{eqnarray}
\end{widetext}
as the gas undergoes a transition into a molecular BEC (MSF). Again,
close to $T_{c2}(\nu)$, the molecular condensate (\ref{Tc2BEC}) grows
linearly with reduced temperature
\begin{equation}
n_{20}(T,\nu)\sim n\left(1-\frac{T}{T_{c2}(\nu)}\right),
\ \ \text{for}\ \ \nu < 0,\ \ T \rightarrow T_{c2}^{-}(\nu),
\end{equation}
consistent with the same order parameter exponent $\beta = 1/2$.

Clearly, the above expressions for the weakly interacting limit are
quite close to standard ones for a single-component Bose gas,
reducing to a purely atomic and molecular BEC for the far detuned
cases, $|\nu|/k_B T \gg 1$. There are, however, nonstandard
contributions to $T_{c\sigma}(\nu)$ and $n_{\sigma 0}(\nu,T)$
arising from the contribution of the secondary, off-resonance,
bosonic component that is gapped out for $\nu \neq 0$. For example,
for $\nu < 0$, upon warming toward $T_{c2}(\nu)$, the molecular
condensate is reduced due to both the conventional mechanism of
thermal excitations of molecules out of the molecular condensate, as
well as the depairing of molecules into thermally excited bosonic
atoms, with the latter special to a Feshbach-resonant system.

Because of the suppression of $T_{c\sigma}(\nu)$ near $\nu = 0$ for
$T_{c0} < T < T_{c2}^\infty(\nu)$, the gas is expected to undergo a
sequence of ${\rm ASF} \rightarrow {\rm N} \rightarrow {\rm MSF}$
transitions upon lowering of $\nu$ (see Fig.\
\ref{phasediagramNuT}). For $T < T_{c0}$ the transition is a direct
${\rm ASF} \rightarrow {\rm MSF}$ one, that for this noninteracting
limit is first order. The condensate densities are undefined right
on the critical line $\nu = 0$, $T < T_{c0}$, and the noninteracting
approximation becomes particularly questionable there.

\subsection{N--ASF and N--MSF BEC transitions in a trap}
\label{transitionsTrap}

The above results are straightforwardly extended to the
experimentally more relevant case of a harmonic trap.  The
modifications due to the trapping potential can all be incorporated
through the change in the density of states.  For an isotropic
harmonic trap (easily extendable to an anisotropic trap) the single
particle energy spectrum
\be
\eps_{n}=\hbar\wt (n_1 + n_2 + \ldots + n_d)
\label{spectrumHO}
\ee
is linear in $n = \sum_i^d n_i$ and exhibits a well-known degeneracy
that, for large quantum numbers of interest to us, in the
macroscopic limit $k_B T_{c\s} \gg \hbar \wt$, is given by
\be
D(\eps)=\frac{1}{(d-1)!}\frac{\eps^{d-1}}{(\hbar\wt)^d}.
\label{g_e}
\ee
Note that $\eps_n$ is actually $\sigma$ independent, i.e., the trap
frequency $\omega_0$ is the same for atoms ($\sigma=1$) and
molecules ($\sigma=2$).  This is a good approximation in the
physically relevant limit of the size $d_0$ of the closed-channel
molecule being much smaller than the trapped cloud size.

In the thermodynamic limit the sum over single-particle states
appearing in (\ref{fBEC}) and (\ref{nBEC}) can be replaced by
integration over energies $\eps$ weighted by above density of
states, giving
\bea
N = \left(\frac{k_B T}{\hbar\wt}\right)^d
\sum_\sigma \sigma g_d(z_\sigma).
\label{Nnormal_trap}
\eea
Paralleling the above calculations for the uniform system, from this
one obtains all the relevant quantities for the trapped system.
Specifically, the transition temperatures $T_{c\sigma}(\nu)$ are
implicitly given by
\bea
\label{implicitTctrap}
N &= &\left(\frac{k_B\T_{c1}}{\hbar\wt}\right)^d
\left[\zeta(d) + 2 g_d
\left(e^{-\tilde\beta_{c1}\nu}\right)\right],
\\
N &= &\left(\frac{k_B\T_{c2}}{\hbar\wt}\right)^d
\left[2\zeta(d) + g_d
\left(e^{-\tilde\beta_{c2}|\nu|/2}\right)\right].
\eea
These can be solved in the asymptotic regimes of small and large
detuning, giving
\begin{equation}
\tilde{T}_{c\sigma}(\nu)\approx
\begin{cases}
\T_{c0} \left[1 + a_\sigma \frac{|\nu|}{k_B \T_{c0}}\right],
& |\nu| \ll k_B \T_{c0} \\
\T_{c\sigma}^\infty\left[1-b_\sigma
e^{-|\nu|/\sigma\T_{c\sigma}^\infty}\right],
& |\nu| \gg k_B \T_{c0},
\end{cases}
\label{trapTc}
\end{equation}
with $a_\sigma = 2\zeta(d-1)/3\sigma^2d\zeta(d)$ and $b_\sigma =
2/\sigma^2 d \zeta(d)$.  The transition temperatures
$\tilde{T}_{c\sigma}^\infty$ in the limit of asymptotically large
positive ($\sigma = 1$) and negative ($\sigma = 2$) detuning
($|\nu|/k_B\T_{c\sigma} \gg 1$), and at the tricritical point
$\tilde{T}_{c0}$ ($\nu=0$), are given by
\begin{eqnarray}
\label{trapTcoo}
\tilde{T}_{c\sigma}^\infty
&=&\hbar\wt\left[\frac{N}{\sigma\zeta(d)}\right]^{1/d},\\
\label{trapTc0}
\tilde{T}_{c0}
&=&\hbar\wt\left[\frac{N}{3\zeta(d)}\right]^{1/d}.
\end{eqnarray}
The latter is approached {\em linearly} with reduced detuning from
either side, in any dimension $d \geq 2$.

As in the bulk case above, and in the well studied single component
trapped Bose gas, here too one can easily compute the number of
condensed atoms and molecules below the transition into the ASF and
MSF states. This is determined by the extension of the total atom
number constraint (\ref{Nnormal_trap}) to include the condensates
$N_{\s0}$:
\be
N = \sum_{\sigma=1,2} \sigma\left[N_{\s0}
+ \left(\frac{k_B T}{\hbar\wt}\right)^d
g_d(z_\sigma)\right].
\label{Nbec_trap}
\ee
The analysis of these equations closely follows that for the bulk
BEC of the previous subsection. Below the transition into the ASF
and MSF phase one can straightforwardly compute the number of atoms
$N_{\sigma 0} = \int d\rv |\Psi_{\s0}(\rv)|^2$ in the corresponding
condensate. For $\nu > 0$ the atomic chemical potential $\mu_1 =
\mu$ vanishes before the molecular one $\mu_2 = 2\mu - \nu$, and for
$T < \T_{c1}(\nu)$ the molecular condensate $N_{20}=0$ and a finite
atomic condensate $N_{10}(T,\nu)$ develops, given by
\begin{eqnarray}
N_{10} &=& N\left[1-\left(\frac{T}{\T_{c1}}\right)^{d}
\frac{\zeta(d)+2 g_{d}(e^{-\nu/k_B T})}
{\zeta(d)+2 g_{d}(e^{-\nu/k_B \T_{c1}})}\right],
\nn \\
&&\ \ \ \ \ \ \ \ \ \ \ \ \ \
\text{for}\ \ \nu > 0,\ \ T < \T_{c1}(\nu).
\label{N10}
\end{eqnarray}

For $\nu < 0$ the molecular chemical potential $\mu_2 = 2\mu - \nu$
vanishes before the atomic one $\mu_1 = \mu$, and for $T <
\T_{c2}(\nu)$ the atomic condensate $N_{10} = 0$ and a finite
molecular condensate $N_{20}(T,\nu)$ develops, given by
\begin{eqnarray}
N_{20} &=& \frac{1}{2} N
\left[1-\left(\frac{T}{\T_{c2}}\right)^{d}
\frac{2\zeta(d)+g_{d}(e^{\nu/2k_B T})}
{2\zeta(d)+g_{d}(e^{\nu/2k_B \T_{c2}})}\right],
\nn \\
&&\ \ \ \ \ \ \ \ \ \ \ \ \ \
\text{for}\ \ \nu < 0,\ \ T < \T_{c2}(\nu),
\label{N20}
\end{eqnarray}
Just below the transition temperatures $\T_{c\s}$ the condensate
growth is of the expected linear in $T$ form, characteristic of the
order parameter exponent $\beta = 1/2$. Also, for a far detuned gas,
$|\nu|/k_BT \gg 1$, the above results reduce to the standard single
component BEC behavior.

The advantage of a trapped system is that, as in the case for an
ordinary single-component trapped condensate that exhibits a
striking narrow BEC peak,\cite{cornellBEC,ketterleBEC} here too we
expect ASF and MSF condensates in a trap to display clearly
identifiable BEC peaks. As discussed in the Introduction and
illustrated in Fig.\ \ref{densityProfiles}, provided that atoms and
molecules can be imaged separately, the ASF should be easily
identified by atomic and molecular BEC peaks,\cite{decoupledBEC}
while the MSF is identified by the presence only of a molecular BEC
peak. In a harmonic trap at low temperature, $T \ll \T_{c\s}$, the
density profile of the cloud is dominated by a narrow Gaussian
$\s$-condensate peak, with the width given by the quantum oscillator
length
\begin{equation}
r_{\s0} = \sqrt{\frac{\hbar}{m_\s\omega_0}}.
\label{rs0}
\end{equation}
This should be easily distinguishable from the high-temperature,
classical Gaussian density profile (coming from the Boltzmann
distribution) with the much wider width set by the thermal
oscillator length
\begin{equation}
r_{\s T} = \sqrt{\frac{k_B T} {\frac{1}{2}m_\s\omega_0^2}}
= r_{\s0} \sqrt{\frac{2k_B T}{\hbar\omega_0}} \gg r_{\s0}.
\label{rsT}
\end{equation}

The full atomic (whether free or bound into molecules) density
profile $n(r)$ at arbitrary temperature is easily calculated for a
noninteracting gas. It is given by
\begin{equation}
n(r) = \sum_\s \s n_\s(r),
\end{equation}
consisting of atomic and molecular contributions, in the BEC limit
tied only by a common chemical potential $\mu$ determined by the
overall particle number constraint (\ref{Nbec_trap}). As derived and
analyzed for a single Bose component in App.\ \ref{app:density},
these in turn are given by
\begin{eqnarray}
n_\s(\rv) &=& \sum_{\nv=0}^\infty
\frac{|\phi_{\nv\s}(\rv)|^2}{e^{\beta(\eps_\nv - \mu_\s)}-1},
\\
&=& \sum_{p=1}^\infty e^{p\beta\mu_\s}\rho^{\rm osc.}_\s(\rv,\rv;
p\beta\hbar\omega_0),
\nonumber
\end{eqnarray}
where $\phi_{\nv\s}(\rv)$ are harmonic oscillator eigenstates and
$\rho^{\rm osc.}_\s(\rv,\rv;\beta\hbar\omega_0)$ is the diagonal
element of the single-particle density matrix for a harmonic
oscillator with mass $m_\s$. In 3d it is given by
\begin{equation}
\rho^{\rm osc.}_\s(\rv,\rv;\beta\hbar\omega_0)
=\left(\frac{m_\s\wt e^{\beta\hbar\wt}}
{2\pi\hbar\sinh(\beta\hbar\wt)}\right)^{3/2}
e^{-r^2/\rts^2(\beta)},
\end{equation}
where
\begin{eqnarray}
\rts^2(\beta) &=& \frac{\hbar}{m_\s\wt}
\coth\left(\beta\hbar\wt/2\right),
\\
&\approx&
\begin{cases}
\frac{\hbar}{m_\s\wt}, & \hbar\wt/k_B T\gg 1,\\
\frac{k_B T}{\frac{1}{2}m_\s\wt^2}, & \hbar\wt/k_B T\ll 1,
\end{cases}
\label{eq:r0T}
\end{eqnarray}
is the finite-temperature ``oscillator length'' that reduces to the
quantum one $\sqrt{\hbar/(m_\s\wt)}$, Eq.\ (\ref{rs0}), at low $T$,
and the classical (thermal) one, Eq.\ (\ref{rsT}), at high $T$.

The spatial profile of the $\sigma$-density, $n_\sigma(r)$, is
determined by the ratio of the chemical potential $\mu_\s$ to the
trap level spacing $\hbar\omega_0$, with former in turn determined
by the temperature through the total atom number constraint. At high
$T \gg T_{c\s}$ (where the gas is nondegenerate), such that $0 <
-\mu_\s\approx -k_B T\ln[\left(\frac{\hbar\omega_0}{k_B T}\right)^3
N] \approx 3 k_B T\ln (T/T_{c\s})\gg k_B T$, the result is a purely
classical thermal (Boltzmann) distribution,
\begin{equation}
n_\sigma(r) \approx \left(\frac{k_B T}{\pi\hbar\omega_0}\right)^{3/2}
\frac{1}{r_{\s 0}^3}e^{-r^2/r_{\s T}^2 - |\mu_\s|/k_B T},\ \ \ T \gg T_c,
\label{nHigh_trap}
\end{equation}
with only of order unity occupation of the lowest oscillator $n = 0$
state and a vanishing ``condensate'' density $n_0(r) = \pi^{-3/2}
r_{\s0}^{-3}e^{-r^2/r_{\s0}^2 - |\mu_\s|/k_B T}$.

As $T$ is lowered further, approaching $T_{c\s}$ from above, the
magnitude of the chemical potential drops below $T$ (remaining
negative) and the boson density profile develops a small $r$
non-Boltzmann peak structure even above $T_{c\s}$:
\begin{widetext}
\begin{eqnarray}
n_\s(r) &\approx&
\left(\frac{k_B T}{\pi\hbar\omega_0}\right)^{3/2}
\frac{1}{r_{\s 0}^3}g_{3/2}
\left[e^{-r^2/r_{\s T}^2 - |\mu_\s|/k_B T}\right],\ \
\text{for}\ \ T \gtrsim T_c,
\\
&\approx& \left(\frac{k_B T}{\pi\hbar\omega_0}\right)^{3/2}
\frac{1}{r_{\s 0}^3}
\begin{cases}
e^{-r^2/r_{\s T}^2 - |\mu_\s|/k_B T},\ \ &r\gg r_{\s T}\cr
\zeta(3/2)-2\pi^{1/2}\left(\frac{r^2}{r_{\s T}^2}
+ \frac{|\mu_\s|}{k_B T}\right)^{1/2},\ \ & r\ll r_{\s T}\ .
\end{cases}
\label{eq:nMed_trap}
\end{eqnarray}
The linear in $r$ cusp is rounded on the length scale below $r_{\s
0} \sqrt{\mu_\s/\hbar\omega_0}$.

Finally at an even lower $T < T_{c\s}$, $|\mu_\s|$ drops below the
level spacing, $|\mu_\sigma|\lesssim\hbar\omega_0$, and the density
profile changes dramatically, developing a bimodal distribution
$n_\s(r) = n_{\s T}(r) + n_{\s0}(r)$ (see Fig.\
\ref{densityProfiles}), that consists of a broad (width $r_{\s T}$)
thermal part
\begin{equation}
n_{\s T}(r) \approx
\left(\frac{k_B T}{\pi\hbar\omega_0}\right)^{3/2}
\frac{1}{r_{\s 0}^3}
\tilde{g}_{3/2}\left(e^{-r^2/r_{\s T}^2 - |\mu_\s|/k_BT},
\frac{k_B T}{\hbar\omega_0}\right),
\ \ \text{for}\ \ T < T_{c\s},
\label{nTs}
\end{equation}
\end{widetext}
with a small $r$ cusp (rounded by $r_{\s0}$) and large $r$ Gaussian
tails, together with a narrow (width $r_{\s0}$) condensate part
\begin{equation}
n_{\s0}(r)\approx \frac{N_{\sigma 0}(T)}{\pi^{3/2}r_{\s0}^{3}}
e^{-r^2/r_{\s0}^2}.
\label{eq:n0s}
\end{equation}
In (\ref{nTs}),
\begin{eqnarray}
\tilde{g}_\alpha(x,p_c) &=& \sum_{p=1}^{p_c} \frac{x^p}{p^\alpha},
\end{eqnarray}
has been defined, while in (\ref{eq:n0s})
\begin{equation}
N_{\s0}(T) \approx \sum_{p=p_c}^\infty e^{-p|\mu|/k_B T}
\approx \frac{e^{-(p_c-1)|\mu|/k_B T}}{e^{|\mu|/k_B T}-1},
\end{equation}
is the number of condensed bosons, given by (\ref{N10}) and
(\ref{N20}) when the total atom number constraint is taken into
account.


\section{Elementary excitations}
\label{sec:excite}

Having established the approximate nature of atomic and molecular
superfluids, consider next the study of their excitations. On
general grounds, as required by the Goldstone's theorem, one expects
{\em one} collective {\em gapless} (sound) mode in each of the ASF
and MSF phases, associated with spontaneous breaking of global
$U(1)$ charge (phase-``rotation'') symmetry. In the MSF it is
associated with the phase $\theta_2$ of the molecular (two-atom)
condensate, $\Psi_{20}$, while in ASF it corresponds to in-phase
fluctuations of the phases $\theta_1$ and $\theta_2$ of the atomic
and molecular condensates.

In addition, there are three\cite{3vs1comment} gapped excitations in
each of the superfluids. In the MSF these are associated with
atom-like (squeezed by Feshbach resonance coupling to the molecular
condensate) quasiparticle excitations (accounting for two modes,
$\psi_1,\psi_1^\dagger$) and molecular density fluctuations
(fluctuations in the order parameter magnitude $|\Psi_{20}|$).  In
the ASF one gapped mode corresponds to out-of-phase fluctuations
$2\theta_1-\theta_2$ of the atomic and molecular condensates (gapped
by the Feshbach resonance coupling $\alpha$), and two others are
atomic and molecular condensate densities (fluctuations in the order
parameter magnitudes $|\Psi_{10}|$ and $|\Psi_{20}|$).

As will be seen below, the MSF-to-ASF transition is accompanied by
closing of the gap for atom-like quasiparticle excitations. However,
this mode remains gapless only at the MSF--ASF critical point, and
is replaced by another gapped mode (associated with out-of-phase
phase fluctuations of the two order parameters) that emerges inside
the ASF. As discussed in Sec.\ \ref{sec:symmetries}, this is
consistent with the Goldstone theorem as (due to the Feshbach
resonance coupling) it is only a {\em discrete} (${\cal Z}_2$)
symmetry that is being broken at the MSF-to-ASF transition and as
such leads to no new gapless modes.

\subsection{Bogoliubov diagonalization}
\label{sec:bog}

Bogoliubov theory provides an asymptotically exact description of
the low energy excitations in a dilute Bose fluid, not too close to
the transition lines. Focusing on quadratic fluctuations, it ignores
interactions between quasiparticles and, among other things, misses
the possibility for their decay. The method proceeds by expanding
the field operators about the mean field solution (equivalently, a
coherent state of $\kv=0$ fields labeled by $\Psi_{\s 0}$):
\bea
\hat{\psi}_\s(\xv) = \Psi_{\s 0} + \hat{\phi}_\s(\xv),
\label{displacement}
\eea
and keeping terms in the Hamiltonian only to quadratic order in
the small deviations $\hat{\phi}_\s$.  In the molecular superfluid
state $\Psi_{10} = 0$, so $\hat \phi_1 = \hat \psi_1$.
Substituting (\ref{displacement}) into (\ref{H2channel}) one obtains
\begin{equation}
\hat H = H_{\rm mf} + \hat H_2
+ O(\hat \phi_\s^3,\hat \phi_\s^4)
\label{4.2}
\end{equation}
in which $H_{\rm mf} \equiv H[\Psi_{\s 0}]$ is the mean field
approximation (\ref{HmuMFT}) to the ground state energy. The absence
of terms linear in excitations $\hat{\phi}_\s$ is guaranteed by the
condition that $\Psi_{\s0}$ is an extremum of the mean field free
energy $\partial H_{\rm mf}/\partial \Psi_{\s 0}^* = 0$.  To
quadratic order, this is equivalent to the requirement
$\av{\hat{\phi}_\s(\xv)} = 0$.\cite{foot:cubiccorrections,W88} For a
homogeneous system [generalization to the trapped case may then be
accomplished through a local density approximation (LDA)] the
quadratic Hamiltonian, $H_2$ governing the dynamics of fluctuations,
can be represented in terms of momentum space operators
\begin{equation}
\hat \phi_\s(\xv) = \frac{1}{\sqrt{V}}
\sum_\kv \hat a_{\kv \s} e^{i\kv \cdot \xv},\
\hat \phi^\dagger_\s(\xv) = \frac{1}{\sqrt{V}}
\sum_\kv \hat a^\dagger_{\kv \s} e^{-i\kv \cdot \xv}.
\label{4.3}
\end{equation}
One obtains
\bea
\hat H_2 &=& \sum_{\kv,\s}
\left[\tilde \eps_{\kv\s} \hat{a}^\dagger_{\kv\s} \hat{a}_{\kv\s}
+ \frac{1}{2}\left(\lambda_\s \hat{a}_{\kv\s} \hat{a}_{-\kv\s}
+ h.c. \right)\right]
\nn\\
&&+\ \sum_\kv \left(t_1 \hat a^\dagger_{\kv 1} \hat a_{\kv 2}
+ t_2 \hat a^\dagger_{\kv 1} \hat a^\dagger_{-\kv 2} + h.c. \right),
\label{Boghamiltonian}
\eea
where the coefficients are given by
\bea
\tilde\eps_{\kv 1} &=& \eps_{\kv 1} - \mu_1
+ 2g_1 |\Psi_{10}|^2 + g_{12} |\Psi_{20}|^2
\nn \\
\tilde\eps_{\kv 2} &=& \eps_{\kv 2} - \mu_2
+ 2g_2 |\Psi_{20}|^2 + g_{12} |\Psi_{10}|^2
\nn \\
\lambda_1 &=& g_1 \Psi_{10}^2 - \alpha \Psi_{20},\
\lambda_2 = g_2 \Psi_{20}^2
\nn \\
t_1 &=& g_{12} \Psi_{10} \Psi_{20}^* - \alpha \Psi_{10}^*,\
t_2 = g_{12} \Psi_{10} \Psi_{20},
\label{4.5}
\eea
with single particle energies $\eps_{\kv\s} = \hbar^2
k^2/2m_\sigma$.

\subsubsection{MSF phase}
\label{subsec:msf}

Consider first the excitations in the molecular superfluid,
characterized by a finite $\Psi_{20}$ and vanishing $\Psi_{10}$. As
a result, the cross terms $t_1$ and $t_2$ vanish, and the atomic and
molecular terms can be diagonalized independently. From
(\ref{mu2>0}), the mean field order parameter is given by $\Psi_{20}
= \sqrt{\mu_2/g_2} = \sqrt{(2\mu-\nu)/g_2}$ (chosen real and
positive for simplicity---more generally any phase $\theta_2$ can be
absorbed into the operators via the redefinition $a_{\kv\s} \to
e^{-i\s\theta_2/2} a_{\kv \s}$). It is straightforward to verify
that the Bogoliubov canonical transformation
\bea
\hat a_{\kv\s} &=& u^*_{\kv\s} \hat \gamma_{\kv\s}
- v_{\kv\s} \hat \gamma^\dagger_{-\kv\s}
\nn\\
\hat \gamma_{\kv\s} &=& u_{\kv\s} \hat a_{\kv\s}
+ v_{\kv\s} \hat a_{-\kv\s}^\dagger
\label{quasiparticle}
\eea
to new bosonic creation and annihilation operators
$\hat\gamma^\dagger_{\kv\s}$, $\hat \gamma_{\kv\s}$, with real,
positive coefficients given by
\bea u_{\kv\s}^2 &=& 1 + v_{\kv\s}^2 = \frac{1}{2}
\left(\frac{\tilde \eps_{\kv\s}}{E_{\kv\s}} + 1 \right)
\label{uv} \\
E^{\rm MSF}_{\kv\s} &=&
\sqrt{\tilde \eps_{\kv\s}^2 - |\lambda_\s|^2},
\label{EkMSF}
\eea
leads to the diagonal form
\begin{equation}
\delta\hat{H}_{\rm MSF}
= \sum_{\kv,\sigma} E^{\rm MSF}_{\kv\s}
\left(\hat \gamma^\dagger_{\kv\s}
\hat \gamma_{\kv\s} - v_{\kv\s}^2 \right).
\label{deltaHmsf}
\end{equation}
The diagonalized Hamiltonian, $\delta\hat{H}_{\rm MSF}$, governing
excitations in the MSF naturally separates into ``atom-like'' ($\s=1$)
and ``molecule-like'' ($\s=2$) contributions, with corresponding
(explicitly positive) excitation energies $E^{\rm MSF}_{\kv\s}$ and
condensation energy
\begin{equation}
\delta E^{\rm MSF}_{\rm cond} \equiv
-\sum_{\kv,\s} E^{\rm MSF}_{\kv\s} v_{\kv\s}^2,
= -\sum_{\kv,\s} \frac{1}{2}
(\tilde\eps_{\kv\s}-E^{\rm MSF}_{\kv\s}).
\end{equation}
The latter lowers the energy of the MSF below that given by the
mean field condensation energy value, $H_{\rm mf}$.

In the normal phase, $\Psi_{20} = 0$, one obtains $\lambda_\sigma =
0$, $\tilde{\eps}_{\kv\s} = \eps_{\kv\s}-\mu_\s$, yielding
$v_{\kv\s} = 0$ and $u_{\kv\s} = 1$. One therefore recovers the
original atomic ($a_{\kv 1}$) and the molecular ($a_{\kv 2}$)
operators as true (to quadratic order) excitations in the normal
state, with corresponding free single-particle spectra $\eps_{\kv\s}
- \mu_\s$.

\subsubsection{ASF phase}
\label{subsec:amsf}

It is clear from the structure of the Hamiltonian $\delta \hat
H_{\rm ASF}$ in the ASF phase (most notably the finite values of the
$t_1$ and $t_2$ couplings), that in addition to the usual Bogoliubov
mixing between particles and holes, a true excitation is also a
mixture of an atom and a molecule. Physically, this is a reflection
of a coherent scattering (by the Feshbach and atom-molecule density
interactions) of atoms and molecules mediated by their respective
condensates. The Bogoliubov theory for the ASF phase is handled most
simply by first converting from creation and annihilation operators
to corresponding ``position'' and ``momentum'' operators
(canonically conjugate ``coordinates'', that are Fourier transforms
of Hermitian field operators):\cite{W88}
\bea
\hat a_{\kv\s} &=& \frac{1}{\sqrt{2}}
(\hat q_{\kv\s} + i \hat p_{\kv\s})
\nn \\
\hat a^\dagger_{\kv\s} &=& \frac{1}{\sqrt{2}}
(\hat q_{-\kv\s} - i \hat p_{-\kv\s})
\nn \\
\hat q_{-\kv\s} &=& \hat q_{\kv\s}^\dagger,\
\hat p_{-\kv\s} = \hat p_{\kv\s}^\dagger
\label{4.9}
\eea
with the only nonvanishing commutation relations being
\begin{equation}
[\hat q_{\kv\s}, \hat p_{-\kv'\s'}]
= i \delta_{\kv\kv'} \delta_{\s\s'}.
\label{4.10}
\end{equation}
By substituting (\ref{4.10}) into (\ref{Boghamiltonian}) one
obtains
\bea
\delta\hat H_{\rm ASF} &=& \sum_{\kv}
\left[\delta\hat H_{\kv}^{\rm ASF}
- \frac{1}{2}(\tilde \eps_{\kv 1}
+ \tilde \eps_{\kv 2}) \right]
\nn \\
\delta\hat H_{\kv}^{\rm ASF} &\equiv& \frac{1}{2}
{\bf \hat p}_\kv^\dagger {\cal P}_\kv {\bf \hat p}_\kv
+ \frac{1}{2}
{\bf \hat q}_\kv^\dagger {\cal Q}_\kv {\bf \hat q}_\kv
\label{4.11}
\eea
in which the $2\times 2$ matrix structure is defined by
\bea
{\bf \hat q}_\kv &=& \left(\begin{array}{c} \hat q_{\kv 1}
\\
\hat q_{\kv 2} \end{array} \right),\
{\bf \hat p}_\kv = \left(\begin{array}{c} \hat p_{\kv 1}
\\
\hat p_{\kv 2} \end{array} \right)
\nn \\
{\cal P}_\kv &=& \left(\begin{array}{cc}
\tilde \eps_{\kv 1} - \lambda_1 & t_1 - t_2 \\
t_1 - t_2 & \tilde \eps_{\kv 2} - \lambda_2
\end{array} \right)
\nn \\
{\cal Q}_\kv &=& \left(\begin{array}{cc}
\tilde \eps_{\kv 1} + \lambda_1 & t_1 + t_2 \\
t_1 + t_2 & \tilde \eps_{\kv 2} + \lambda_2
\end{array} \right).
\label{4.12}
\eea
In deriving (\ref{4.12}) the symmetry $\eps_{-\kv \s} = \eps_{\kv
\s}$ has been used, and $t_1, t_2, \lambda_1, \lambda_2$ have been
taken to be all real (or, equivalently, their phases absorbed into
redefinitions of $\hat a_{\kv \s}$).

One seeks a (real) linear transformation
\begin{equation}
{\bf \hat p}_\kv = {\cal A}_\kv {\bf \hat P}_\kv,\
{\bf \hat q}_\kv = {\cal B}_\kv {\bf \hat Q}_\kv
\label{4.13}
\end{equation}
which diagonalizes $\delta\hat H_{\kv}^{\rm ASF}$.  The canonical
requirement that the transformation preserve the commutation
relations (\ref{4.10}), i.e., that
\begin{equation}
[{\bf \hat Q}_\kv, {\bf \hat P}_{\kv'}^\dagger]
=[{\bf \hat q}_\kv, {\bf \hat p}_{\kv'}^\dagger]
= i \delta_{\kv \kv'} \openone,
\label{4.14}
\end{equation}
implies that
\begin{equation}
{\cal B}_\kv^T = {\cal A}_\kv^{-1}.
\label{4.15}
\end{equation}
Thus, the transformation ${\cal B}_\kv$ should simultaneously
diagonalize ${\cal P}_\kv^{-1}$ and ${\cal Q}_\kv$.  Without loss of
generality this is equivalent to demanding that
\begin{equation}
{\cal B}_\kv^T {\cal P}_\kv^{-1} {\cal B}_\kv = \openone,\
{\cal B}_\kv^T {\cal Q}_\kv {\cal B}_\kv = {\cal E}_\kv,
\label{4.16}
\end{equation}
in which ${\cal E}_\kv = \mathrm{diag}[(E_{\kv 1}^{\rm
ASF})^2,(E_{\kv 2}^{\rm ASF})^2]$ is diagonal, containing the
squares of the Bogoliubov energies (see below). It follows that
\begin{equation}
{\cal B}_\kv^{-1} {\cal P}_\kv {\cal Q}_\kv {\cal B}_\kv
= {\cal E}_\kv,
\label{4.17}
\end{equation}
so that ${\cal E}_\kv$ is obtained by diagonalizing ${\cal P}_\kv
{\cal Q}_\kv$. The squared energies are therefore solutions to the
eigenvalue equation
\begin{equation}
\det[(E_{\kv \s}^{\rm ASF})^2 \openone
- {\cal P}_\kv {\cal Q}_\kv] = 0.
\label{4.18}
\end{equation}
The solutions to the resulting quadratic equation in $(E_{\kv
\s}^{\rm ASF})^2$ are
\begin{equation}
(E_{\kv \s}^{\rm ASF})^2
= e_\kv \pm \sqrt{d_\kv^2 + c^{(1)}_\kv c^{(2)}_\kv}
\label{EkASF}
\end{equation}
in which the upper sign corresponds to $\s = 1$, the lower sign to
$\s = 2$, and the various parameters are defined by
\bea
{\cal P}_\kv {\cal Q}_\kv &=& \left(\begin{array}{cc}
e_\kv + d_\kv & c_\kv^{(2)} \\
c_\kv^{(1)} & e_\kv - d_\kv
\end{array} \right)
\nn \\
e_\kv &=& \frac{1}{2}(\tilde \eps_{\kv 1}^2 - \lambda_1^2
+ \tilde \eps_{\kv 2}^2 - \lambda_2^2) + t_1^2 - t_2^2
\nn \\
d_\kv &=& \frac{1}{2} (\tilde \eps_{\kv 1}^2 - \lambda_1^2
- \tilde \eps_{\kv 2}^2 + \lambda_2^2)
\label{4.20} \\
c_\kv^{(1)} &=& (t_1-t_2)(\tilde \eps_{\kv 1} + \lambda_1)
+ (t_1+t_2)(\tilde \eps_{\kv 2} - \lambda_2)
\nn \\
c_\kv^{(2)} &=& (t_1+t_2)(\tilde \eps_{\kv 1} - \lambda_1)
+ (t_1-t_2)(\tilde \eps_{\kv 2} + \lambda_2).
\nn
\eea
It is easy to check that the MSF results (\ref{uv}) and
(\ref{EkMSF}) are recovered when $t_1 = t_2 = 0$.

The columns ${\bf b}_{\kv \s}$ of ${\cal B}_\kv \equiv ({\bf
b}_{\kv 1} {\bf b}_{\kv 2})$ are the eigenvectors of ${\cal
P}_\kv{\cal Q}_\kv$ and take the form
\begin{equation}
{\bf b}_{\kv \s} = \frac{1}{{\cal N}_{\kv \s}}
\left(\begin{array}{c}
-c_\kv^{(2)} \\ e_\kv + d_\kv - E_{\kv\s}^2
\end{array} \right),
\label{4.21}
\end{equation}
in which the normalization ${\cal N}_{\kv \s}$ is chosen so that ${\bf
b}^T_{\kv \s} {\cal P}_\kv^{-1} {\bf b}_{\kv \s} = 1$.

The quadratic Hamiltonian takes the form
\bea
\delta\hat H_{\rm ASF} &=& \frac{1}{2} \sum_{\kv,\s}
(\hat P_{-\kv \s} \hat P_{\kv \s}
+ (E_{\kv \s}^{\rm MSF})^2 \hat Q_{-\kv \s} \hat Q_{\kv \s}
- \tilde \eps_{\kv\s})
\nn \\
&=& \sum_{\kv \s} \left[ E_{\kv \s}^{\rm ASF}
\hat \gamma_{\kv \s}^\dagger \hat \gamma_{\kv \s}
+ \frac{1}{2}(E_{\kv \s}^{\rm ASF} - \tilde \eps_{\kv \s}) \right]
\label{4.22}
\eea
in which the new bosonic raising and lowering operators are given by
\bea
\hat \gamma_{\kv\s} &=& \frac{1}{\sqrt{2}}
\left(E^{1/2}_{\kv\s} \hat Q_{\kv\s}
+ \frac{i}{E^{1/2}_{\kv\s}} \hat P_{\kv\s} \right)
\nn \\
\hat \gamma_{\kv\s}^\dagger &=& \frac{1}{\sqrt{2}}
\left(E^{1/2}_{\kv\s} \hat Q_{-\kv\s}
- \frac{i}{E^{1/2}_{\kv\s}} \hat P_{-\kv\s} \right).
\label{4.23}
\eea
These may be reexpressed in terms of the original raising and
lowering operators via
\bea
\hat {\bf a}_\kv &=& \frac{1}{2}
\left({\cal B}_\kv {\cal E}_\kv^{-1/4}
+ {\cal B}_\kv^{-T} {\cal E}_\kv^{1/4} \right)
\hat {\bm \gamma}_\kv
\nn \\
&&+\ \frac{1}{2} \left({\cal B}_\kv {\cal E}_\kv^{-1/4}
- {\cal B}_\kv^{-T} {\cal E}_\kv^{1/4} \right)
\hat {\bm \gamma}_{-\kv}^\dagger
\nn \\
\hat {\bm \gamma}_\kv &=& \frac{1}{2}
\left({\cal E}_\kv^{1/4} {\cal B}_\kv^{-1}
+ {\cal E}_\kv^{-1/4} {\cal B}_\kv^T  \right)
\hat {\bf a}_\kv
\nn \\
&&+\ \frac{1}{2} \left({\cal E}_\kv^{1/4}
{\cal B}_\kv^{-1}
- {\cal E}_\kv^{-1/4} {\cal B}_\kv^T  \right)
\hat {\bf a}_{-\kv}^\dagger
\label{4.24}
\eea
in which $\hat {\bf a}_\kv, \hat {\bf a}_{-\kv}^\dagger, \hat {\bm
\gamma}_\kv, \hat {\bm \gamma}_{-\kv}^\dagger$ are all column
vectors defined in the natural way, consistent with (\ref{4.12}).

\subsection{Acoustic and gapped modes}
\label{subsec:modes}

Consider now the Bogoliubov excitation spectra, (\ref{EkMSF}) and
(\ref{EkASF}) in more detail. It will be shown that in both phases
there is indeed one acoustic mode and one gapped mode (in addition
to two other less interesting gapped modes\cite{3vs1comment}), as
required by general principles discussed in the beginning of this
section and in Sec.\ \ref{sec:symmetries}.  As previously indicated,
the MSF phase the acoustic mode corresponds to long wavelength
fluctuations in the phase of $\Psi_{20}$, while the gapped mode is
associated with pair-breaking fluctuations of molecules into two
atom-like excitations, with spectral gap corresponding to a
renormalized molecular binding energy. In the ASF phase the acoustic
and gapped modes correspond to in-phase and out-of-phase
fluctuations of $\Psi_{10}$ and $\Psi_{20}$, respectively, with the
gap in the latter governed by the Feshbach resonance coupling
$\alpha$.

\subsubsection{MSF phase}
\label{subsec:msfspec}

The MSF quasiparticle spectrum (\ref{EkMSF}) appearing in
(\ref{deltaHmsf}) (and summarized in Fig.\ \ref{fig:spectrum}), may
be written in the form
\begin{equation}
E_{\kv \s} = \sqrt{(\eps_{\kv \s} + \eps_{\s-})
(\eps_{\kv \s} + \eps_{\s+})}
\label{spectrum}
\end{equation}
in which the (positive) energies $\eps_{\s\pm}$ are given by
\bea
\eps_{1\pm} &=& -\nu/2 + (g_{12} - g_2/2) |\Psi_{20}|^2
\pm \alpha |\Psi_{20}|
\nn \\
\eps_{2+} &=& 2g_2 |\Psi_{20}|^2,\ \eps_{2-} = 0.
\label{4.26}
\eea
The molecule-like branch ($\s = 2$) is gapless (consistent with
Goldstone's theorem), having an acoustic spectrum $E_{\kv 2}^{\rm
MSF} \approx \hbar c_2^{\rm MSF} k$ at small $k$ with sound speed
\begin{equation}
c_2^{\rm MSF} = |\Psi_{20}| \sqrt{g_2/m_2},
\label{4.27}
\end{equation}
corresponding to collective, long wavelength oscillations of the
molecular condensate. The spectrum crosses over to a particle-like
$E^{\rm MSF}_{\kv 2} \approx \eps_{\kv 2}$, for $k
\xi_\mathrm{coh}^{\rm MSF} \gg 1$, where
\begin{equation}
\xi_\mathrm{coh}^{\rm MSF} = \frac{\hbar}{2m_2 c_2^{\rm MSF}}
\label{4.28}
\end{equation}
is a coherence length beyond which superfluid behavior sets in: the
collective superfluid response\cite{foot:trap} dominates
disturbances with wavelength longer than $\xi_\mathrm{coh}$, while
the microscopic single molecule response\cite{foot:roton} dominates
those with shorter wavelength. This length $\xi_\mathrm{coh}^{\rm
MSF} \propto 1/|\Psi_{20}| \propto 1/\sqrt{\mu_2}$ diverges as the
normal phase boundary, $\mu_2 = 0$, is approached.

In contrast, the atomic-like branch has a gap
\begin{equation}
E_{\rm gap}^{\rm MSF} = E_{{\bf 0}1}^{\rm MSF}
= \sqrt{\eps_{1+} \eps_{1-}},
\label{msfgap}
\end{equation}
which closes with increasing $\nu$ precisely on the ASF--MSF
transition line, the latter being equivalent to the condition
$\eps_{1-} = 0$. This leads to the critical detuning
\bea
\nu_c &=& -(g_2 - 2g_{12}) |\Psi_{20}|^2
- 2\alpha |\Psi_{20}|
\nn \\
&=& 2\mu - \frac{g_2}{2g_{12}^2}
\left(\alpha^2 + 2\mu g_{12}
\pm \alpha \sqrt{\alpha^2 + 4\mu g_{12}} \right)
\nn \\
&\to& 2\mu - \frac{g_2 \mu^2}{\alpha^2},\ g_{12} \to 0,
\label{nucrit1}
\eea
where the second line requires $\mu > -\alpha^2/4g_{12}$, and
follows by substituting $|\Psi_{20}|^2 = (2\mu-\nu)/g_2$ and solving
for $\nu$. The existence of two solutions reflects the reentrant
behavior as a function of chemical potential seen in Fig.\
\ref{phasediagramNuT}.

At low temperature and for weak interactions, the condensate depletion
is minimal, $n_{20} = |\Psi_{20}|^2 \approx n/2$, and the critical
detuning for the quantum MSF--ASF transition is given by
\begin{equation}
\nu_c(T=0) \approx -\frac{1}{2}(g_2 - 2g_{12})n
- \alpha \sqrt{2n}
\label{nucrit0}
\end{equation}
The behavior of $\nu_c(T)$ for high temperature [as well as the
corresponding temperature dependence of the condensate $n_{20}(T)$
at fixed density $n$], illustrated in Fig.\ \ref{phasediagramNuT},
will be discussed in Sec.\ \ref{sec:td} below.

\subsubsection{ASF phase}
\label{subsec:amsfspec}

In the ASF the extremum conditions (\ref{saddlepoint}) allow $\tilde
\eps_{\kv \s}$ to be reduced to the forms
\bea
\tilde \eps_{\kv 1} &=& \eps_{\kv 1} + \lambda_1 + \alpha_1
\nn \\
\tilde \eps_{\kv 2} &=& \eps_{\kv 2} + \lambda_2 + \alpha_2,
\label{4.31}
\eea
and from (\ref{4.5}) one has
\begin{equation}
t_2 - t_1 = \alpha_3,
\label{4.32}
\end{equation}
with the definitions,
\bea
\alpha_1 &=& 2\alpha |\Psi_{20}|
\nn \\
\alpha_2 &=& \frac{1}{2}\alpha |\Psi_{10}|^2/|\Psi_{20}|
\nn \\
\alpha_3 &=& \alpha |\Psi_{10}| = \sqrt{\alpha_1 \alpha_2}.
\label{4.33}
\eea
Substituting (\ref{4.31})--(\ref{4.33}) into (\ref{4.20}) one
obtains,
\bea
e_\kv + d_\kv &=& -\alpha_3 (t_1+t_2)
+ (\eps_{\kv 1} + \alpha_1)(\eps_{\kv 1} + \alpha_1 + 2\lambda_1)
\nn \\
e_\kv - d_\kv &=& -\alpha_3 (t_1+t_2)
+ (\eps_{\kv 2} + \alpha_2)(\eps_{\kv 2} + \alpha_2 + 2\lambda_2)
\nn \\
c_\kv^{(1)} &=& (\eps_{\kv 2} + \alpha_2)(t_1+t_2)
- \alpha_3(\eps_{\kv 1} + \alpha_1 + 2\lambda_1)
\nn \\
c_\kv^{(2)} &=& (\eps_{\kv 1} + \alpha_1)(t_1+t_2)
- \alpha_3(\eps_{\kv 2} + \alpha_2 + 2\lambda_2).
\nn \\
\label{4.34}
\eea
At $\kv = {\bf 0}$ it is easy to verify that
\bea
e_{\bf 0} + d_{\bf 0} &=&
-\sqrt{\frac{\alpha_1}{\alpha_2}} c_{\bf 0}^{(1)}
= -\frac{2|\Psi_{20}|}{|\Psi_{10}|} c_{\bf 0}^{(1)}
\nn \\
e_{\bf 0} - d_{\bf 0} &=&
-\sqrt{\frac{\alpha_2}{\alpha_1}} c_{\bf 0}^{(2)}
= -\frac{|\Psi_{10}|}{2|\Psi_{20}|} c_{\bf 0}^{(2)},
\label{4.35}
\eea
and therefore that
\begin{equation}
e_{\bf 0}^2 = d_{\bf 0}^2 + c_{\bf 0}^{(1)} c_{\bf 0}^{(2)}.
\label{4.36}
\end{equation}
Substituting these results into (\ref{EkASF}) one obtains the
excitation energies at zero momentum, i.e., the gaps:
\bea
E_{{\bf 0} 1}^{\rm ASF} &=& \sqrt{2 e_{\bf 0}}
\nn \\
&=& \sqrt{\alpha_1(\alpha_1 + 2\lambda_1)
+ \alpha_2(\alpha_2 + 2\lambda_2)
- 2 \alpha_3(t_1 + t_2)}
\nn \\
E_{{\bf 0} 2}^{\rm ASF} &=& 0,
\label{gapsASF}
\eea
which confirms the existence of one gapped and one gapless
mode\cite{3vs1comment} in the ASF state. From (\ref{4.5}) and
(\ref{4.33}) one sees that $\alpha_2$, $\alpha_3$, $\alpha_1 +
2\lambda_1$, and hence $e_{\bf 0}$, vanish on the MSF--ASF phase
boundary where $\Psi_{10} = 0$.  The gap therefore closes on the
transition line, as expected.  Note also that if $\alpha = 0$ one
has $e_{\bf 0} = 0$, and the atomic-like gap remains closed
throughout the ASF phase, as expected from the additional
spontaneously broken $U(1)$ symmetry (separate atom and molecule
number conservation) and associated Goldstone modes, as discussed at
the beginning of this section and in Sec.\ \ref{sec:symmetries}.

The small $k$ (low-energy) behavior of the excitation spectra are
now examined in the ASF phase, $|\Psi_{10}| > 0$, and in the
neighborhood of the MSF--ASF transition, where $|\Psi_{10}| \to 0$,
but $\Psi_{20}$ remains finite. To this end, the $|\Psi_{10}|$ and
$k$ dependencies are isolated by writing
\bea
e_{\bf 0} &=& 2\alpha |\Psi_{10}|^2
\left(g_e + \alpha \frac{|\Psi_{10}|^2}{16|\Psi_{20}|^2} \right)
\nn \\
d_{\bf 0} &=& 2\alpha |\Psi_{10}|^2
\left(g_d - \alpha \frac{|\Psi_{10}|^2}{16|\Psi_{20}|^2} \right)
\nn \\
c_{\bf 0}^{(1)} &=& -\alpha \frac{|\Psi_{10}|^3}{|\Psi_{20}|} g_c^{(1)}
\nn \\
c_{\bf 0}^{(2)} &=& -4\alpha |\Psi_{10}| |\Psi_{20}|
\left(g_c^{(2)} + \alpha \frac{|\Psi_{10}|^2}{8|\Psi_{20}|^2} \right)
\nn \\
\delta e_\kv &\equiv& e_\kv - e_{\bf 0}
\nn \\
&=& |\Psi_{20}| \eps_{\kv 1} \left(\gamma_e
+ \frac{|\Psi_{10}|^2}{|\Psi_{20}|^2} \delta_e
+ \frac{5 \eps_{\kv 1}}{8 |\Psi_{20}|} \right)
\nn \\
\delta d_\kv &\equiv& d_\kv - d_{\bf 0}
\nn \\
&=& |\Psi_{20}| \eps_{\kv 1} \left(\gamma_d
+ \frac{|\Psi_{10}|^2}{|\Psi_{20}|^2} \delta_d
+ \frac{3 \eps_{\kv 1}}{8 |\Psi_{20}|} \right)
\nn \\
\delta c_\kv^{(1)} &\equiv& c_\kv^{(1)} - c^{(1)}_{\bf 0}
= |\Psi_{10}| \eps_{\kv 1} \gamma_c^{(1)}
\nn \\
\delta c_\kv^{(2)} &\equiv& c^{(2)}_\kv - c^{(2)}_{\bf 0}
= |\Psi_{10}| \eps_{\kv 1} \gamma_c^{(2)}
\label{4.38}
\eea
in which the coefficients
\bea
g_e &=& (g_1 - g_{12} + g_2/4) |\Psi_{20}| + \alpha/2
\nn \\
g_d &=& (g_1 - g_2/4) |\Psi_{20}|
\nn \\
g_c^{(1)} &=& (2g_1 - g_{12}) |\Psi_{20}| + \alpha/2 = g_e + g_d
\nn \\
g_c^{(2)} &=& (g_2/2 - g_{12}) |\Psi_{20}| + \alpha/2 = g_e - g_d
\nn \\
\gamma_e &=& g_2 |\Psi_{20}|/2 + \alpha
\nn \\
\delta_e &=& g_1 |\Psi_{20}| + \alpha/4
\nn \\
\gamma_d &=& -g_2 |\Psi_{20}|/2 + \alpha
\nn \\
\delta_d &=& g_1 |\Psi_{20}| - \alpha/4
\nn \\
\gamma_c^{(1)} &=& g_{12} |\Psi_{20}| - 3\alpha/2
\nn \\
\gamma_c^{(2)} &=& 2 g_{12} |\Psi_{20}| - 3\alpha/2
\label{4.39}
\eea
are all finite for $|\Psi_{10}| = 0$ and $k = 0$.

\paragraph{At the ASF--MSF critical point, $|\Psi_{10}| = 0$:}

On the ASF--MSF transition line, $|\Psi_{10}| = 0$ (and also for a
vanishing Feshbach resonance coupling, $\alpha = 0$, when the order
parameter phases are decoupled), the zero momentum
coefficients---the first four lines of (\ref{4.38})---vanish
identically and one obtains two {\em gapless} spectra
\begin{equation}
(E_{\kv \s}^{\rm crit})^2 = \delta e_\kv \pm \sqrt{\delta d_\kv^2
+ \delta c_\kv^{(1)} \delta c_\kv^{(2)}},
\label{4.40}
\end{equation}
which lead to two acoustic critical modes, $E_{\kv \s}^{\rm crit}
\approx \hbar c_\s^{\rm crit} k$ at small $k$. For $|\Psi_{10}| =
0$, $\delta c_\kv^{(1)}, \delta c_\kv^{(2)}$ vanish and the sound
speeds are given by
\begin{equation}
(c_\s^{\rm crit})^2 = \frac{|\Psi_{20}|}{2m_1}
(\gamma_e \pm \gamma_d)
= \left\{\begin{array}{ll}
\frac{\alpha}{m_1} |\Psi_{20}|, & \sigma = 1 \\
\frac{g_2}{2m_1}|\Psi_{20}|^2, & \sigma = 2.
\end{array} \right.
\label{4.41}
\end{equation}

\paragraph{In the ASF phase, $|\Psi_{10}| > 0$:}

As found above, Eq.\ (\ref{gapsASF}), in the ASF phase the spectrum of
out-of-phase excitations, labeled by $\s=1$, is gapped, while that for
$\s=2$ excitations, corresponding to in-phase fluctuations of the two
condensates is given by
\bea
(E_{\kv 2}^{\rm ASF})^2 &=& \frac{e_{\kv}^2 - d_\kv^2
- c_\kv^{(1)} c_\kv^{(2)}}{E_{\kv 1}^2}
\label{4.43} \\
&\approx& \frac{1}{2 e_{\bf 0}}\left[(e_{\bf 0} + d_{\bf 0})
\left(\delta e_\kv - \delta d_\kv
+ \frac{|\Psi_{10}|}{2|\Psi_{20}|} \delta c_\kv^{(2)} \right)
\right.
\nn \\
&&\left. +\ (e_{\bf 0} - d_{\bf 0}) \left(\delta e_\kv + \delta d_\kv
+ \frac{2|\Psi_{20}|}{|\Psi_{10}|} \delta c_\kv^{(1)} \right) \right],
\nn
\eea
in which (\ref{4.35}) has been used.  As expected from the general
symmetry arguments discussed in Sec.\ \ref{sec:symmetries} and
earlier in this section, the in-phase excitations are
acoustic, $E_{\kv 2}^{\rm ASF} \approx \hbar c_2^{\rm ASF} k$ at
small $k$, with sound speed given by
\be
(c_2^{\rm ASF})^2 = |\Psi_{20}|
\frac{(e_{\bf 0} + d_{\bf 0}) f^{(1)}_{\bf 0}
+ (e_{\bf 0} - d_{\bf 0}) f_{\bf 0}^{(2)}}{4e_{\bf 0}m_1},
\label{4.44}
\ee
with the constants $f_0^{(\s)}$ defined by
\bea
f_{\bf 0}^{(1)} &\equiv& g_2 |\Psi_{20}|
+ (g_{12}|\Psi_{20}| - \alpha/4)
\frac{|\Psi_{10}|^2}{|\Psi_{20}|^2}
\nn \\
f_{\bf 0}^{(2)} &\equiv& 2g_{12} |\Psi_{20}| - \alpha
+ 2g_1|\Psi_{20}| \frac{|\Psi_{10}|^2}{|\Psi_{20}|^2}.
\label{f0s}
\eea

Note in passing, that, as expected, for a vanishing Feshbach
resonance coupling, $\alpha = 0$, {\em both} in-phase and out-phase
modes become acoustic, with sound speeds
\bea
&&2m_1 c_{\s0}^2 = \frac{1}{2} g_2 |\Psi_{20}|^2
+ g_1 |\Psi_{10}|^2
\label{4.42} \\
&&\pm\ \sqrt{\left(\frac{1}{2} g_2 |\Psi_{20}|^2
- g_1 |\Psi_{10}|^2 \right)^2
+ 2g_{12}^2 |\Psi_{10}|^2 |\Psi_{20}|^2}.
\nn
\eea
that are real and positive for $g_1 g_2 > g_{12}^2$.

\paragraph{Scaling form for small $k$ and $|\Psi_{10}|$:}

It is easy to check that the $|\Psi_{10}| \to 0$ limit of
(\ref{4.44}) is very different from (\ref{4.41}), which therefore
does not commute with the $k \to 0$ limit.  In order to show more
carefully the distinction between these two limits, a scaling form
that is valid when $k$, $|\Psi_{10}|$ are both small, but have
arbitrary ratio, is derived.  By keeping only leading terms in
$|\Psi_{10}|^2$ and $\eps_{\kv 1}$, one obtains
\bea
E_{\kv \s}^{\rm ASF} &=& 2 \alpha |\Psi_{10}|^2
\bigg[g_e + \gamma_e y
\label{4.45} \\
&&\pm\ \sqrt{(g_d + \gamma_d y)^2
+ (g_c^{(1)} - 2 \gamma_c^{(1)} y) g_c^{(2)}} \bigg]
\nn
\eea
in which the dimensionless scaling variable is
\begin{equation}
y = \frac{|\Psi_{20}| \eps_{\kv 1}}{2\alpha |\Psi_{10}|^2}.
\label{4.46}
\end{equation}
It is easily checked that for large $y$ (\ref{4.41}) is recovered,
while for small $y$ (\ref{4.44}) is recovered.

\subsection{MSF paired ground-state wave function}
\label{sec:gswavefn}

The zero temperature molecular superfluid ground state is
constructed by requiring that it be the quasiparticle vacuum:
\begin{equation}
\hat{\gamma}_{\kv\s} \ket{\mathrm{MSF}} = 0,\
\mbox{for all } \kv \neq 0,\s.
\label{4.47}
\end{equation}
The additional constraint
\be
\hat a_{{\bf 0}2} \ket{\mathrm{MSF}} = \sqrt{V}
\Psi_{20}\ket{\mathrm{MSF}},
\label{4.48}
\ee
where $\hat a_{{\bf 0}2}= V^{-1/2} \int d\xv \hat \psi_2(\xv)$,
ensures that the MSF is a coherent state for the lowest single
particle trap state ${\bf k} = {\bf 0}$ and thereby has the correct
amplitude $\Psi_{20}$ corresponding to molecular superfluid order.

Using the commutation relations
\begin{equation}
[\hat a, e^{\lambda \hat a^\dag}]
= \lambda e^{\lambda \hat a^\dag},\ \
[\hat a, e^{\lambda \hat a^\dag \hat b^\dag}]
= \lambda b^\dag e^{\lambda \hat a^\dag \hat b^\dag},
\label{4.49}
\end{equation}
where $\hat a$, $\hat b$ are any two independent harmonic oscillator
operators, it follows that the state
\begin{equation}
\ket{\rm MSF} = \exp\bigg({\Psi_{20} \sqrt{V} \hat a^\dag_{{\bf 0}2}
-\frac{1}{2} \sum_{\kv \neq 0,\s} \chi_{\kv\s}
\hat{a}^\dag_{\kv\s} \hat{a}^\dag_{-\kv\s}} \bigg) \ket{0}
\label{MSF}
\end{equation}
indeed obeys (\ref{4.47}) and (\ref{4.48}) with the choice
\begin{equation}
\chi_{\kv\s} = \frac{v_{\kv\s}}{u_{\kv\s}}
= \frac{\tilde \eps_{\kv\s} - E_{\kv\s}^{\rm MSF}}{\lambda_\s}.
\label{pairwf1}
\end{equation}
The factor of $1/2$ in front of the sum is required because each
term actually appears twice, once for ${\bf k}$ and once for $-{\bf
k}$.

The quantity $\chi_{\kv\s}$ may be identified as the Fourier
transform of the atomic ($\s=1$) and molecular ($\s=2$) pair
wavefunctions with zero center of mass momentum.  The asymptotic
long-distance behavior of its Fourier transform $\chi_\s(\rv)$,
which is now computed, is governed by the singularity of
$\chi_{\kv\s}$ nearest $k=0$. Since $\chi_{\kv\s}$ depends only on
the magnitude $k$, one may use the Bessel function identity
\begin{equation}
\int d\Omega_{\kv} e^{i\kv \cdot \rv}
= 2\pi^{d/2} \left(\frac{2}{kr} \right)^{\frac{d-2}{2}}
J_{\frac{d-2}{2}}(kr)
\label{4.51a}
\end{equation}
to perform the $d-1$-dimensional angular integration in its Fourier
transform, yielding
\begin{eqnarray}
\chi_\s(\rv) &=&\int \frac{d\kv}{(2\pi)^d}
e^{i\kv\cdot\rv}\chi_{\kv\s},
\nn \\
&=& \int_0^\infty \frac{2 k^{d-1} dk}{(4\pi)^{d/2}}
\chi_{\kv\s} \left(\frac{2}{kr} \right)^{\frac{d-2}{2}}
J_{\frac{d-2}{2}}(kr).\hspace{1cm}
\label{4.51b}
\end{eqnarray}
Since the right hand side of (\ref{4.51a}) is an even function of
$k$, the integration may be extended to the full real line, avoiding
the branch cut along $k < 0$ by shifting the contour an
infinitesimal distance into the upper half plane, and simultaneously
dividing by the factor $1+e^{i\pi(d-1)}$.  Since $k^{d-1}
(2/kr)^{(d-2)/2} J_{-(d-2)/2}(kr)$ is analytic through the origin,
and an odd function of $k$, its integral vanishes, and one may write
$\chi_\s({\bf r})$ in the form
\begin{equation}
\chi_\s(\rv) = \int_{-\infty+i\eta}^{\infty+i\eta}
\frac{k^{d-1} dk}{(4\pi)^{d/2}}
\chi_{\kv\s} \left(\frac{2}{kr} \right)^{\frac{d-2}{2}}
H^{(1)}_{\frac{d-2}{2}}(kr),
\label{4.51c}
\end{equation}
in which $\eta$ is a positive infinitesimal and $H^{(1)}_\nu(x)$ is
a Hankel function of the first kind.

From (\ref{spectrum}), one observes that $E_{\kv\s}^{\rm MSF}$ has
finite branch cuts along the imaginary $k=|\kv|$ axis over the
intervals $\pm i(k_{\s-},k_{\s+})$, where $k_{\s\pm} =
\sqrt{2m\eps_{\s\pm}}/\hbar$. To evaluate $\chi_\s(\rv)$ the
integration contour is deformed into the upper half plane to run
down, around the origin, and then back up the imaginary axis,
avoiding the upper branch cut.  Since the $H_\nu^{(1)}(x)$ decays
exponentially in the upper-half plane, one can close the contour and
then shrink it around the upper branch cut of $E_{\kv\s}^{\rm MSF}$.
Because the integrand is finite near the branch points, the
infinitesimal circular parts of the contour integral, and the
complete integrals of the analytic parts of $\chi_{\kv\s}$, both
vanish. The remaining parts on the left and right sides running
along the branch cut double up, giving
\begin{eqnarray}
\chi_\s(\rv) &=&
\frac{\hbar^2}{\pi (2\pi)^{d/2} \lambda_\s m_\s} r^{\frac{2-d}{2}}
\int_{k_{\s-}}^{k_{\s+}} d\kappa \kappa^{d/2}
K_{\frac{d-2}{2}}(\kappa r)
\nonumber \\
&& \ \ \ \ \ \ \ \ \
\times\ \sqrt{(\kappa^2-k_{\s-}^2)(k_{\s+}^2 - \kappa^2)}
\label{4.51d}
\end{eqnarray}
with $K_\nu(z) = (\pi i/2) e^{i \nu \pi/2} H_\nu^{(1)}(iz)$ the
modified Bessel function.  For large $k_{\s+} r$ the integral is
dominated by the region near $k_{\s-}$, and one may safely (with
exponential accuracy) extend the upper limit to infinity and
approximate the square root factor by the form
\begin{equation}
\left\{\begin{array}{ll}
\sqrt{2k_{\s-}(k_{\s+}^2-k_{\s-}^2)} \sqrt{\kappa - k_{\s-}},
& \kappa - k_{\s-} k_{\s-} r \gg 1 \\
k_{\s+} \kappa, & k_{\s-}r \ll 1,
\end{array} \right.
\label{4.51e}
\end{equation}
the lower relation being especially required for $\s=2$ where
$k_{2-} = 0$. One therefore obtains
\begin{widetext}
\begin{equation}
\chi_\s(\rv) \approx \left\{\begin{array}{ll}
\frac{\hbar^2}{2\lambda_\s m_\s}
\sqrt{k_{\s+}^2 - k_{\s-}^2} \left(\frac{k_{\s-}}{2\pi} \right)^{d/2}
\frac{e^{-k_{\s-} r}}{r^{\frac{d+2}{2}}}, & k_{\s-} r \gg 1 \\
\frac{\hbar^2 k_{\s+}}{2 \lambda_\s m_\s}
\frac{\Gamma(\frac{d+1}{2})}{\pi^{\frac{d+1}{2}}} \frac{1}{r^{d+1}},
& k_{\s-}r \ll 1,
\end{array} \right.
\label{4.51f}
\end{equation}
\end{widetext}
in which the asymptotic form $K_\nu(x) \approx \sqrt{\pi/2x}
e^{-x}$, $|x| \gg 1$, has been used to obtain the first line, and
the identity\cite{GR}
\begin{equation}
\int_0^\infty x^\mu K_\nu(x) dx = 2^{\mu-1}
\Gamma\left(\frac{1+\mu+\nu}{2}\right)
\Gamma\left(\frac{1+\mu-\nu}{2}\right)
\label{4.51g}
\end{equation}
to obtain the second.

It thus follows that in the MSF phase the relative atomic wavefunction
decays exponentially according to $\chi_1(\rv) \sim e^{-r/\xi_\s}$,
with a decay length
\begin{equation}
\xi_1 = \frac{1}{k_{1-}} = \frac{\hbar}{\sqrt{2m_1 \eps_{1-}}},
\label{4.52}
\end{equation}
reflecting the confinement of (gapped) atomic excitations, and the
corresponding absence of atomic long-range order inside the MSF.
Since $\eps_{1-} \sim \nu - \nu_c$, $\xi_1 \sim (\nu -
\nu_c)^{-1/2}$ has a square root divergence as the ASF phase
boundary is approached. On the other hand, since $k_{2-} = 0$, the
molecular wavefunction $\chi_2(\rv) \sim 1/r^{d+1}$ has a power law
decay, reflecting the existence of molecular long-range order inside
the MSF.

Note that the ground state (\ref{MSF}), in addition to being a
molecular coherent state, is also an (atomic and molecular) {\em
pair} coherent state. It thus makes explicit that within the
molecular superfluid state, a molecular $\kv = 0$ condensation,
$\Psi_{20} \neq 0$, is accompanied by a nonzero BCS-like atomic
pairing at finite relative $\kv$, with an anomalous correlation
function,
\begin{equation}
\av{\hat{a}_{\kv 1} \hat{a}_{-\kv 1}} = -u_{\kv 1} v_{\kv 1}
= \frac{\alpha \Psi_{20}}{2E_{\kv 1}}.
\label{4.53}
\end{equation}
Exactly the same branch cut structure as described above applies to
the right hand side of (\ref{4.53}), and its Fourier transform, the
BCS-type atom pair correlation function, falls off exponentially at
the same rate $e^{-r/\xi_1}$.  The correlation length $\xi_1$ (that is
finite inside the MSF, but diverges as the transition into ASF is
approached) characterizes the size of the virtual cloud of atom pairs
surrounding each closed-channel molecule (whose size, $d_0$,
characterized by the microscopic range of the interatomic potential,
remains finite throughout).

On the other hand, the molecular anomalous pair correlation function
\begin{equation}
\av{\hat{a}_{\kv 2} \hat{a}_{-\kv 2}} = -u_{\kv 2} v_{\kv 2}
= \frac{g_2 \Psi_{20}^2}{2E_{\kv 2}}.
\label{4.54}
\end{equation}
exhibits a $1/k$ divergence near the origin [on top of the
$\Psi_{20}^2 V (2\pi)^d \delta(\kv)$ condensate contribution due to
the long-range order], so that its Fourier transform approaches the
$\Psi_{20}^2$ asymptote via a slow $1/r^{d-1}$ power law decay.
This is a signature of quantum fluctuations in the low energy
molecular Goldstone mode.

\subsection{Thermodynamics}
\label{sec:td}

As is clear from (\ref{deltaHmsf}) and (\ref{4.22}), within the
Bogoliubov approximation a superfluid (be it MSF or ASF) is a
coherent state with excitations described by a gas of noninteracting
bosonic Bogoliubov quasiparticles, $\hat \gamma_{\kv \s}$,
respectively given by (\ref{quasiparticle}) and (\ref{4.24}).
Thermodynamics is therefore easily computed in a standard way.

\subsubsection{MSF phase}
\label{subsec:tdmsf}

The free energy density in the MSF consists of the ground state
condensate energy ${\cal H}_{\rm mf} = {\cal H}[\Psi_{20}]$, plus a
contribution from the noninteracting Bogoliubov quasiparticles,
governed by $\delta\hat H_{\rm MSF}$, Eq.\ (\ref{deltaHmsf}). A
standard free boson computation gives
\begin{widetext}
\be
f_\mathrm{MSF}[h_2] = -\mu_2|\Psi_{20}|^2
+ \frac{1}{2} g_2|\Psi_{20}|^4
- \mathrm{Re}[h_2^* \Psi_{20}]
+\ \sum_\s \int \frac{d\kv}{(2\pi)^d}
\left[\frac{1}{\beta} \ln(1-e^{-\beta E_{\kv\s}^\mathrm{MSF}})
+ \frac{1}{2}(E_{\kv\s}^\mathrm{MSF} - \tilde{\eps}_{\kv\s}) \right],
\label{FreeEnergy}
\ee
\end{widetext}
where a complex molecular ``source field'' $h_2$ (that vanishes for
a physical system) has been included. As usual, derivatives of
$f_\M[h_2]$ with respect to $h_2$ generate correlation functions of
the molecular field. In interpreting this quantity, it is important
to emphasize that $\Psi_{20}$ here (in an unfortunate abuse of
notation) is the \emph{mean field} order parameter, an explicit
function of the Hamiltonian parameters $\mu,\nu,h_2$, etc., that
does not include any fluctuation corrections.\cite{W88} The leading
Bogoliubov corrections are provided by the $h_2$ derivatives of
$f_\M[h_2]$. For the molecular condensate order parameter, corrected
by quantum and thermal fluctuations this gives:
\begin{equation}
\Psi_{2}\equiv \Psi_{20} + \delta \Psi_{2}
= -2 \left(\frac{\partial f_\M}{\partial h_2^*}
\right)_{h_2=0},
\label{4.56}
\end{equation}
in which $h_2$ enters through its explicit appearance in the first
line of (\ref{FreeEnergy}) as well as implicitly through $\Psi_{20}$.
The extremum property of $\Psi_{20}$ with respect to
$H_\text{mf}[\Psi_{20}]$ therefore gives
\bea
\delta \Psi_{2}
&=& -2 \left(\frac{\partial f_\M}
{\partial \Psi_{20}^*} \right)
\left(\frac{\partial \Psi_{20}^*}{\partial h_2^*} \right)_{h_2=0}
\nn \\
&=& -\frac{\Psi_{20}}{2\mu_2}
[I_{d,1}(\mu,\nu) + 2I_{d,2}(\mu,\nu)],
\label{4.57}
\eea
where the mean field longitudinal susceptibility is $(\partial
\Psi_{20}^*/\partial h_2^*)_{h_2=0} = 1/(-2\mu_2 +
6g_2|\Psi_{20}|^2) = 1/4\mu_2$. Consistency requires that the
original forms (\ref{4.5}) be used for the $\Psi_{20}$ dependence,
and
\bea
I_{d,1} &=& \int \frac{d\kv}{(2\pi)^d}
\left[\left(n_{\kv 1} + \frac{1}{2} \right)
\frac{2 g_{12} \tilde{\eps}_{\kv 1}
- \alpha^2}{E^\M_{\kv 1}} - g_{12} \right]
\nn\\
I_{d,2} &=& \int \frac{d\kv}{(2\pi)^d}
\left[ \left(n_{\kv 2} + \frac{1}{2} \right)
\frac{2 g_2\tilde{\eps}_{\kv 2} - g_2^2 |\Psi_{20}|^2}
{E^\M_{\kv 2}} - g_2 \right],
\nn \\
\label{4.58}
\eea
where $n_{\kv\s} = (e^{\beta E_{\kv\s}}-1)^{-1}$ are the standard
Bose occupation factors for Bogoliubov quasiparticles.  The number
density to this same order is
\begin{equation}
n = -\left(\frac{\partial f_\M}{\partial \mu}\right)_{T,\nu}
= 2|\Psi_2|^2 + \delta n_\M(T,\nu),
\label{density}
\end{equation}
where the density of bosons not condensed into the lowest $\kv =0$
single particle state (i.e., the condensate depletion) is given by
\begin{equation}
\delta n_\M(T,\nu) = \sum_\s \s \int\frac{d\kv}{(2\pi)^d}
\left[v_{\kv\s}^2 + (u_{\kv\s}^2 + v_{\kv \s}^2)
n_{\kv\s} \right],
\label{4.60}
\end{equation}
The depletion density $\delta n_\M(T,\nu)$ comes from the explicit
$\mu$-dependence in $E^\M_{\kv \s}$ and remains finite even at zero
temperature due to the interaction-induced zero-point contribution
$v_{\kv\s}^2$.\cite{W88} The remaining implicit $\mu$-dependence
entering through the condensate $\Psi_{20}$ gives rise to the term
$|\Psi_2|^2$ in (\ref{density}), in place of the mean field
condensate density $|\Psi_{20}|^2$.

Evaluating (\ref{4.60}) at $T=0$ and $\nu = \nu_c$ one obtains
\begin{widetext}
\be
\delta n_\M(0,\nu_c) = \sum_\s \frac{\s}{2}
\int\frac{d\kv}{(2\pi)^d} \left[\frac{\eps_{\kv\s} + \eps_{\s+}/2}
{\sqrt{\eps_{\kv\s}(\eps_{\kv\s} + \eps_{\s+})}} - 1 \right]
= \frac{B_d}{2 \Gamma(d/2)} \left(\frac{2\pi m}{h^2} \right)^{d/2}
(\eps_{1+}^{d/2} + 2^{(d+2)/2} \eps_{2+}^{d/2})
\label{4.61}
\ee
in which $\eps_{1+}(\nu_c) = 2\alpha|\Psi_{20}|$, $\eps_{2+}(\nu_c)
= 2g_2 |\Psi_{20}|^2$, and the coefficient is given by
\be
B_d = \int_0^\infty dv v^{(d-2)/2}
\left[\frac{v+1/2}{\sqrt{v(v+1)}} - 1 \right]
= \frac{1}{d\sqrt{\pi}} \Gamma\left(\frac{d-1}{2}\right)
\Gamma\left(\frac{4-d}{2}\right).
\label{4.62}
\ee
In $d=3$ one finds $B_3 = 1/3$ and Eq.\ (\ref{max_depletion}) quoted
in the Introduction immediately follows.  Since
$\varepsilon_{1+}^{d/2} \propto n_{20}^{d/4}$, this ``correction''
term becomes much larger than $n_{20}$ close to the MSF--N
transition line. This is a sign of the breakdown of the mean field
description of criticality, and (\ref{4.62}) ceases to valid in this
nontrivial critical regime.\cite{WRFS}

For $\nu < \nu_c$ (i.e., inside MSF phase) one obtains:
\be
\delta n_\M(0,\nu) = \frac{B_d}{2 \Gamma(d/2)}
\left(\frac{2\pi m}{h^2} \right)^{d/2}
\left\{[1 + b_d(\delta)] \eps_{1+}^{d/2}
+ 2^{(d+2)/2} \eps_{2+}^{d/2} \right\}
\label{4.63}
\ee
in which $\delta = \eps_{1-}(\nu)/\eps_{1+}(\nu)$, and
\begin{equation}
b_d(\delta) = \int_0^\infty dv\frac{v^{(d-2)/2}}{B_d}
\left[\frac{v+(1+\delta)/2}{\sqrt{(v+\delta)(v+1)}}
- \frac{v+1/2}{\sqrt{v(v+1)}} \right].
\label{4.64}
\end{equation}
\end{widetext}
Of interest is the behavior of this integral near $\nu_c$, i.e.,
for small $\delta$.  The singular behavior can be obtained by
first computing the derivative
\begin{equation}
\frac{d b_d}{d\delta} = -\frac{1-\delta}{4B_d}
\int_0^\infty \frac{v^{(d-2)/2} dv}
{(v+\delta)^{3/2}\sqrt{v+1}}.
\label{4.65}
\end{equation}
For $d < 3$ the integral diverges as $\delta \to 0$, and one
obtains
\begin{equation}
\frac{d b_d}{d\delta}= -\frac{1-\delta}{4B_d}
[\beta_{d,s} \delta^{(d-3)/2} + \beta_{d,1} + O(\delta)],
\label{4.66}
\end{equation}
where the singular coefficient $\beta_{d,s}$ is obtained from the
small $v$ part of the (infrared divergent) integral by scaling out
$\delta$ via the change of variable $u = v/\delta$:\cite{GR}
\begin{equation}
\beta_{d,s} = \int_0^\infty \frac{u^{(d-2)/2}du}{(u+1)^{3/2}}
= \frac{2}{\sqrt{\pi}} \Gamma\left(\frac{d}{2}\right)
\Gamma\left(\frac{3-d}{2}\right),
\label{4.67}
\end{equation}
The linear term is obtained by first subtracting this small $v$
singular (in $\delta$) part of the integral, and then letting $\delta
\to 0$:\cite{GR}
\bea
\beta_{d,1} &=& \int_0^\infty v^{(d-5)/2} dv
\left(\frac{1}{\sqrt{v+1}} - 1 \right)
\nn \\
&=& \frac{1}{\sqrt{\pi}} \Gamma\left(\frac{d-3}{2}\right)
\Gamma\left(\frac{4-d}{2}\right).
\label{4.68}
\eea
On the other hand, for $3 < d < 4$, $db_d(0)/d\delta$ is
finite,\cite{foot:d4} and one finds the leading term $\beta_{d,1}$
simply by setting $\delta = 0$.  Related to this, the singular term
no longer diverges, and it is obtained by first subtracting the
$\beta_{d,1}$ (the $\delta = 0$) term, and then again simply scaling
$\delta$ out of the integral. One may verify that the final results
for both coefficients are identical to (\ref{4.67}) and
(\ref{4.68}). Integrating (\ref{4.66}) with respect to $\delta$, one
finally obtains
\begin{equation}
b_d(\delta) = -\frac{1}{4B_d}
\left[\frac{2}{d-1} \beta_{d,s} \delta^{(d-1)/2}
+ \beta_{d,1} \delta \right] [1 + O(\delta)].
\label{4.69}
\end{equation}
In $d=3$ both $\beta_{d,s}$ and $\beta_{d,1}$ separately diverge.
However the sum is finite, giving rise to a logarithmic dependence on
$\delta$:
\begin{equation}
b_3(\delta) = -\frac{3}{4}\delta \left\{\ln(1/\delta) + 4 \ln(2)
- 1 \right\} [1 + O(\delta)].
\label{4.70}
\end{equation}
This same result also follows from a direct asymptotic evaluation of
the integral (\ref{4.65}) in $d = 3$.

Defining a $T=0$ critical exponent $\tilde\alpha$ via $\delta
n_\M(T=0,\delta) \sim \delta^{1-\tilde\alpha}$ (i.e., the
zero-temperature quantum transition analog of a specific heat
exponent), one finds
\begin{equation}
\tilde\alpha = \left\{\begin{array}{ll}
\frac{3-d}{2}, & d \neq 3 \\
0 \mbox{ (log)}, & d=3
\end{array} \right. .
\label{4.71}
\end{equation}
This result will be modified by critical fluctuations sufficiently
close to the MSF--ASF quantum phase transitions.\cite{WRFS}  The
resulting behavior of the condensate depletion $\delta n_\M(T,\nu)$
is illustrated in Fig.\ \ref{fig:depletion}.

Before ending this subsection, the order of magnitude of the MSF
zero-temperature depletion (\ref{4.63}) is examined in light of the
identification in Sec.\ \ref{sec:2channelmodel} of the small
parameter $\gamma$, Eq.\ (\ref{eq:gamma}).  In order for the
fluctuation correction (\ref{4.63}) to be accurate, it is necessary
that it be much smaller than the mean field value $n_{20}$ (and the
same should be true of the correction $\delta \Psi_2$ relative to
$\Psi_{20}$). Using forms (\ref{4.26}) for the energy gaps, one
obtains
\begin{eqnarray}
\frac{(2m \varepsilon_{1+}/\hbar^2)^{d/2}}{n_{20}}
&\approx& n_{20}^{(d-4)/4} (4m \alpha/\hbar^2)^{d/2},
\nonumber \\
&\propto& (n_{20} r_0^d)^{(d-4)/4},
\nonumber \\
&\propto& \gamma^{d(4-d)/4},
\label{eq:relcorr1} \\
\frac{(2m \varepsilon_{2+}/\hbar^2)^{d/2}}{n_{20}},
&=& n_{20}^{(d-2)/2} (2m g_2/\hbar^2)^{d/2},
\nonumber \\
&\propto& (n_{20} a_2^d)^{(d-2)/2},
\label{eq:relcorr2}
\end{eqnarray}
in which $g_2 \propto (\hbar^2/2m) a_2^{d-2}$ and $\alpha \propto
(\hbar^2/2m) |r_0|^{(d-4)/2}$ relate the Hamiltonian parameters to
the molecular scattering length and effective range [see
(\ref{Eq:width}) and (\ref{eq:r0})] in $d$ dimensions.  In
(\ref{eq:relcorr1}) it has been assumed that $\epsilon_{1+}$ is of
the same order of magnitude as it is on the MSF--ASF phase boundary,
where $\epsilon_{1+} = 2\alpha \sqrt{n_{20}}$.

It is seen that the two terms in (\ref{4.63}) are very different in
character. The second term, estimated via (\ref{eq:relcorr2}), is
the standard result for a monatomic Bose gas, and is (for $d > 2$)
small in the dilute limit, $n^{1/d} a_2 \ll 1$.  The first term,
estimated via (\ref{eq:relcorr1}), is small (for $d < 4$) only if
$\gamma \ll 1$.  However, this requires $n^{1/d} r_0 >> 1$, which
places a \emph{lower bound} on the density.  The expansion about
mean field theory presented in this paper therefore requires a
sufficiently narrow Feshbach resonance (small $\alpha$) such that
the separation of scales $r_0 \gg a_2$ exists, and its validity is
limited to densities in the intermediate regime
\begin{equation}
\frac{1}{r_0^d} \ll n \ll \frac{1}{a_2^d}.
\label{eq:densityrange}
\end{equation}
This confirms, within an explicit perturbation calculation, the
claims made in Sec.\ \ref{sec:2channelmodel}.

\subsubsection{ASF phase}

Next consider the $\nu > \nu_c$ case where both $\Psi_{\sigma 0}
\neq 0$. Computations in the ASF phase are most conveniently
performed by taking $h_{\sigma 0}$ and $\Psi_{\sigma 0}$ real and
positive at the outset. Expressions for thermodynamic quantities are
quite long and involved, and they will only be sketched here.

The Bogoliubov free energy density in the ASF is given by
\bea
f_\A &=& {\cal H}_\mathrm{mf}[\Psi_{10},\Psi_{20}]
\nn \\
&&+\ \sum_\s \int \frac{d\kv}{(2\pi)^d}
\left[\frac{1}{\beta} \ln(1-e^{-\beta E_{\kv\s}^\A}) \right.
\nn \\
&&\ \ \ \ \ \ \ \
\left.+\ \frac{1}{2}(E_{\kv\s}^\A - \tilde{\eps}_{\kv\s}) \right]
\label{eq:fbogasf}
\eea
where ${\cal H}_\text{mf}[\Psi_{10},\Psi_{20}]$ takes the form of
(\ref{HmuMFT}), but with, as in (\ref{FreeEnergy}), additional
ordering field terms $-\sum_\sigma \mathrm{Re}[h_\sigma^*
\Psi_{\sigma 0}]$ now included.

The Bogoliubov corrections to the mean field order parameter are
then found in the form
\bea
\Psi_\s \equiv \Psi_{\sigma 0} + \delta \Psi_{\sigma}
= -\left(\frac{\partial f_\A}{\partial h_\sigma}
\right)_{h_\sigma=0}
\label{eq:psi0asf}
\eea
with
\bea
\delta \Psi_\s &=& -\sum_{\s'} \left(\frac{\partial f_\A}
{\partial \Psi_{\sigma' 0}} \right)
\left(\frac{\partial \Psi_{\sigma' 0}}{\partial h_\sigma}
\right)_{h_\sigma=0}
\label{eq:delpsi}
\eea
in which, due to the mean field conditions (\ref{saddlepoint}), only
the non-mean field part of $f_\A$ actually contributes to
(\ref{eq:delpsi}).

Self-consistency, via the mean field equations, but with $h_1/2$,
$h_2/2$ replacing the zeroes on the left hand sides of
(\ref{saddlepointA}), (\ref{saddlepointB}), respectively, determines
the $h_\sigma$ and $\mu$-dependence of $\Psi_{\sigma 0}$. The four
(complex) equations for $(\partial \Psi_{\s' 0}/\partial
h_\s)_{h_\s=0}$ decouple into a pair of separate equations for
$(\partial \Psi_{\s 0}/\partial h_1)_{h_1 = h_2 = 0}$ and $(\partial
\Psi_{\s 0}/\partial h_2)_{h_1 = h_2 = 0}$:
\bea
\left[
\begin{array}{c}
(\partial \Psi_{1 0}/\partial h_\s)_{h_\s = 0} \\
(\partial \Psi_{2 0}/\partial h_\s)_{h_\s = 0}
\end{array} \right]
=\left(\begin{array}{cc}
A & B   \\
C & D
\end{array} \right)^{-1}
\ket{\sigma}
\label{eq:dpsidh}
\eea
where
\bea
A &=& 4 g_1 \Psi_{10}^2
\nn \\
B &=& 2 \Psi_{10} \left(2 g_{12} \Psi_{20} - \alpha \right)
\nn\\
C &=& 2 \Psi_{10} (2 g_{12} \Psi_{10} - \alpha)
\nn\\
D &=& 4 g_2 \Psi_{20}^2 + \frac{\alpha \Psi_{10}^2}{\Psi_{20}}
\eea
and $\ket{1}=\left(\begin{array}{c}
1 \\
0
\end{array}\right)$ and $\ket{2}=\left(\begin{array}{c}
0 \\
1
\end{array}\right)$.

The free energy derivatives in (\ref{eq:delpsi}), taken at constant
values of the Hamiltonian parameters $\mu_\s,g_\s,\alpha,g_{12}$,
are given by
\be
\frac{\partial f_\A}{\partial \Psi_{\sigma 0}}
= \sum_\sigma \int \frac{d\kv}{(2\pi)^d}
\left[\left(n_{\kv\sigma}+\frac{1}{2} \right)
\frac{\partial E_{\kv\sigma}^\A} {\partial \Psi_{\sigma 0}}
- \frac{1}{2} \frac{\partial\tilde{\eps}_{\kv\sigma}}
{\partial \Psi_{\sigma 0}} \right]
\label{eq:dfpsi0}
\ee
with the two energies, and the parameters entering them, given by
(\ref{4.5}), (\ref{EkASF}), and (\ref{4.20}). These will not be
evaluated any further here, except to note that, as in the MSF
phase, at zero temperature the leading behavior of the integrand at
small $k$ is proportional to $1/E_{\kv \s}$, while at finite
temperature it is proportional to $1/E_{\kv \s}^2$. As expected, the
acoustic mode therefore generates divergent fluctuation corrections
for $d \leq 1$ at $T = 0$, and for $d \leq 2$ for $T >
0$.\cite{Hohenberg,MerminWagner}.

The total density is given by
\bea
n &=& -\left(\frac{\partial f_\A}{\partial \mu}\right)_{T,\nu}
\nonumber \\
&=& \Psi_{10}^2 + 2 \Psi_{10} \delta \Psi_1
+ 2 \Psi_{20}^2 +  4 \Psi_{20} \delta \Psi_2
+ \delta n_\A(T,\nu)
\nonumber \\
&\approx& \Psi_{1}^2 + 2 \Psi_{2}^2 + \delta n_\A(T,\nu)
\eea
where, in the second line, the two terms linear in $\delta \Psi_\s$
subsume the implicit dependence of $\Psi_{\sigma 0}$ on $\mu$.  This
result follows from the fact that, via (\ref{saddlepoint}),
$\partial \Psi_{\s 0}/\partial \mu$ obeys (\ref{eq:dpsidh}), but
with $2\Psi_{10} \ket{1} + 4 \Psi_{20} \ket{2}$ replacing $\ket{\s}$
on the right hand side. The depletion $\delta n_\A(T,\nu)$ may be
derived either from the derivative of the non-mean field part of
(\ref{eq:fbogasf}) with respect to the \emph{explicit}
$\mu$-dependence (which appears only additively in $\tilde
\varepsilon_{\kv\s}$)---yielding a form identical to the right hand
side of (\ref{eq:dfpsi0}), but with $\partial/\partial \mu$
(performed at constant $\Psi_{\s0}$) replacing $\partial/\partial
\Psi_{\s0}$---or as the total number of uncondensed particles,
\be
\delta n_\A(T,\nu) = \frac{1}{V} \sum_{\kv,\s} \s
\langle \hat a_{\kv \s}^\dagger \hat a_{\kv \s} \rangle,
\label{eq:asfexc}
\ee
in which (\ref{4.24}) connects the $\hat a$ and $\hat \gamma$
operators. Using either approach, one obtains
\begin{widetext}
\bea
\delta n_\A(T,\nu) &=& \sum_\s \int \frac{d\kv}{(2\pi)^d} \Bigg\{
\left(n_{\kv\s} + \frac{1}{2} \right)
\Bigg[\frac{2\varepsilon_{\kv 1} + \lambda_1 + \alpha_1
+ 2(\lambda_2 + \alpha_2)}{2 E_{\kv \s}}
\nonumber \\
&&-\ (-1)^\s \frac{2 d_\kv [\lambda_1 + \alpha_1
- 2(\lambda_2 + \alpha_2)] + (3t_1-t_2) c_\kv^{(1)}
+ (3t_1+t_2) c_\kv^{(2)}}{2 E_{\kv \s}
\sqrt{d_\kv^2 + c_\kv^{(1)} c_\kv^{(2)}}} \Bigg]
- \frac{\s}{2} \Bigg\}
\label{eq:dnt0}
\eea
\end{widetext}
The terms involving $c_\kv^{(\s)}$ are of order $t_1^2, t_2^2, t_1
t_2$, and therefore vanish along the MSF--ASF transition line.
Because these expressions involve a number of parameters, such as
scattering lengths and the Feshbach resonance coupling, their final
integrated expressions are not very enlightening without additional
(e.g., experimental) input. Thus these predictions are not
explicitly evaluated further. It is noted only that estimates
similar to (\ref{eq:densityrange}) (but now involving all three
scattering lengths on the right hand side) may be derived for the
range of validity of (\ref{eq:dnt0}).

\subsection{Superfluid density}
\label{sec:sfdensity}

The superfluid (number) density $n_s$ is a measure of the stiffness
$\Upsilon_s$ of the order parameter against a long-wavelength spatial
gradient in its phase $\theta(\rv)$, defined by the corresponding
change in the free energy
\begin{equation}
\Delta F_s = \frac{1}{2} \Upsilon_s(T)
\int d\xv |\nabla \bar \theta|^2.
\label{F_sGamma_s}
\end{equation}
Expressing the free energy in terms of the superfluid velocity ${\bf
v}_s = (\hbar/m) \nabla \theta$, one obtains $\Delta F_s =
\frac{m}{2} n_s \int d\xv |{\bf v}_s|^2$ with the standard
relation\cite{MEFisher1}
\begin{equation}
n_s = \frac{m}{\hbar^2} \Upsilon_s.
\label{nsUpsilon}
\end{equation}

For a two-component Bose (atomic and molecular) gas that is
considered here, at long length scales the two phases are locked by
the Feshbach coupling to be $\theta_2 = 2\theta_1$, i.e.,
$\theta_\sigma = \sigma\theta$. As discussed previously, this is
obvious in the ASF state, where the gas is a superfluid with respect
to both atoms and molecules and out-of-phase fluctuations
$\theta_1-\theta_2/2$ are gapped. It is also valid more generally,
coming from the requirement that the imposed velocities of atoms
($\s=1$) and molecules ($\s=2$), $\vv_{s\s} = (\hbar/m_\s)
\nabla\theta_\s$, relative to a stationary boundary are the same.

As outlined above $n_s$ is calculated by computing the free energy
change $\Delta F_s(k_0)$ in the presence of a uniform phase gradient
$\theta_0(\rv) = \kv_0\cdot\rv$, corresponding to a superfluid
component with uniform velocity $\vv_0 = \hbar\kv_0/m$ (and
stationary normal component). To this end one imposes phase twist
boundary conditions on the field operators:
\begin{equation}
\hat \psi_\s(\xv + L{\bf \hat n})
= e^{i\s\theta_0} \hat \psi_\s(\xv),
\label{phase_twistBC}
\end{equation}
where $L$ is the system length along a chosen direction ${\bf \hat
n}$ of the phase gradient and $\theta_0 \equiv \theta_0(L\nv) =
\kv_0\cdot\hat \nv L$.  From this definition, one
obtains\cite{MEFisher1}
\begin{equation}
\Upsilon_s \equiv \lim_{L \to \infty}
\frac{2 L^2}{\theta^2_0}(f_{\theta_0} - f_0)
= \left(\frac{\partial^2 f_{\theta_0}}{\partial k_0^2}
\right)_{k_0=0}
\label{4.76}
\end{equation}
where $f_{\theta_0}$ is the free energy density in the presence of the
twist $\theta_0$, and $L \to \infty$ includes the thermodynamic limit
and is to be taken here at fixed $\theta_0$.

Reexpressing the Hamiltonian in terms of periodic field operators
$\tilde \psi_\s = e^{-i\s \kv_0 \cdot \xv} \hat \psi_\s$, one
obtains
\begin{eqnarray}
\hat H[\hat \psi_\s] &=& \hat H[\tilde\psi_\s]
+ \sum_\s \left(\frac{\hbar^2k_0^2}{2m_1}\s\hat N_\s
+ \frac{\hbar\kv_0}{m_1}\cdot {\bf \hat P}_\s \right)
\nn \\
&=& \hat H[\tilde\psi_\s]
+ \frac{\hbar^2k_0^2}{2m_1} \hat N
+ \vv_0 \cdot {\bf \hat P},
\label{4.77}
\end{eqnarray}
where
\bea
{\bf \hat P}_\s &=& -i\hbar \int d\xv \tilde \psi_\s^\dagger(\xv)
\nabla \tilde \psi_\s(\xv)
\nn \\
\hat N_\s &=& \int d\xv \tilde \psi_\s^\dagger(\xv) \tilde \psi_\s(\xv)
\label{4.78}
\eea
are momentum and number operators for component $\s$ and ${\bf \hat
P} \equiv \sum_\s{\bf \hat P}_\s$.

From the above form for $\hat H$, and defining equations
(\ref{nsUpsilon}), (\ref{4.76}) for $n_s$ and $\Upsilon_s$, one
observes that the superfluid density is also given by
\begin{equation}
n_s = \left. m\frac{\partial j_s}{\partial(\hbar k_0)} \right|_{k_0=0},
\label{ns_js}
\end{equation}
or equivalently defined by the relation
\begin{equation}
\lim_{v_0 \to 0}{\bf j}_s = n_s \vv_0,
\end{equation}
where the supercurrent density ${\bf j}_s$ is the expectation value of
the (number) current density operator
\begin{eqnarray}
\hat{\bf j}_s &=& \frac{1}{V}
\frac{\partial\hat H}{\partial\hbar\kv_0},
\nn \\
&=& \frac{1}{V} \left(\vv_0 \hat N +  \frac{1}{m} {\bf \hat P} \right).
\label{js}
\end{eqnarray}
To compute $n_s$ one expands $\hat H[\tilde \psi_\s]$ to quadratic order
in the fluctuations $\tilde\phi_\s = \tilde \psi_s - \Psi_{\s0}$ and
diagonalizes it at a finite $k_0$.  Because it is odd under ${\bf k}
\to -{\bf k}$ the new momentum term remains diagonal under the $k_0 = 0$
Bogoliubov transformation:\cite{W88}
\begin{equation}
{\bf \hat P}_\s
= \sum_\kv \hbar \kv \hat a_{\kv \s}^\dagger \hat a_{\kv \s}
= \sum_\kv \hbar \kv \hat \gamma_{\kv \s}^\dagger \hat \gamma_{\kv \s}.
\label{4.79}
\end{equation}
Thus the Bogoliubov transformation at finite $k_0$ is unchanged from
that at $k_0 = 0$, except for a shift in the chemical potential $\mu
\rightarrow \tilde \mu = \mu - \hbar^2k_0^2/2m$. The spectrum,
however, does change, but in a simple way
\begin{equation}
\tilde E_{\kv \s} = E_{\kv \s} + {\bf v}_0 \cdot \hbar{\bf k},
\label{Ek_change}
\end{equation}
that is in accord with a general requirement for a
Galilean-invariant system. Computing the expectation value of
$\hat{\bf j}_s$ in (\ref{js}) and using (\ref{ns_js}) [or,
equivalently computing the free energy and using (\ref{4.76})] one
finds
\bea
n_s(T) &=& n-n_n(T)
\nn \\
n_n(T) &=& -\frac{2}{d} \sum_{\s} \s \int \frac{d\kv}{(2\pi)^d}
\eps_{\kv\s} \frac{d n_{\kv\s}}{d E_{\kv\s}},
\label{superfluiddensity}
\eea
where clearly $n_s \le n$, i.e., the normal fluid density $n_n \ge
0$. At zero temperature all excitations are exponentially suppressed
and the normal fluid density $n_n$ vanishes, giving $n_s(T=0) = n$
independent of interactions, as required by Galilean invariance.  In
the normal phase, where $E_{\kv \s} = \eps_{\kv \s} - \mu_\s$, an
integration by parts yields $n_n = n$, and $n_s$ vanishes as
expected for a normal fluid.

In contrast, in the MSF phase, despite the absence of atomic
long-range order, at finite temperature there is a nontrivial atomic
contribution $(\s = 1)$ to the superfluid density (though not to the
condensate).  The corresponding reduction in $n_s$ is due to
thermally excited, unpaired atoms with a gapped spectrum
$E_{k1}^{\rm MSF}$ (due to Feshbach coupling to condensed
molecules), that is not simply the free spectrum $\eps_{k1}$ of the
normal state. In the weakly interacting limit, away from both $T=0$
and $T=T_c$, $n_s(T)/n \approx 1 - (T/T_c)^{d/2}$ is well
approximated by the ideal gas form, which in turn coincides with the
condensate fraction $n_{20}(T)/n$.  As usual, sufficiently close
$T_c$ this result must be modified by critical fluctuations which
strongly modify (\ref{superfluiddensity}).\cite{W88,WRFS} On the
other hand, deviations near $T=0$, where the reduction in $n_s$ is
dominated by gapless molecular excitations that are sound-like with
$E_{k2}^{\rm MSF} \sim \sqrt{\eps_{k2}\eps_{2+}}$, are accurately
described by (\ref{superfluiddensity}), which implies that $n_n(T)
\sim T^{d+1}$.  The low temperature crossover from $T^{d+1}$ to
$T^{d/2}$ takes place when the temperature is high enough that
excitation of the higher energy quasiparticles with quadratic
dispersion, $E_{\kv \s} \approx \eps_{\kv \s}$, dominate the
thermodynamics.\cite{W88} For specific model parameters, the full
detailed form of $n_s(T)$ can be straightforwardly evaluated
numerically.

\section{MSF--ASF phase transition}
\label{ASF_MSFtransition}

As has already been seen in Sec.\ \ref{sec:symmetries}, many of the
properties of the phase transitions appearing in the phase diagram,
Fig.\ \ref{phasediagramNuT} can be deduced based on the nature of
the underlying symmetry that is spontaneously broken in the MSF and
ASF phases.  In particular, there it was argued that because the MSF
exhibits a discrete residual $\hat \psi_1 \rightarrow -\hat \psi_1$
(global phase rotation by $\pi$) symmetry associated with the
diatomic nature of the molecule, the MSF--ASF transition at the
level of mean field theory is of the Ising type.\cite{commentZ2}
However, it is important to stress that the $\mathbb{Z}_2$ symmetry
that is broken at the MSF--ASF transition does {\em not} necessarily
imply that the critical properties of the transition (beyond a mean
field approximation) are of Ising type. One condition for this is
that, in addition to the local interactions of Ising symmetry, the
derivative terms associated with kinetic energy (spatial gradients)
and (in quantum theory) the Berry phase (``$|\psi|^2
\partial_t\theta$'' time derivative) terms must reduce to a standard
rotationally-invariant (in Euclidean space) $d+1$ dimensional
gradient term. The other condition is that additional fields (if
any) coupled to the Ising order parameter must not modify the Ising
critical behavior, i.e., must be irrelevant in the renormalization
group sense.

Below these issues are explored in more detail.  It will shown that
although the first condition is indeed satisfied (i.e., for the
quantum MSF--ASF transition the scalar Ising order parameter indeed
has a standard gradient $d+1$-dimensional Lorentz-invariant
``elasticity''), the existence of the Goldstone mode (the phase of
the molecular order parameter) that couples to the Ising field can
have nontrivial effects and (as was first pointed out by Lee and
Lee,\cite{LL04} based on an earlier study by Frey and Balents in a
different context\cite{FB97}) likely drives the MSF--ASF transition
first order.  For the extremely dilute gases of experimental
interest, this first order behavior is weak and may only be visible
very close to the transition.

Focusing on a homogeneous trap (a box), the $T=0$ and finite $T$
MSF--ASF transitions will now be studied in more detail. This can be
most easily done working with the coherent-state action, $S$, Eq.\
(\ref{S}), corresponding to the two-channel Hamiltonian
(\ref{H2channel}). The MSF--ASF transition will be studied from the
MSF side, where the molecular field $\psi_2 = |\psi_2|
e^{i\theta_2}$ exhibits massless Goldstone mode phase fluctuations
in $\theta_2$, and small, gapped fluctuations in the magnitude
$|\psi_2|$ about the molecular condensate $\langle \psi_2 \rangle =
\Psi_{20}$. Integrating out the latter leads to a standard
superfluid hydrodynamic action\cite{Popov}
\begin{equation}
S_2[\theta_2] = \frac{1}{2} \Upsilon_s
\int_0^{\beta\hbar} d\tau \int d{\xv}
\left[c_{\rm MSF}^{-2}(\partial_\tau\theta_2)^2 +
(\nabla\theta_2)^2\right],
\label{S_2}
\end{equation}
that controls the (acoustic) fluctuations of $\theta_2$, with sound
speed $c_{\rm MSF}$ given by (\ref{4.27}), and helicity
modulus/superfluid density $\Upsilon_s$ given by (\ref{nsUpsilon})
and (\ref{superfluiddensity}).

The atomic contribution to the action, together with the key
Feshbach resonant atom-molecule coupling, is given by
\begin{widetext}
\begin{equation}
S_1 = \int_0^{\beta\hbar} d\tau \int
d{\xv} \left[\psi^*_1\hbar\partial_\tau\psi_1
- \psi^*_1 \left(\frac{\hbar^2}{2m}\nabla^2 + \mu_1 \right)\psi_1
-\alpha\text{Re}\left(|\Psi_2| e^{-i\theta_2}\psi_1\psi_1
\right) \right] + S_\text{nonlinear},
\label{S_1a}
\end{equation}
in which $S_\text{nonlinear}$ contains the conventional quartic
scattering terms. As discussed earlier, the latter locks the
molecule and atom phase fluctuations such that low energy
excitations are governed by $\theta_1 = \theta_2/2$. Thus it is
convenient to define ``dressed'' atomic fields $\tilde \psi_1$
according to
\begin{equation}
\psi_1 \equiv e^{i\theta_2/2} \tilde \psi_1,
\end{equation}
which leads to
\begin{equation}
S_1 = \int_0^{\beta\hbar} d\tau \int
d{\xv}\left\{\tilde \psi^*_1 \hbar\partial_\tau \tilde \psi_1
+ \frac{i}{2}|\tilde\psi_1|^2 \hbar \partial_\tau \theta_2
+ \frac{\hbar^2}{2m}\psi^*_1 \left[\left(-i\nabla
+ \frac{1}{2}\nabla\theta_2 \right)^2 - \mu_1 \right] \psi_1
- \alpha\text{Re}(|\Psi_2|\tilde\psi_1\tilde\psi_1)\right\}
+ S_\text{nonlinear}.
\label{S_1}
\end{equation}
Straightforward analysis\cite{commentIrrelevant} shows that near the
MSF--ASF transition the minimal coupling to the induced gauge-like
field $\nabla \theta_2$ above is irrelevant near a Gaussian fixed
point. Dropping this subdominant contribution and writing $\tilde
\psi_1 = \tilde \psi_R + i \tilde \psi_I$ in terms of its real and
imaginary parts one finds
\begin{equation}
S_1 = \int_0^{\beta\hbar} d\tau \int d{\xv}
\left[\frac{i}{2}(\tilde\psi_R^2 + \tilde\psi_I^2)
\hbar \partial_\tau \theta_2
- 2i \tilde\psi_I \hbar\partial_\tau \tilde\psi_R
- \tilde \psi_R \left(\frac{\hbar^2}{2m} \nabla^2
+ \mu_R \right) \tilde \psi_R
- \tilde\psi_I \left(\frac{\hbar^2}{2m}\nabla^2
+ \mu_I \right) \tilde \psi_I \right]
+ S_{\rm nonlinear}
\end{equation}
\end{widetext}
where the shifted chemical potentials are given by
\begin{equation}
\mu_R = \mu + 2\alpha|\Psi_2|,\ \
\mu_I = \mu - 2\alpha|\Psi_2|.
\label{muRI}
\end{equation}
This form of $S_1$ makes it clear that in the presence of the
molecular condensate, $|\Psi_{2}| > 0$, positive $\alpha$ reduces
the $O(2) = U(1)$ symmetry down to $\mathbb{Z}_2$, and with $\mu_R >
\mu_I$ results in $\tilde \psi_R$ reaching criticality {\em before}
$\tilde \psi_I$. Because the canonically conjugate field
$\tilde\psi_I$ remains ``massive'' (noncritical) at the MSF--ASF
critical point [defined by where the coefficient $\mu_R$ of
$\tilde\psi_R^2$ vanishes, consistent with (\ref{nuc})], it can be
safely integrated out and leads to a $d+1$-dimensional
(Lorentz-invariant) action which is even in the scalar order
parameter $\phi \equiv \tilde\psi_R$, and whose relevant part is
given by
\begin{equation}
S_\mathrm{eff}[\theta_2,\phi] = S_\mathrm{SF}[\theta_2]
+ S_\mathrm{Ising}[\phi] + S_\mathrm{int}[\theta_2,\phi],
\label{Seff}
\end{equation}
in which
\begin{eqnarray}
S_\mathrm{SF}[\theta_2] &=& \frac{1}{2} \Upsilon_s
\int_0^{\beta\hbar} d\tau \int d{\xv}
\left[c_{\rm MSF}^{-2} (\partial_\tau \theta_2)^2
+ (\nabla\theta_2)^2 \right]
\nn \\
S_\mathrm{Ising}[\phi] &=& \int_0^{\beta\hbar} d\tau \int d{\xv}
\left[\frac{1}{2} K_\tau (\partial_\tau \phi)^2
+ \frac{\hbar^2}{2m}(\nabla\phi)^2 \right.
\nn \\
&&\ \ \ \ \ \ \ \ \ \ \ -\ \left.\mu_R \phi^2 + g \phi^4
\phantom{\frac{1}{2}} \right]
\nn \\
S_\mathrm{int}[\theta_2,\phi]
&=& \frac{i}{2} \int_0^{\beta\hbar} d\tau \int d{\xv}
\phi^2 \hbar \partial_\tau \theta_2,
\label{Spieces}
\end{eqnarray}
represent separate superfluid hydrodynamic and Ising actions,
together with a Berry phase-like term that couples them. The
coefficient $K_\tau \approx 1/(\alpha|\Psi_2|)$ to lowest order in
$1/\alpha$, and the leading $\phi^4$ nonlinearity comes from
$S_\mathrm{nonlinear}$.

Thus as advertised, if the coupling of the Ising order parameter
$\phi$ to the molecular Goldstone mode $\theta_2$ is neglected, the
$T=0$ MSF--ASF transition (near $\mu_R = 0$) is indeed in the
$(d+1)$-dimensional Ising universality class; at finite $T$ it
crosses over to the $d$-dimensional Ising transition. The Ising
transition is well studied, and leads to the following predictions.
\cite{ZinnJustin} At $T=0$, for $d=3$, up to logarithmic corrections,
the mean field theory derived above remains an accurate description.
On the other hand in $d = 2$, the MSF--ASF exponents are nontrivial
but are well-known. For example, standard scaling arguments predict:
\begin{eqnarray}
n_{10} \sim |\nu - \nu_c|^{2\beta_I},\ \
E_{\rm gap}^{(1)} \sim |\nu - \nu_c|^{z_I \nu_I},
\end{eqnarray}
where $\beta_I \approx 0.31$, $z_I = 1$, and $\nu_I \approx 0.63$ are
3d classical Ising exponents. These, together with the relevance of
$T$ at this quantum critical point, also imply a {\em universal} shape
of the MSF--ASF phase boundary $\nu_c(n,T) \sim \nu_c(n,0) + a\,
T^{1/\nu_I}$, as shown in Fig.\ \ref{phasediagramNuT}. One may hope
that when long-lived molecular condensates are produced, nontrivial
behavior of $E_{\rm gap}^{(1)}(\nu)$ and the full excitation spectra
may be observed in Ramsey fringes\cite{DCTW02}, and in Bragg and RF
spectroscopy experiments
\cite{BraggKetterle,RFChin,RFKetterle,BraggBruunBaym}.

However, as first emphasized and studied by Lee and
Lee\cite{LL04,FB97} the existence of $\theta_2$ fluctuations can
modify this conclusion sufficiently close to the MSF--ASF
transition---intuitively this follows from the fact that if one
integrates out the superfluid fluctuations, a long-range power law
$\phi^2$-$\phi^2$ interaction (highly anistropic in space-imaginary
time) is generated. Indeed, at $T=0$ and $d+1 < 4$ the $\phi^2
\partial_\tau \theta_2$ coupling term is relevant around the Gaussian
fixed point, scaling like $b^{(3-d)/2}$ with increasing
renormalization length scale $b$, and thus competes with the Ising
$\phi^4$ nonlinearity. The resulting theory embodied in
$S_\mathrm{eff}$ has, in fact, a form very similar to that of an Ising
model on a compressible lattice,\cite{BH76} with $\theta_2$ playing
the role of a phonon $\vec{u}$ and the $\theta_2-\phi$ coupling
scaling similarly to the magneto-elastic coupling $\phi^2 \nabla \cdot
\vec{u}$.\cite{commentConnection} The latter model (as well as its
Heisenberg generalizations) have been extensively studied.\cite{BH76}
The conclusion of that work is that for $\alpha_I > 0$ (with
$\alpha_I$ the specific heat exponent of the uncoupled model) the
magneto-elastic coupling leads to runaway flows that has traditionally
been interpreted as a signature of a fluctuation-driven first order
transition.\cite{HLM74} For $\alpha_I < 0$ Goldstone mode fluctuations
(lattice elasticity) is in fact irrelevant and the transition is in
the universality class of the usual (elastically rigid) Ising model.

Based on these results, since the $d$-dimensional Ising specific heat
exponent is positive for $d > 2$,\cite{FisherRMP,ZinnJustin} one thus
concludes that here too, sufficiently close to the MSF--ASF critical
point, the transition is driven first order.  It should be emphasized
that many results in this paper, namely those that refer to
thermodynamic and elementary excitation properties of the different
{\em phases}, not too close to the transition lines, remain valid and
are unaffected in any way by this issue.

\section{Bose-BCS Model}
\label{sec:BoseBCS}

As is shown in this section, the analysis of the Bose atom-molecule
system via the two-channel model, presented in the previous
subsections can be complemented by Bose-BCS variational approach of
the one-channel model (\ref{Ha}). Similar analyses have been presented
previously,\cite{Girardeau,Nozieres} but it is worth presenting, and
generalizing them somewhat, here in a form that can be compared to the
results of the two-species model (\ref{H2channel}) that has so far
been the focus of this paper. A description entirely in terms of Bose
atomic constituents (single-channel model) lends further physical
insight into the microscopic nature of the phases and phase
transitions, and facilitates comparisons with BEC--BCS crossover in
Fermi systems.\cite{GRJ04,Z04,GR07,L80,NS85,E69,MRE93,CSL05}
Furthermore, the atom-only model should be more appropriate for
describing the case of a broad resonance.

The phase diagram of the system is explored as a function of the gas
parameter $n^{1/3} a$, where $n$ is the atomic density and $a$ is
atomic scattering length.  As discussed in Sec.\
\ref{sec:2channelmodel}, experimentally $a$ is controlled by
magnetic field-tuned proximity to the Feshbach resonance (diverging
at the resonance) as well as by the atom specific background
scattering length, $a_{\rm bg}$. However, within the one-channel
model (\ref{Ha}), this is encoded into the tunable pseudo-potential
amplitude $g_1$, characterizing atom-atom microscopic interaction
and connected to experiments via the atomic scattering length,
$a(g_1)$.\cite{GR07} Focusing on a dilute gas, the gas parameter
away from the Feshbach resonance (where $a \rightarrow a_{\rm bg}
\approx d_0$, with $d_0$ the microscopic range of the interatomic
potential) is taken to be small.

For detuning far below the resonance, the attractive interaction is
strong enough to lead to a deep two-body molecular bound state, that
corresponds to the appearance of a dilute gas of strongly bound
compact molecules. In addition, for a large atomic scattering length
the system is known to exhibit Efimov states of trimers,\cite{E71}
leaving the questions of the stability and the nature of the condensed
state open.\cite{GurarieDiscuss} However, for a sufficiently large
repulsive three-body interaction, and/or away from the Feshbach
resonance (short scattering length), the system is expected to be
stable. In this case, at low energies it will be governed by an
effective low energy s-wave molecule-molecule scattering length $a_m$,
corresponding to a repulsive interaction characterized by a molecular
pseudo-potential $g_2$. Thus, the appropriate effective Hamiltonian is
given by (\ref{H2channel}) with all terms containing the atomic field
$\hat \psi_1$ dropped. The theory is identical to an atomic theory of
a dilute gas of composite bosons with mass $m_2 = 2 m$, and chemical
potential $\mu_2 = 2\mu - E_{\rm b}$, where $E_{\rm b}$ is the binding
energy. Thus a Bogoliubov analysis provides an essentially exact
description. In particular, at $T = 0$ the system is a vacuum for
$\mu_2 < 0$ and Bose condensed for $\mu_2 > 0$ with order parameter
$\Psi_{20} = \sqrt{\mu_2/g_2}$, density $n_2 = n_{20} = |\Psi_{20}|^2$
and low energy acoustic excitation spectrum $E({\bf k}) = c_2 \hbar k$
with $c_2 = \sqrt{n_2 g_2/m_2} = \hbar \sqrt{\mu_2/m_2}$.

In the opposite limit of repulsive interactions, no molecules are
present and the system is, conversely, described by the Hamiltonian
(\ref{H2channel}) with all terms containing the molecular field
$\hat \psi_2$ set to zero.  The phenomenology is again that of a
dilute, single component (this time atomic) Bose gas as described
above, with constituents of mass $m$, chemical potential $\mu_1$ and
interaction $g_1$.

The focus here is on the interesting intervening region around the
transition between these two atomic and molecular superfluid phases.
Thus, the behavior of the single species fluid is explored from the
conventional $g_1 > 0$ (positive scattering length) atomic BEC
limit, through the small $g_1 < 0$ (such that the molecular bound
state, or the resonance, is of spatial extent comparable to
intermolecular separation) Bose BCS limit, to the larger $g_1 < 0$
(indicating a two-body bound state that is much smaller than the
intermolecular separation) molecular BEC limit. As shown in Fig.\
\ref{fig:feshbach}, over this region the atom-atom scattering length
goes from positive, to negative, and then back to positive.

The one-channel grand canonical Hamiltonian is given by
\bea
\hspace{-1cm}\hat{H} &=& \int d\xv
\bigg\{-\hat \psi^\dagger(\xv)
\left(\frac{\hbar^2}{2m}\nabla^2 + \mu \right)\hat \psi(\xv)
\nn \\
&+& \frac{1}{2} g \hat\psi^\dagger(\xv)^2 \hat \psi(\xv)^2
+ \frac{1}{6} w \hat \psi^\dagger(\xv)^3 \hat \psi(\xv)^3
\bigg\},
\label{HaVar}
\eea
where for simplicity the subscript ``1'' on the atomic fields has
been dropped. Though $g$ can have either sign, and the new physics
of primary interest here enters for $g < 0$, $w$ remains positive to
ensure thermodynamic stability. The relation of (\ref{HaVar}) to the
more general two-channel model (that reduces to it in the wide
resonance limit) was summarized in Sec.\ \ref{sec:2channelmodel},
and discussed in detail in Ref.\ \onlinecite{GR07}.

\subsection{Variational mean field approximation}

In the dilute limit, and away from any phase transitions, the
variational approach to be presented, essentially equivalent to the
Bogoliubov approximation, provides asymptotically exact results. In
a standard treatment\cite{FeynmanVariational} the approach relies on
the inequality
\begin{equation}
F \leq F_v \equiv F_\mathrm{MF}
+ \langle \hat H - \hat H_\mathrm{MF} \rangle_\mathrm{MF},
\label{5.2}
\end{equation}
between the true free energy $F = -k_B T\ln\Tr[e^{-\beta\hat H}]$
(where $\hat H$ is the system's full interacting Hamiltonian) and
the variational free energy $F_v$ defined in terms of an arbitrary
Hamiltonian $H_\text{MF}$ and its corresponding free energy
$F_\text{MF} = -k_B T\ln\Tr[e^{-\beta\hat H_\text{MF}}]$. Here
$\langle \cdot \rangle_\mathrm{MF}$ is the thermodynamic average
with respect to $\hat H_\mathrm{MF}$. Since $F_v$ is an upper-bound
for $F$, Eq.\ (\ref{5.2}) guarantees that the better the choice of
$\hat H_\text{MF}$ the closer one can approximate the true free
energy with $F_v$.  On the other hand to take advantage of the
variational method one needs to pick a simple enough $H_\text{MF}$
that the thermodynamic averages appearing in $F_v$ may be calculated
explicitly. Thus, $H_\text{MF}$ is chosen here to be a quadratic
Hamiltonian
\begin{eqnarray}
\hat H_\mathrm{MF} &=& \sum_\kv
\left[N_\kv \hat a_\kv^\dag \hat a_\kv
+ \frac{1}{2} P_\kv \left(\hat a_\kv^\dag \hat a_{-\kv}^\dag
+ \mathrm{H.c.} \right) \right]
\nn \\
&&-\ \frac{1}{2} Q_0 \sqrt{V}
(\hat a_{\bf 0} + \hat a_{\bf 0}^\dag),
\label{5.1}
\end{eqnarray}
with variational parameters $N_\kv, P_\kv, Q_0$ (chosen real, by
absorbing any extra phase factors into the Bose operators if
necessary) to be selected to minimize $F_v$.

The linear term (required to deal properly with the possibility of
an atomic condensate) is removed via a zero momentum shift
\begin{equation}
\tilde \psi(\xv) = \hat \psi(\xv) - \psi_0,\
\tilde a_\kv = \hat a_\kv - \sqrt{V}\psi_0 \delta_{\kv,{\bf 0}},
\label{5.3}
\end{equation}
with
\begin{equation}
\psi_0 = \frac{Q_0}{2(N_{\bf 0} + P_{\bf 0})}.
\label{5.4}
\end{equation}
Following this, the Bogoliubov transformation
\bea
\tilde a_\kv &=& u_\kv \hat \gamma_\kv
- v_\kv \hat \gamma_{-\kv}^\dag
\nn \\
\tilde a_\kv^\dag &=& u_\kv \hat \gamma_\kv^\dag
- v_\kv \hat \gamma_{-\kv}
\label{5.5}
\eea
with the choices
\bea
u_\kv^2 &=& 1 + v_\kv^2 = \frac{1}{2}
\left(\frac{N_\kv}{E_\kv} + 1 \right)
\nn \\
E_\kv &\equiv& \sqrt{N_\kv^2 - P_\kv^2},
\label{5.6}
\eea
leads to the diagonal quadratic form
\bea
\hat H_\mathrm{MF} &=& \sum_\kv
\left[E_\kv \hat \gamma_\kv^\dag \hat \gamma_\kv
+ \frac{1}{2}(N_\kv - E_\kv) \right]
\nn \\
&&+\ [(N_{\bf 0} + P_{\bf 0}) \psi_0^2 - Q_0 \psi_0]V.
\label{5.7}
\eea
The two-point averages are easily computed, and are given by
\bea
\langle \tilde a_\kv^\dag \tilde a_\kv \rangle_\mathrm{MF}
&=& \frac{N_\kv}{E_\kv} \left(n_\kv + \frac{1}{2} \right)
- \frac{1}{2}
\label{5.8} \\
\langle \tilde a_\kv^\dag \tilde a_{-\kv}^\dag \rangle_\mathrm{MF}
&=& \langle \tilde a_{-\kv} \tilde a_\kv \rangle_\mathrm{MF}
= -\frac{P_\kv}{E_\kv} \left(n_\kv + \frac{1}{2} \right),
\nn
\eea
in which $n_\kv = (e^{\beta E_\kv} - 1)^{-1}$ is again the Bose
occupation factor. The atomic number density and molecular order
parameter defined by $\hat H_\mathrm{MF}$ are therefore given by
$n_\mathrm{MF} = \psi_0^2 + \tilde n_\mathrm{MF}$,
$\Phi_\mathrm{MF} = \psi_0^2 + \tilde \Phi_\mathrm{MF}$, in which
\bea
\tilde n_\mathrm{MF} &\equiv&
\langle \tilde \psi^\dagger \tilde \psi \rangle_\mathrm{MF}
= \int \frac{d\kv}{(2\pi)^d}
\left[\frac{N_\kv}{E_\kv} \left(n_\kv + \frac{1}{2} \right)
- \frac{1}{2} \right] \ \ \ \ \ \ \
\label{5.9a} \\
\tilde \Phi_\mathrm{MF} &\equiv&
\langle \tilde \psi \tilde \psi \rangle_\mathrm{MF}
= -\int \frac{d\kv}{(2\pi)^d}
\frac{P_\kv}{E_\kv} \left(n_\kv + \frac{1}{2} \right).
\label{5.9b}
\eea

Using (\ref{5.3}) to shift the operators in $\hat H$, together with
Wick's theorem and Eqs.\ (\ref{5.9a}), (\ref{5.9b}), the variational
free energy density (\ref{5.2}) takes the form
\bea
f_v &=& \int \frac{d\kv}{(2\pi)^d}
\bigg\{k_B T \left[n_\kv \ln(n_\kv)
- (1 + n_\kv) \ln\left(1+n_\kv \right) \right]
\nn \\
&&\ \ \ \ \ \ \ \ \ \ \ \ \
+ \epsilon_\kv\frac{N_\kv}{E_\kv}
\left(n_\kv + \frac{1}{2} \right)
- \frac{1}{2} \epsilon_\kv \bigg\}
\nn \\
&&+\ {\cal F}_0(\tilde n_\mathrm{MF},\tilde \Phi_\mathrm{MF},\psi_0),
\label{5.10}
\eea
in which
\bea
{\cal F}_0 &\equiv& f_0 - \tilde \mu \psi_0^2
+ \frac{1}{2} \tilde g \psi_0^4 + \frac{1}{6} w \psi_0^6
\nn \\
f_0 &=& -\mu \tilde n_\mathrm{MF}
+ \frac{1}{2} g (2\tilde n_\mathrm{MF}^2
+ \tilde \Phi_\mathrm{MF}^2)
\nn \\
&&+\ \frac{1}{2} w \tilde n_\mathrm{MF}
(2 \tilde n_\mathrm{MF}^2 + 3 \tilde \Phi_\mathrm{MF}^2)
\nn \\
\tilde \mu &=& \mu
- g(2 \tilde n_\mathrm{MF} + \tilde \Phi_\mathrm{MF})
\nn \\
&&-\ \frac{3}{2} w(2 \tilde n_\mathrm{MF}^2
+ 2 \tilde n_\mathrm{MF} \tilde \Phi_\mathrm{MF}
+ \tilde \Phi_\mathrm{MF}^2)
\nn \\
\tilde g &=& g
+ w(3\tilde n_\mathrm{MF} + 2\tilde \Phi_\mathrm{MF}).
\label{5.11}
\eea

In minimizing (\ref{5.10}) one may treat the ratio $R_\kv =
P_\kv/N_\kv$, $n_\kv$, and $\psi_0$ as independent variables.  The
derivative with respect to $R_\kv$ yields the simple result
\begin{equation}
R_\kv = \frac{\gamma_\mathrm{MF}}
{\epsilon_\kv - \mu_\mathrm{MF}},
\label{5.12}
\end{equation}
in which
\bea
\mu_\mathrm{MF} &\equiv& -\frac{\partial {\cal F}_0}
{\partial \tilde n_\mathrm{MF}}
\nn\\
&=& \mu - 2g \tilde n_\mathrm{MF}
- \frac{3}{2} w(2 \tilde n_\mathrm{MF}^2
+ \tilde \Phi_\mathrm{MF}^2)
\nn \\
&&-\ [2g + 3w(2 \tilde n_\mathrm{MF}
+ \tilde \Phi_\mathrm{MF})] \psi_0^2
- \frac{3}{2} w \psi_0^4\ ,
\label{mu_MF}\ \ \ \ \ \\
\gamma_\mathrm{MF} &\equiv& \frac{\partial {\cal F}_0}
{\partial \tilde \Phi_\mathrm{MF}}
\nn \\
&=& (g + 3w \tilde n_\mathrm{MF}) \tilde \Phi_\mathrm{MF}
\nn \\
&&+\ [g + 3w (\tilde n_\mathrm{MF}
+ \tilde \Phi_\mathrm{MF})] \psi_0^2 + w \psi_0^4.
\label{5.13}
\eea
All $\kv$-dependence therefore resides in the energy denominator
of (\ref{5.12}).

The derivative with respect to $n_\kv$ yields
\begin{equation}
-\frac{1}{\beta} \ln\left(\frac{n_\kv}{1+n_\kv} \right)
= E_\kv = \frac{(\epsilon_\kv - \mu_\mathrm{MF}) N_\kv
- \gamma_\mathrm{MF} P_\kv}{E_\kv},
\label{5.14}
\end{equation}
which, upon substitution of (\ref{5.6}) and (\ref{5.12}), yields the
single particle excitation spectrum
\begin{equation}
E_\kv = \sqrt{(\epsilon_\kv - \mu_\mathrm{MF})^2
- \gamma_\mathrm{MF}^2},
\label{5.15}
\end{equation}
with an energy gap
\begin{equation}
E_{\rm gap} \equiv E_{\kv = {\bf 0}}
= \sqrt{\mu_\mathrm{MF}^2 - \gamma_\mathrm{MF}^2}.
\label{5.16}
\end{equation}
One identifies in addition,
\begin{equation}
N_\kv = \epsilon_\kv - \mu_\mathrm{MF},\
P_\kv = \gamma_\mathrm{MF}.
\label{5.17}
\end{equation}
By substituting these results into (\ref{5.9a}) and (\ref{5.9b}),
one finally obtains the self-consistency conditions,
\bea
\tilde n_\mathrm{MF} &=& \int \frac{d\kv}{(2\pi)^d}
\left[\frac{\epsilon_\kv - \mu_\mathrm{MF}}{E_\kv}
\left(n_\kv + \frac{1}{2} \right) - \frac{1}{2} \right]
\ \ \ \ \ \ \ \ \ \
\label{5.18} \\
\tilde \Phi_\mathrm{MF}
&=& -\gamma_\mathrm{MF}
\int \frac{d\kv}{(2\pi)^d} \frac{1}{E_\kv}
\left(n_\kv + \frac{1}{2} \right).
\label{5.19}
\eea
In the normal phase, $\psi_0 = 0$, $\Phi_\mathrm{MF} = 0$, Eq.\
(\ref{5.19}) is automatically satisfied, and (\ref{5.18}) determines
$n_\mathrm{MF}$. In the MSF phase, $\psi_0 = 0$ but
$\Phi_\mathrm{MF} \neq 0$, giving $\gamma_\mathrm{MF} = (3g + w
\tilde n_\mathrm{MF}) \tilde \Phi_\mathrm{MF}$ and leading to a
self-consistent (``gap''-like) equation (\ref{5.19}), that together
with (\ref{5.18}), determines $\tilde \Phi_\mathrm{MF}$ and $\tilde
n_\mathrm{MF}$.  Since the integral in (\ref{5.19}) is positive,
clearly, there is a nontrivial solution, $\Phi_\mathrm{MF} \neq 0$
only if $\gamma_\mathrm{MF} < 0$, i.e., for sufficiently attractive
atomic interactions ($g$ sufficiently negative).

Finally, the derivative with respect to $\psi_0$ yields
\bea
0 &=& -\tilde \mu + \tilde g \psi_0^2 + \frac{1}{2} w \psi_0^4,
\label{5.20}
\eea
that, together with (\ref{5.18}) and (\ref{5.19}), self-consistently
determines $\psi_0$ inside the ASF phase.  The limit $\psi_0 \to 0$
then yields the ASF--MSF phase boundary at the critical value
$\tilde \mu_c = 0$.  At this point (and only at this point)
$\mu_\mathrm{MF} = \gamma_\mathrm{MF}$, and the energy gap
(\ref{5.16}) therefore vanishes at the ASF--MSF transition. This is
consistent with the results of the two-channel (atom-molecule) model
[see (\ref{msfgap}) and (\ref{gapsASF})], where in both the ASF and
MSF phases the atomic branch $E_{\kv 1}$ (that corresponds to
spectrum $E_{\kv}$, above) remains gapped except at the MSF--ASF
transition point.  Because, in the one-channel model, molecular
excitations do not explicitly appear in the Hamiltonian [though
molecular superfluid order clearly does appear, via anomalous
averages (\ref{5.9b})], neither does the gapless spectrum of the
corresponding (molecular) superfluid mode. However, as will be seen
in the discussion in Sec.\ \ref{sec:bcs_sfdensity}, the presence of
these gapless molecular modes will appear in the calculation of the
superfluid density.

With the above substitutions, the free energy density (\ref{5.10})
simplifies to
\begin{eqnarray}
f_v &=& \int\frac{d\kv}{(2\pi)^d}
\left\{k_B T \ln\left(1-e^{-\beta E_\kv} \right)
+ \frac{\gamma_\mathrm{MF}^2}{E_\kv} n_\kv \right.
\nn \\
&&+\ \left. \frac{\epsilon_\kv - \mu_\mathrm{MF}}{2}
\left[\frac{\epsilon_\kv - \mu_\mathrm{MF}}{E_\kv} - 1 \right]
\right\}
\nonumber \\
&&+\ {\cal F}_0(\tilde n_\mathrm{MF},\tilde \Phi_\mathrm{MF},\psi_0)
+ \mu_\mathrm{MF} \tilde n_\mathrm{MF}.
\label{5.21}
\end{eqnarray}
The similarity of (\ref{5.21}) to the two-species form
(\ref{FreeEnergy}) is evident.

The extremum conditions $(\partial f_v/\partial \tilde
\Phi_\mathrm{MF})_{\tilde n_\mathrm{MF}} = 0$, $(\partial
f_v/\partial \tilde n_\mathrm{MF})_{\tilde \Phi_\mathrm{MF}} = 0$,
in which the derivatives include all dependence in
$E_\kv,\mu_\mathrm{MF}, \gamma_\mathrm{MF}$, yield precisely the
constraints (\ref{5.18}), (\ref{5.19}).  As a consequence, one also
has the relation $n_\mathrm{MF} = -(\partial f_v/\partial
\mu_\mathrm{MF})_{n_\mathrm{MF}, \Phi_\mathrm{MF}}$, in which the
derivative includes only the dependence from the combination
$\epsilon_\kv - \mu_\mathrm{MF}$ (appearing especially in $E_\kv$).

Equations (\ref{5.18}), (\ref{5.19}) and (\ref{5.20}) are the
fundamental results of this section, providing a set of closed
relations to be solved for $n_\mathrm{MF}$, $\Phi_\mathrm{MF}$ and
$\psi_0$ as functions of $\mu,T$. The stabilizing three-body
repulsion $w$ plays no essential role here: the equations remain
perfectly well defined for $w = 0$. This is because the form
(\ref{5.1}) for the variational Hamiltonian already precludes the
type of real space system collapse against which $w$ is intended to
stabilize.  It should be kept in mind, however, that it is
\emph{only} in the presence of $w$ that the type of superfluid-order
considered here would actually occur. Thus, $w > 0$ motivates the
form of the variational ground state, but once this form is adopted,
$w$ effectively disappears from the calculation.  As observed in
experiments, where $w$ is generally quite small and system collapse
does eventually occur, such states are expected to be dynamically
metastable in the dilute limit even when they do not describe true
equilibrium.

\subsection{Superfluid density}
\label{sec:bcs_sfdensity}

One can infer the existence of gapless molecular excitations in the
single species model from the existence of a nonzero superfluid
(number) density, $n_s = (m/\hbar^2) \Upsilon_s$ defined (as before)
in terms of the change in the free energy, $\Delta F_s$,
(\ref{4.76}) associated with imposition of twisted boundary
conditions on $\hat \psi(\xv)$.

As in (\ref{phase_twistBC})--(\ref{4.77}), one expresses $\hat H$ in
terms of the periodic field operator $\tilde \psi({\xv}) = e^{-i
{\bf k}_0 \cdot {\xv}} \psi({\xv})$, with the result
\begin{equation}
\hat H[\hat \psi] = \hat H[\tilde \psi] + \varepsilon_0 \hat N
+ {\bf v}_0 \cdot {\bf \hat P}
\label{5.22}
\end{equation}
where $\varepsilon_0 = \hbar^2 k_0^2/2m$, ${\bf v}_0 = \hbar \kv_0/m$,
and $\hat N, {\bf \hat P}$ are given by (\ref{4.78}) with $\tilde
\psi$ replacing $\tilde \psi_\sigma$ and dropping $\s$ summations.  In
the presence of twisted boundary conditions one generalizes the
variational Hamiltonian to the form
\bea
\hat H_\mathrm{MF} &=& \sum_\kv \left[(N_\kv + M_\kv)
\tilde a_\kv^\dag \tilde a_\kv
+ \frac{1}{2} P_\kv \left(\tilde a_\kv^\dag \tilde a_{-\kv}^\dag
+ \mathrm{H.c.} \right) \right]
\nn \\
&&-\ \frac{1}{2} Q_0 \sqrt{V} \left(\tilde a_{\bf 0}
+ \tilde a_{\bf 0}^\dag \right),
\label{5.23}
\eea
in which $\tilde a_\kv$ is the Fourier transform of $\tilde \psi$. All
variational coefficients depend on the twist wavevector $\kv_0$, but
$N_\kv$, $P_\kv$ are \emph{even} functions of $\kv$ as before, while
$M_\kv$ is an \emph{odd} function of $\kv$.  In fact, as will seen
shortly, $M_\kv = \hbar {\bf v}_0 \cdot \kv$, but this is not assumed
at the outset.  The shifted Bogoliubov transformation
(\ref{5.3})--(\ref{5.6}) is performed in an identical fashion, with
$\tilde a_\kv$ replacing $\hat a_\kv$, and $u_\kv, v_\kv$ depending
only on the even functions $N_\kv,P_\kv$.  As in (\ref{4.79}), because
the odd part is invariant under the transformation,
\begin{equation}
\sum_\kv M_\kv \tilde a_\kv^\dag \tilde a_\kv
= \sum_\kv M_\kv \hat \gamma_\kv^\dag \hat \gamma_\kv,
\label{5.24}
\end{equation}
the Bogoliubov transformation is independent of ${\bf k}_0$.  After
the diagonalization, $\hat H_{\rm MF}$ takes on the following form:
\bea
\hat H_\mathrm{MF} &=& \sum_\kv \left[(E_\kv + M_\kv)
\hat \gamma_\kv^\dag \hat \gamma_\kv
+ \frac{1}{2} (N_\kv - E_\kv) \right]
\nn \\
&&+\ [N_{\bf 0} - P_{\bf 0}] \psi_0^2 - Q_0 \psi_0] V.
\label{5.25}
\eea
The atom number and molecular order parameter densities follow in
a form identical to (\ref{5.9a}) and (\ref{5.9b}), but with the Bose
occupation factor given by
\begin{equation}
n_\kv = \frac{1}{e^{\beta (E_\kv + M_\kv)} - 1},
\label{5.26}
\end{equation}
that includes the odd function $M_\kv$. The variational free
energy (\ref{5.2}) follows in the form (\ref{5.10}), but (i) with
a single additional term
\begin{equation}
\Delta F_v \equiv \int \frac{d\kv}{(2\pi)^d}
{\bf v}_0 \cdot \hbar\kv n_\kv
\label{5.27}
\end{equation}
arising from the ${\bf v}_0 \cdot {\bf \hat P}$ term in
(\ref{5.22}), and (ii) with $\mu$ replaced by $\mu - \varepsilon_0$
everywhere, arising from the $\varepsilon_0 \hat N$ term in
(\ref{5.22}).

One performs the variational minimization as before, treating
$n_\kv$, $R_\kv = P_\kv/N_\kv$, and $\psi_0$ as independent
variational parameters.  Minimization over $R_\kv$ yields
(\ref{5.12}), but again with $\mu$ replaced by $\mu - \varepsilon_0$
everywhere in (\ref{5.13}). Minimization over $n_\kv$ yields
\begin{equation}
E_\kv + M_\kv = \sqrt{(\epsilon_\kv - \mu_\mathrm{MF})^2
- \gamma_\mathrm{MF}^2} + {\bf v}_0 \cdot \hbar\kv,
\label{5.28}
\end{equation}
and one immediately recovers (\ref{5.15}) for $E_\kv$, and
\begin{equation}
M_\kv = \hbar {\bf v}_0 \cdot {\bf k},
\label{5.29}
\end{equation}
as promised. Minimization over $\psi_0$ recovers (\ref{5.20}),
again with $\mu$ replaced by $\mu - \varepsilon_0$ everywhere.

The superfluid density, $n_s$, Eq.\ (\ref{nsUpsilon}) is
proportional to the second derivative of $f_v$ with respect to
$k_0$, Eq.\ (\ref{4.76}). Since there is an implicit
$k_0$-dependence through all terms in the free energy, the
computation of $\Upsilon_s$ appears overwhelming at first sight.
However, two observations simplify it enormously: (1) All single
$k_0$-derivatives of even functions of $k_0$ are odd functions of
$k_0$, and hence vanish at $k_0 = 0$, and (2) the variational
property implies that all single derivatives of the free energy with
respect to $R_\kv$, $n_\kv$ of $\psi_0$ vanish identically.  Thus,
for example, (1) implies that $(\partial \tilde
n_\mathrm{MF}/\partial k_0)_{k_0 = 0} = (\partial \tilde
\Phi_\mathrm{MF}/\partial k_0)_{k_0 = 0} = (\partial \psi_0/\partial
k_0)_{k_0 = 0} = 0$, and hence that cross terms such as $(\partial^2
f_v/\partial \tilde n_\mathrm{MF} \partial \tilde \Phi_\mathrm{MF})
(\partial \tilde n_\mathrm{MF}/\partial k_0) (\partial \tilde
\Phi_\mathrm{MF}/\partial k_0)$ vanish in the limit $k_0 \to 0$.
Similarly, (2) implies that $(\partial f_v/\partial n_\kv)
(\partial^2 n_\kv/\partial k_0^2) = (\partial f_v/\partial R_\kv)
(\partial^2 R_\kv/\partial k_0^2) = (\partial f_v/\partial \psi_0)
(\partial^2 \psi_0/\partial k_0^2) = 0$.

The result is that there are only two contributions to the
superfluid density.  The first comes from the $\mu -
\varepsilon_0$ combination, and yields a term
\begin{equation}
-\frac{\partial f_v}{\partial \mu}
\frac{\partial^2 \varepsilon_0}{\partial k_0^2}
= \frac{\hbar^2}{m} n_\mathrm{MF},
\label{5.30}
\end{equation}
where $n_\mathrm{MF} = -\partial f_v/\partial \mu = \psi_0^2 + \tilde
n_\mathrm{MF}$ is the number density [the variational conditions again
imply that the only contributions to the $\mu$-derivative come from
the explicit dependence in ${\cal F}_0$---see (\ref{5.11})]. The other
contribution, interpreted as the normal fluid density, comes from the
term (\ref{5.27}) and together these give:
\bea
n_s &=& \frac{m}{\hbar^2} \Upsilon_s = n_\mathrm{MF} - n_n
\nn \\
n_n &=& -\lim_{k_0 \to 0}
\int \frac{d\kv}{(2\pi)^d} (\hat \kv_0 \cdot \kv)
\frac{\partial n_\kv}{\partial k_0}
\nn \\
&=& -\frac{2}{d} \int \frac{d\kv}{(2\pi)^d}
\epsilon_\kv \frac{\partial n_\kv}{\partial E_\kv},
\label{5.31}
\eea
in which $\hat \kv_0$ is the unit vector along $\kv_0$, and $\kv_0$
has been set to zero inside $n_\kv$ in the last line.  The
resemblance to (\ref{superfluiddensity}) is clear.  As promised, the
superfluid density is finite even though $E_\kv$ is gapped in both
the ASF and MSF phases, indirectly indicating the presence of the
gapless molecular excitations.\cite{gapcomment}

\subsection{Solutions to the variational equations}
\label{sec:var_soln}

In what follows $\mu_\mathrm{MF}$ will be treated as the independent
control parameter, and $\gamma_\mathrm{MF}$ viewed as fixed.  If one
wishes, one may use solutions to (\ref{5.13}) and (\ref{5.14})
obtained in this way, together with (\ref{5.10}), to solve in the
end for the behavior as a function of $\mu$ at fixed $g,w$. For
simplicity only $T = 0$ where $n_\kv \equiv 0$ will be considered
here.

\subsubsection{Onset of molecular superfluidity: vacuum-MSF transition}
\label{subsec:msf_onset}

In the MSF phase the order parameter constraint
(\ref{5.19}) reduces to
\begin{equation}
\frac{1}{g_\mathrm{MF}} = -\int \frac{d\kv}{(2\pi)^d}
\frac{1}{E_\kv} \left(n_\kv + \frac{1}{2} \right),
\label{5.32}
\end{equation}
in which $g_\mathrm{MF} \equiv g + 3 w \tilde n_\mathrm{MF}$, and
$\mu_\mathrm{MF}$ is given by the second line of (\ref{mu_MF}). The
onset of molecular superfluidity at $T=0$ takes place at the value
$\mu = \mu_0$ at which particles first begin to enter the system,
i.e., at zero density.  Letting $n_\mathrm{MF}, \Phi_\mathrm{MF} \to
0$, (\ref{5.32}) yields
\begin{equation}
-\frac{2}{g} = \int \frac{d\kv}{(2\pi)^d}
\frac{1}{\epsilon_\kv - \mu_0}
\label{5.33}
\end{equation}
In App.\ \ref{app:twobody} an equation is obtained for the ground
state energy $E$ of a single molecule in the weak-binding limit
where the molecular size is much larger than the diameter $d_0$ of
the attractive potential. Comparing (\ref{5.33}) to (\ref{B3}) (in
the case where $\tilde v({\bf k}) \equiv 1$---note that an effective
cutoff $k_\Lambda \sim \pi/d_0$ is required in this case to
regularize the integral for $d \geq 2$) one sees that $\mu_0 =
-E/2$. Therefore the onset of molecular superfluidity indeed occurs
precisely when the chemical potential rises above the molecular
binding energy per particle.  This confirms that ${\cal H}_{\rm MF}$
correctly captures this limit.

\subsubsection{MSF--ASF phase boundary and the closing of the single
particle gap}

More generally, the MSF phase single particle spectrum
(\ref{5.15}) has a gap
\begin{equation}
E_\text{gap} = E_{\kv=0} = \sqrt{\mu_\mathrm{MF}^2
- g_{\rm MF}^2 \Phi_\mathrm{MF}^2},
\label{5.34}
\end{equation}
representing the minimum energy required to create a single atom
excitation.\cite{Coniglio}

As $\mu$ increases from $\mu_0$ this gap shrinks and vanishes at a
critical value $\mu_c$ such that $\mu^c_\mathrm{MF} = g_\mathrm{MF}
\Phi^c_\mathrm{MF}$. This point identifies the MSF-to-ASF transition,
and at $T=0$ (\ref{5.32}) reduces to
\begin{equation}
-\frac{2}{g_\mathrm{MF}} = \int \frac{d^dk}{(2\pi)^d}
\frac{1}{\sqrt{\epsilon_\kv(\epsilon_\kv - 2 \mu^c_\mathrm{MF})}}
\label{5.35}
\end{equation}
Subtracting (\ref{5.32}) (at $T=0$) from (\ref{5.35}) one obtains
\begin{equation}
0 = \int \frac{d^dk}{(2\pi)^d}
\left[\frac{1}{\sqrt{\epsilon_\kv
(\epsilon_\kv - 2 \mu^c_\mathrm{MF})}}
- \frac{1}{E_\kv} \right],
\label{5.36}
\end{equation}
in which the integral is now fully convergent for $2 < d < 4$, and
the short-scale uv-cutoff may be dropped. The latter is now
effectively subsumed into a nonuniversal value of the critical
chemical potential, $\mu^c_{\rm MF}$, while critical behavior, which
depends only on deviations of $\mu_\mathrm{MF}$ from this value,
remains universal.

\begin{figure*}
\centerline{\includegraphics[width=0.9\textwidth]{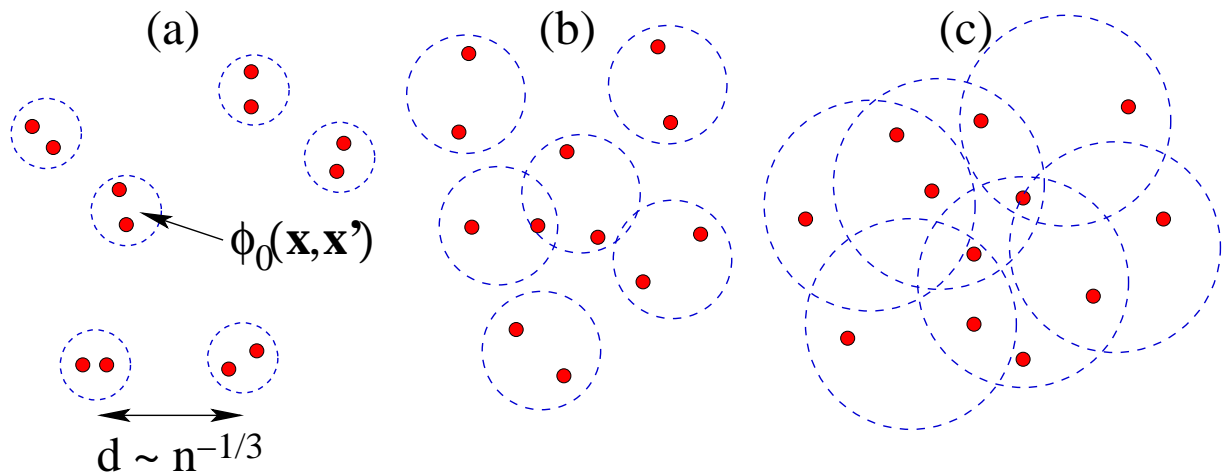}}

\caption{Schematic illustration of the single species MSF--ASF
transition, accompanying the BEC--BCS crossover. (a) For
sufficiently strongly attractive interactions, the extent of the
molecular wavefunction $\phi_0({\bf x},{\bf x}')$ is much smaller
than the intermolecular separation $d$, and is essentially
determined only by two-body physics. Interactions between molecules
are weak, and only rarely do they overlap sufficiently to exchange
atoms. At low temperatures the molecules Bose condense, but coherent
intermolecular hopping that would lead to atomic Bose condensation
is suppressed. (b) The MSF--ASF transition takes place when the size
of the molecular wavefunction (which is strongly renormalized by
many body effects) becomes comparable to $d$ and molecules begin to
significantly overlap. Coherent atomic hopping occurs over ever
larger distances as the overlap increases, and the divergence of
this distance signifies the MSF-to-ASF transition. (c) Deep in the
ASF phase, the size of the molecular wavefunction is much larger
than $d$, and essentially loses its physical meaning: coherent
hopping of atoms over the entire system leads to atomic Bose
condensation, and one can no longer identify any given pair of atoms
with a particular molecule. Instead, the physics is described by a
BCS-type many body wavefunction, embodied in the Hamiltonian
(\ref{5.1}), which maintains strong nonlocal, pair correlations in
the absence of identifiable molecules.}

\label{overlap}
\end{figure*}

The density integral (\ref{5.18}) is already convergent for $d < 4$,
and one obtains the critical value
\begin{equation}
n_{\rm MF}^c = \left(\frac{-2m \mu_\mathrm{MF}^c}{\hbar^2}
\right)^{d/2} I_d
\label{5.37}
\end{equation}
which exhibits a simple power-law relation between $n_{\rm MF}^c$
and $\mu_{\rm MF}^c$.  The substitution $u^2 = -\hbar^2 k^2/2m_A
\mu_{\rm MF}^c$ has been used to define a dimensionless constant
\begin{equation}
I_d \equiv \frac{1}{2} \int \frac{d^du}{(2\pi)^d}
\left[\frac{u^2 + 1}{\sqrt{u^2(u^2+2)}} - 1 \right],
\label{5.38}
\end{equation}
The quantity $r_M = \sqrt{2 m \mu_\mathrm{MF}/\hbar^2}$ may be
interpreted as the background molecular diameter.  Thus (\ref{5.37})
shows that the condition for closing of the atomic gap corresponds
to a criterion of the mean atomic separation $r_A =
n_\mathrm{MF}^{1/d}$ reaching $r_M$.  The MSF--ASF transition
therefore takes place in a regime in which pairs begin to strongly
overlap, which is the condition under which atoms can begin to hop
from one molecule to another as they themselves become delocalized
in a sea of extended pairs (see Fig.\ \ref{overlap}).

\subsubsection{Critical behavior near the ASF phase boundary}

The approach to ASF--MSF critical point may be analyzed by considering
small deviations
\begin{eqnarray}
\tau &=& \frac{\mu^c_{\rm MF} - \mu_{\rm MF}}{\mu_{\rm MF}^c},
\nonumber \\
f &=& \frac{|\Phi^c_{\rm MF}|^2
- |\Phi_{\rm MF}|^2}{|\Phi_{\rm MF}^c|^2},
\nonumber \\
\rho &=& \frac{n^c_{\rm MF} - n_{\rm MF}}{n_{\rm MF}^c}
\label{5.39}
\end{eqnarray}
from criticality, all of which will turn out to be positive in the MSF
phase.  Thus, in fact, although $\mu$ increases as the density
increases, $\mu_{\rm MF}$ \emph{decreases} due to the extra terms in
(\ref{mu_MF}). In terms of these one obtains from (\ref{5.18}) and
(\ref{5.32}):
\begin{widetext}
\begin{eqnarray}
0 &=& J(\tau,f) \equiv \int \frac{d^du}{(2\pi)^d}
\left[\frac{1}{\sqrt{u^2(u^2 + 2)}}
- \frac{1}{\sqrt{(u^2 + 1 - \tau)^2 - 1 + f}} \right]
\nonumber \\
\rho &=& R(\tau,f) \equiv \frac{1}{2 I_d} \int \frac{d^du}{(2\pi)^d}
\left[\frac{u^2 + 1}{\sqrt{u^2(u^2 + 2)}}
- \frac{u^2 + 1 - \tau}{\sqrt{(u^2 + 1 - \tau)^2 - 1 + f}} \right]
\label{5.40}
\end{eqnarray}
\end{widetext}
The detailed analysis of (\ref{5.40}) is relegated to Appendix
\ref{app:bcscrit}.  The results may be summarized as follows.

The critical behavior is found to change dramatically at dimension
$d=3$, which is to be expected because the upper-critical dimension
for a quantum Ising model, controlling the zero-temperature MSF--ASF
transition is $d^I_{uc} = 3$. For $d > 3$ the critical behavior is
Gaussian,
\begin{eqnarray}
f(\tau) &=& \tau \left[f_d^{(1)}  + f_d^{(c)} \tau^{(d-3)/2}
+ O(\tau) \right]
\nonumber \\
\rho(\tau) &=& \tau \left[\rho_d^{(1)} \tau
+ \rho_d^{(c)} \tau^{(d-3)/2} + O(\tau) \right],
\label{5.41}
\end{eqnarray}
where the $d$-dependent coefficients are given in (\ref{C5}),
(\ref{C6}), and one may expect ${\cal H}_{\rm MF}$ to provide an
asymptotically exact description of the MSF--ASF transition.

For $d < 3$ the critical behavior is nontrivial.  The variational
theory predicts
\begin{eqnarray}
f(\tau) &=& \tau \left[f_d^{(1)}
+ f_d^{(c)} \tau^{(3-d)/(d-1)} + O(\tau) \right]
\nonumber \\
\rho(\tau) &=& \tau \left[\rho_d^{(1)}
+ \rho_d^{(c)} \tau^{(3-d)/(d-1)} + O(\tau) \right],
\ \ \ \ \
\label{5.42}
\end{eqnarray}
[see (\ref{C12}), (\ref{C13})], but this approximation must break
down sufficiently close to the transition point, determined by a
Ginzburg criterion that can be worked out in a standard way. For a
dilute gas, relevant to the atomic gas experiments considered here,
the size of the Ginzburg region should be be very small, and
therefore it is unlikely that the resulting asymptotic critical
behavior (which, as discussed in Sec.\ \ref{ASF_MSFtransition}, is
likely actually a fluctuation-driven first order transition) can be
observed. The apparent exponent singularity at $d=1$ is a signature
of the lower-critical dimension below which the ordered phase (ASF),
and therefore the transition to it, is destabilized by quantum
fluctuations.

In $d=3$ there are logarithmic corrections:
\begin{eqnarray}
f(\tau) &=& \tau \left\{f_3^{(1)}
+ \frac{f_3^{(c)}}{\ln(\tau_0/\tau)}
+ O\left[\frac{\ln\ln(\tau_0/\tau)}
{\ln^2(\tau_0/\tau)} \right] \right\}
\nonumber \\
\rho(\tau) &=& \tau \left\{\rho_3^{(1)}
+ \frac{\rho_3^{(c)}}{\ln(\tau_0/\tau)}
+ O\left[\frac{\ln\ln(\tau_0/\tau)}
{\ln^2(\tau_0/\tau)} \right] \right\},
\ \ \ \ \ \
\label{5.43}
\end{eqnarray}
with amplitudes given in (\ref{C17}) and (\ref{C18}).

In all cases, there is a leading analytic dependence, linear in
$\tau$, followed by a subleading singular contribution
$\tau^{1-\tilde\alpha}$, with ``quantum specific heat'' exponent
\begin{equation}
\tilde\alpha = \left\{\begin{array}{ll}
-\frac{d-3}{2}, & d>3 \\
0\ (\log), & d=3 \\
-\frac{3-d}{d-1}, & d<3.
\end{array} \right.
\label{5.44}
\end{equation}
Note that $\tilde\alpha$ here differs from that in (\ref{4.71})
which, although computed for a slightly different quantity,
nevertheless reflects the same universal energy singularity.  The
change is not due to a change in the universality class, but rather
to the extra constraint embodied in the first line of (\ref{5.40}),
which essentially reflects the existence of only a single species.
It is well known that such constraints lead to a ``Fisher
renormalization'' $\alpha \to -\alpha/(1-\alpha)$ whenever $\alpha >
0$, while leaving it unchanged if $\alpha < 0$ (the constraint
therefore always leads to a negative specific heat
exponent).\cite{fishren} If one were to enforce the total density
constraint $n = 2n_2 + n_1$, the condensate depletion (\ref{4.61})
would also display the Fisher-renormalized exponent.

One may use a similar analysis to extend these results into the ASF
phase, and to positive temperatures. However, rather than exploring
(difficult to probe) critical behavior, the aim here is mainly to
demonstrate the existence of the same three phases (N, MSF, ASF)
illustrated in Fig.\ \ref{phasediagramNuT}; the role of the detuning
$\nu$ is played here by the s-wave interaction parameter $g$, or more
properly the corresponding dimensionless measure of scattering length,
i.e., the gas parameter $n^{1/3} a$.  Despite the complete loss of
molecular identity in the broad resonance (single-channel) model as
the ASF phase is approached, the topology of the phase diagram and
critical behavior (accounting appropriately for constraints) is the
same as for the two-channel model.

\section{Topological excitations}
\label{sec:vortex}

In the previous sections a description of a (s-wave) resonantly
interacting atomic Bose gas was presented, formulated in terms the
atomic and molecular superfluid (condensate) order parameters
$\Psi_{10}$ and $\Psi_{20}$, and corresponding fluctuations in the
two ordered, ASF and MSF states were studied, and characterized the
associated $T=0$ and finite $T$ phase transitions.

In this section, a complementary description of this two-component
(atoms and diatomic molecules) Bose gas is presented, in terms of
topological excitations in the MSF and ASF phases. These will be
shown below to be vortices and domain walls. Descriptions of phases
and phase transitions in terms of topological excitations has a long
and successful history, with ordinary vortices in superfluids and
superconductors, dislocations and disclinations in crystalline
solids, and domain walls in Ising ferromagnets being only a few most
prominent examples.\cite{ChaikinLubensky}

The importance of this description is two-fold. Firstly, topological
defects are true nonlinear excitations of the system and thus are
essential for a full characterization of the response of an ordered
state to an external perturbation. For example, a rotated neutral
superfluid (or a superconductor in the presence of a sufficiently
strong magnetic field) responds by nucleating quantized vortices
that carry discrete units of fluid's angular momentum (magnetic
flux). Secondly, fluctuation-induced (quantum or thermal)
topological defects provide a dual characterization of phases and
transitions between them that complements their description in terms
of a Landau-type order-parameter. For example, a
superfluid-to-normal transition can be understood through a dual
model of fluctuation-induced proliferation of vortices, with the
superfluid state acting as a vortex vacuum (or an insulator) and the
normal state as a vortex condensate.\cite{DasguptaHalperin}

In addition to simply playing a complementary role, such dual vortex
``disorder parameter'' descriptions are also a powerful way to
characterize subtly ordered phases that do not allow a direct Landau
order parameter description. The most prominent examples of this are
2d ordered phases with a continuous symmetry that are
(usually\cite{2Dexceptions}) ``forbidden'' by the
Hohenberg-Mermin-Wagner theorem\cite{Hohenberg,MerminWagner} to
exhibit true long-range order and thus cannot be characterized by a
condensate order parameter.\cite{commentOP} Such ``ordered'' phases
(e.g., a 2d superfluid or a 2d crystalline solid) are in fact
disordered, only distinguished from the short-ranged (exponentially)
correlated fully disordered states by a quasi-long-ranged (QLR)
order with correlation functions falling off as a power-law.
Descriptions of such QLR-ordered phases and their transition to
fully disordered states is best done in terms of a proliferation of
topological defects, e.g., vortices in 2d
superfluids.\cite{commentOP} In higher dimensions, a description in
terms of a proliferation of topological defects can also be more
effective, as for example found in disordering of a 3d type-II
superconductors by proliferation of vortex
loops.\cite{DasguptaHalperin,FisherLee}

Such dual topological defect descriptions, in addition to providing
added physical insight, provide important complementary
computational tools for the studying these phenomena. With this
motivation, topological defects in the ASF and MSF will now be
considered.

\subsection{Atomic superfluid}

Since the fully ordered ASF state has two nonzero order parameters
$\Psi_{10},\Psi_{20}$, there are interesting features of the
topological excitations generated by the (Feshbach) coupling between
them. The thermodynamics of the state can be conveniently and
equivalently described in terms of the local magnitudes and phases
of its two (atomic, $\s=1$, and molecular, $\s=2$) coherent-state
fields
\begin{equation}
\psi_\s = \sqrt{n_\s} e^{i\theta_\s}.
\label{T1}
\end{equation}
In terms of these, the real-time coherent-state action corresponding
to the Hamiltonian (\ref{H2channel}) takes the form
\begin{eqnarray}
S &=& S_1 + S_2 + S_{12}
\label{Srealtime} \\
S_1 &=& \int dt d\xv \left[\hbar n_1\partial_t\theta_1
+ \frac{\hbar^2}{2m} n_1 (\nabla\theta_1)^2 \right.
- \mu n_1
\nn \\
&&\ \ \ \ \ \ \ \ \
+\ \left. \frac{1}{2} g_1 n_1^2 \right]
\nn \\
S_2 &=& \int dt d\xv \left[\hbar n_2\partial_t\theta_2 +
\frac{\hbar^2}{4m} n_2(\nabla\theta_2)^2
- (2\mu-\nu) n_2 \right.
\nn \\
&&\ \ \ \ \ \ \ \ \
+\ \left. \frac{1}{2} g_2 n_2^2 \right]
\nn \\
S_{12} &=& \int dt d\xv \left[g_{12} n_1 n_2
- \alpha n_1 n_2^{1/2}\cos(2\theta_1-\theta_2) \right],
\nn
\end{eqnarray}
where terms involving $\nabla n_\s$, that are less important than the
finite compressibility $g_\s$ terms, have been dropped. From the
action $S$ the phase diagrams of Sec.\ \ref{sec:mft} and Bogoliubov
modes of Sec.\ \ref{sec:excite} can be straightforwardly reproduced
in terms of the canonically conjugate densities $n_\s$ and phases
$\theta_\s$.

Note now that the mean field equations of motion for the phases
$\theta_\s$, namely the Euler-Lagrange equations $\frac{\delta
S}{\delta\theta_\s} = 0$, are given by
\begin{eqnarray}
\partial_t n_1 + \frac{\hbar}{m} \nabla \cdot(n_1 \nabla\theta_1)
&=& 2 \frac{J}{\hbar} \sin(2\theta_1-\theta_2),
\label{ELtheta1} \\
\partial_t n_2 + \frac{\hbar}{2m} \nabla\cdot(n_2 \nabla\theta_2)
&=& -\frac{J}{\hbar} \sin(2\theta_1-\theta_2),
\label{ELtheta2}
\end{eqnarray}
where the internal Josephson coupling between atomic and molecular
superfluids,
\begin{equation}
J = \alpha n_1 \sqrt{n_2},
\label{J}
\end{equation}
is proportional to the Feshbach resonance amplitude $\alpha$. These
can be combined to derived the total boson number $n = n_1 + 2n_2$
conservation (continuity) equation
\begin{equation}
\partial_t n + \nabla \cdot {\bf j} = 0,
\label{Jcontinuity}
\end{equation}
where the total number current ${\bf j} = {\bf j}_1 + 2{\bf j}_2$
naturally consists of the atomic and molecular contributions,
\begin{eqnarray}
{\bf j}_1 = \frac{\hbar}{m} n_1 \nabla\theta_1,
\label{j1} \\
{\bf j}_2 = \frac{\hbar}{2m} n_2 \nabla\theta_2.
\label{j2}
\end{eqnarray}
Observe that, due to the atom-molecule Feshbach resonant
interconversion captured by the ``current'' $J
\sin(2\theta_1-\theta_2)$ on the right hand sides of
(\ref{ELtheta1}) and (\ref{ELtheta2}), as expected, $n_1$ and $n_2$
are not independently conserved.

The microscopic action $S$ in (\ref{Srealtime}) is not completely
generic. A more general model (that can be obtained either based on
symmetry or by incorporating quantum and thermal fluctuations)
includes an atomic-molecular current-current interaction of the form
\begin{equation}
\delta S_{12} = \frac{1}{2} K_{12}
|\nabla(2\theta_1-\theta_2)|^2,
\label{S12}
\end{equation}
arising from coarse-graining of the action $S$ in the presence of
the Feshbach resonance cosine nonlinearity. The form of this term
ensures that total atom conservation embodied in the continuity
equation (\ref{Jcontinuity}) remains satisfied.

Fluctuations lead to corrections to the mean field equations of
motion.  At the hydrodynamic level, in which only the dynamics of
slow, large scale distortions of the fields are considered, these
corrections may be embodied simply in renormalization of the terms
appearing in $S$.  Most significantly, in this limit phase
fluctuations dominate, and fluctuations in $n_\s$ may be subsumed
into renormalized stiffness coefficients.  Thus, the squared phase
gradient terms undergo the replacement
\begin{equation}
\frac{\hbar^2}{2m_\sigma} n_\s |\nabla \theta_\s|^2
\to \frac{1}{2} K_\s |\nabla \theta_\s|^2,\ \
K_\s \equiv \frac{\hbar^2}{m} n_{s\s},
\label{Ksigma}
\end{equation}
which replaces the fluctuating number density $n_\s$ by the atomic
and molecular superfluid (number) densities, $n_{s\s}$, to be
distinguished from the corresponding, quite distinct (see Sec.\
\ref{sec:excite}) condensate fractions $n_{0\s}$. As in a
single-component superfluid at $T=0$, Galilean invariance enforces
the condition
\begin{equation}
n_{s1} + 2 n_{s2} = n,\ \ {\rm for}\ \ T=0,
\label{GInvce}
\end{equation}
namely that the total superfluid density equals the total density.

With this preface, the focus will now be on the finite temperature
classical limit, ignoring quantum dynamics that are left to future
investigation. The model to be studied is defined by the
``hydrodynamic'' energy density
\begin{widetext}
\begin{equation}
\label{Etheta12}
{\cal E} = \frac{1}{2} K_1(\nabla\theta_1)^2
+ \frac{1}{2} K_2(\nabla\theta_2)^2
+ \frac{1}{2} K_{12}|\nabla(2\theta_1-\theta_2)|^2
- J \cos(2\theta_1 - \theta_2),
\end{equation}
\end{widetext}
and is valid in any region where the phases $\theta_\s$ are well
defined, but must be supplemented by core energies in regions where an
order parameter magnitude vanishes.  The four coefficients are all
renormalized hydrodynamic parameters that depend on the chemical
potentials, and other microscopic parameters, and have their own
nontrivial critical behavior.\cite{MEFisher1} The detailed knowledge
of their exact values (some of which have been computed in the dilute
limit in earlier sections of this paper) is not required to understand
general features of topological excitations.

\subsubsection{Vortices in the ASF}
\label{vorticesASF}

\begin{figure}

\includegraphics[width=\columnwidth]{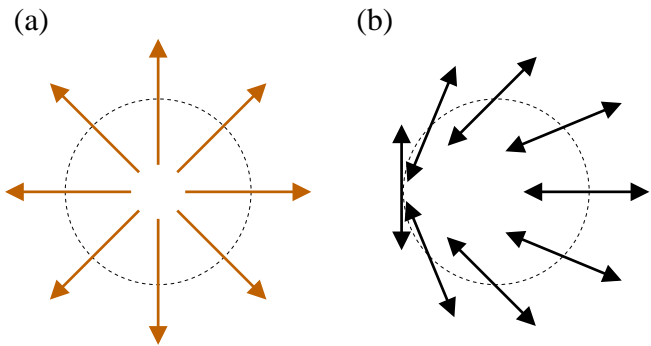}

\caption{(a) A unit vortex in the order parameter field
$\Psi_{10}(\xv)$ represented as a $2\pi$-rotation in the vector
field ${\bf M}_0(\xv)$. (b) In the $\Psi_{10}$ complex plane, a unit
vortex in $\Psi_{20}(\xv)$ is represented by a $\pi$-rotation in the
double headed vector field ${\cal Q}_0(\xv)$. One identifies the
Feshbach resonance coupling $-\alpha{\rm Re}[\Psi^*_{20}
\Psi^2_{10}]$ as the energetic tendency for spatial alignment of
${\bf M}_0(\xv)$ and ${\cal Q}_0(\xv)$.  Details of the mapping
between these two representations are provided in Appendix
\ref{mapNematic}.}

\label{unitvortex}
\end{figure}

In the absence of the atom-molecule couplings, $K_{12}$ and $J$, the
superfluid admits independent atomic and molecular
vortices---pictured in Fig.\ \ref{unitvortex}. Focusing for
simplicity on $d=2$, these are point defects in the atomic and
molecular superfluid order parameters, around which their respective
phases wind by an integer-multiple of $2\pi$ as the point is
encircled. As usual, this quantization condition arises from the
single-valuedness of the superfluid order parameter away from the
vortex core (located at position ${\bf r}_{0\s}$):
\begin{equation}
\oint_{{\bf r}_{0\s}} \nabla\theta_\s\cdot d{\bf r} = 2\pi p_\s,
\end{equation}
with ``charge'' $p_\s$.

Imposing this topological constraint and minimizing the energy
${\cal E}$ at $K_{12} = J = 0$, one obtains independent atomic and
molecular superfluid velocities around the corresponding vortices
\begin{equation}
{\bf v}_1 = p_1\frac{\hbar}{m} \frac{{\hat{\bf\varphi}}}{r},\ \
{\bf v}_2 = p_2\frac{\hbar}{2m} \frac{{\hat{\bf\varphi}}}{r},
\end{equation}
with integer charges $p_\s$. Equivalently, the superfluid phases
$\theta_\s$ are given simply by integer multiples of the azimuthal
coordinate angles $\varphi_\s$ (measured with respect to an origin
chosen at the vortex core positions ${\bf r}_{0\s}$), with
$\theta_\s = p_\s \varphi_\s$. As usual in 2d, vortex energies grow
logarithmically with system size $L$,
\begin{equation}
E_\s^{(v)} = p_\s^2 K_\s\pi\ln(L/\xi_\s),
\label{Evortex}
\end{equation}
where $\xi_\s$ is the vortex core size set by the corresponding
coherence lengths.

In the presence of the inter-superfluid couplings $K_{12}$, $J$ no
general solution is available. However, considerable insight can be
obtained by analyzing limiting regimes. It is clear from the energy
density $\E$, Eq.\ (\ref{Etheta12}), that to avoid extensive
(scaling with system size) energy cost proportional to $J$ (and/or
$K_{12}$), the two phases are on average locked together according
to
\begin{equation}
\theta_2 = 2\theta_1.
\label{lockThetas}
\end{equation}
Hence the energy is minimized when positions of the atomic and
molecular vortices coincide, and their topological charges (winding
numbers) are related by $p_2 = 2 p_1$. Thus an elementary $p_1 = 1$
vortex in the atomic order parameter will be accompanied by a
spatially coincident molecular vortex of topological charge that is
double its elementary value, $p_2 = 2$.

\begin{figure}

\centerline{\includegraphics[width=\columnwidth]{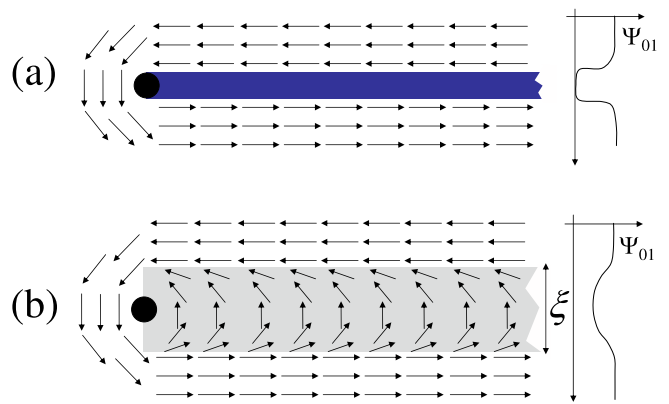}}

\caption{A schematic illustration of a $2\pi$ (elementary unit)
molecular vortex that induces a $\pi$ (half-unit) atomic vortex,
that in turn induces a domain-wall ray. In (a) the wall width, $\xi$
is smaller than the coherence length and the energy cost per unit of
wall length exceeds that of the condensation energy, thus leading to
a ``normal'' ($\Psi_{10}=0$) domain wall. In (b) the superfluid
stiffness is large and Feshbach resonance is narrow (small $\alpha$)
leading to a wide domain wall (width $\xi$ exceeding the coherence
length), with an interface that is in the ASF state and $\Psi_{10}$
only slightly suppressed below its bulk value. The wall structure is
illustrated in more detail in Fig.\ \ref{domainWalldetail}.}

\label{domainWallAB}
\end{figure}

In contrast, an elementary $p_2 = 1$ molecular vortex is
energetically significantly more costly due to incompatibility of
the atomic order parameter single-valuedness constraint and the
Feshbach resonance constraint (\ref{lockThetas}). It is clear that
energetically there are two competing, least costly, ways to
accommodate this frustration. One, illustrated in Fig.\
\ref{domainWallAB}(a), is with a spatially coincident half-integer
($p_1 = 1/2$) atomic $\pi$ vortex, that requires that $\Psi_{10}$
vanish (atomic component of the gas is normal) along a ray emanating
from the location of the vortex core, and across which $\theta_1$
exhibits a $\pi$ jump discontinuity. The cost of such a defect
clearly scales linearly with the length, $L$ of the defect ray (more
generally, as $L^{d-1}$ in $d$ dimensions) and is dominated by the
loss of the condensation energy along the linear defect.

Another competing possibility, illustrated in Fig.\
\ref{domainWallAB}(b), is an atomic $p_1 = 1$ vortex localized on
the elementary molecular vortex, but (in contrast to the
noninteracting case where the vortex is isotropic with
$\theta_1(\varphi) = \varphi$) the atomic phase winding is highly
anisotropic, with
\begin{eqnarray}
\theta_1(\varphi) &\approx& \theta_2(\varphi)/2,
\nn \\
&\approx& \varphi/2,
\end{eqnarray}
outside a narrow domain wall strip.  The atomic phase makes up the
remaining $\pi$ deficit angle (required by single-valuedness of the
atomic order parameter), by rapidly winding across the domain wall of
width set by $J$ and a combination of phase stiffnesses $K_\s$.

It is clear that the first scenario is the limiting case of the
second configuration, with large $J$ and small $K_\s$, such that the
wall width is microscopic (formally smaller than $\xi_\s$) and the
corresponding energy comparable to condensation energy, thus driving
the discontinuity ray normal. The resulting energy in both cases
clearly grows linearly (as $L^{d-1}$ in $d$ dimensions) with the
length of the domain wall ray, and the energy scale is set by the
minimum of the condensation energy or line tension, with the latter
given by the geometrical mean of $J$ and a combination of the $K_\s$
(see below).

\subsubsection{Domain walls in the ASF}
\label{domainwallASF}

\begin{figure}

\centerline{\includegraphics[width=\columnwidth]{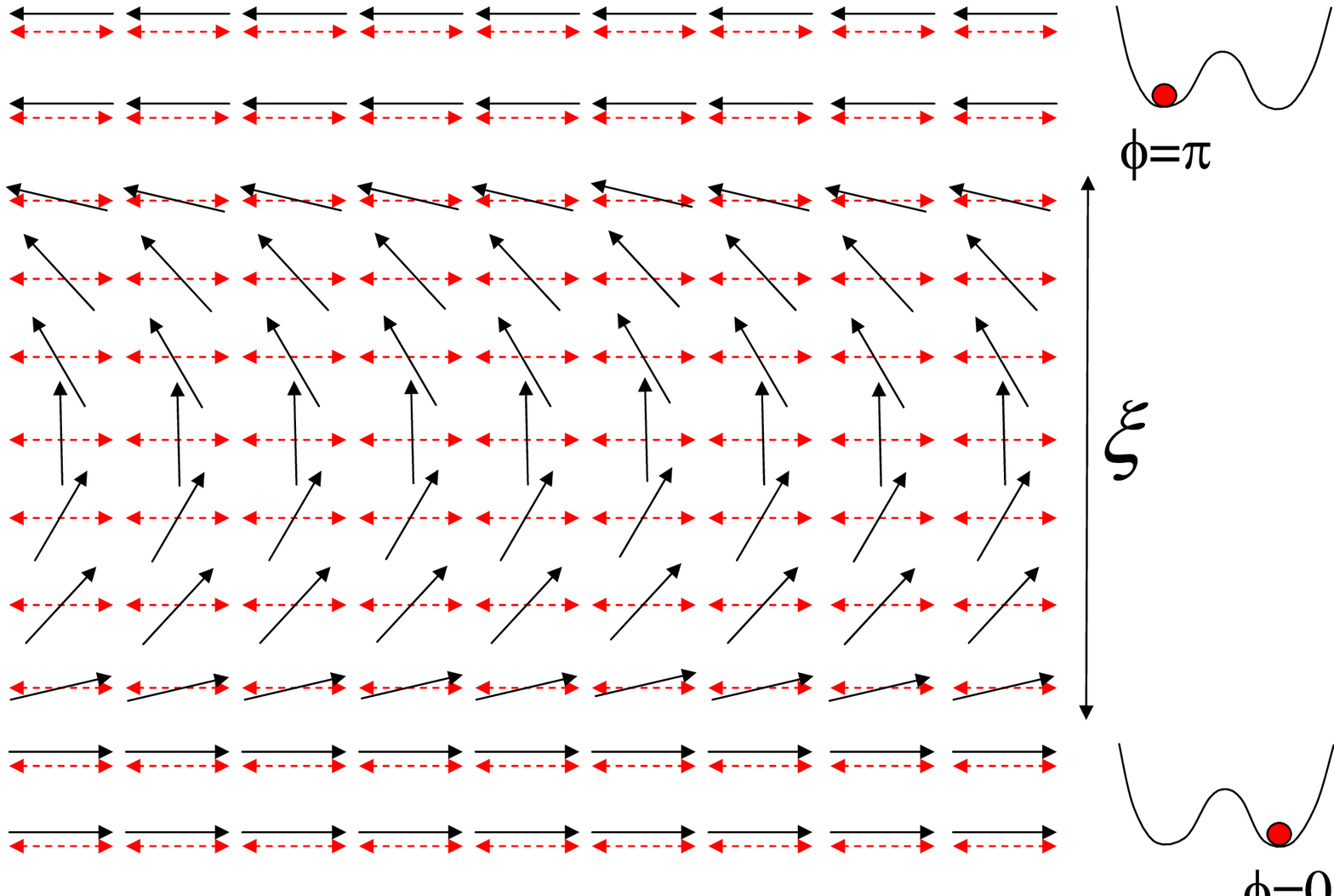}}

\caption{Details of the domain wall structure, separating
$\Psi_1^{(0,1)} = \pm 1$ Ising domains of the ASF corresponding to
$\theta_1^{(0)} = \theta_2/2$ and $\theta_1^{(1)} = \theta_2/2 +
\pi$, respectively (illustrated here for $K_{\theta\phi}/K_\theta =
1$). Across the wall the atomic condensate phase, $\theta_1$, winds
by $\pi$ relative to the molecular condensate phase (double-headed
arrow angle), $\theta_2$.}

\label{domainWalldetail}
\end{figure}

It has been argued that the domain wall with energy linear in its
length $L$ (more generally, growing as its surface area $L^{d-1}$ in
$d$ dimensions) is another type of a topological excitation in the
ASF. Although in the previous subsection it emerged as a necessary
string component of a $p_2 = 1$ molecular vortex, the existence of a
domain wall excitation can be understood on more general grounds.
Quite similar to domain walls in an Ising ferromagnet, here too it
is a defect that separates ordered domains associated with two
physically distinct configurations of the Ising order parameter in
the ASF that spontaneously breaks the ${\cal Z}_2$ symmetry of the
MSF state. In terms of phases $\theta_\s$ the two domains correspond
to two solutions
\begin{equation}
\theta_1^{(n)} = \theta_2/2 + n \pi, \ n=0,1,
\label{solnDomain}
\end{equation}
of the constraint in (\ref{lockThetas}), that are associated with
two values of the atomic order parameter $\Psi_1^{(0,1)} =
e^{i\theta_1^{(0,1)}} = \pm e^{i\theta_2/2}$, pictorially
illustrated in Fig.\ \ref{domainWalldetail}.

A detailed solution for a domain wall can be straightforwardly worked
out. To this end, the key (internal) Josephson nonlinearity associated
with the Feshbach resonance is isolated by a convenient change of
phase variables to new phase fields $\theta$ and $\phi$:
\begin{eqnarray}
\theta &=& \frac{1}{2} (2\theta_1 + \theta_2)
\label{theta} \\
\phi &=& \frac{1}{2} (2\theta_1 - \theta_2),
\label{phi}
\end{eqnarray}
corresponding to the in-phase and out-of-phase fluctuations of
$\theta_1 = (\theta + \phi)/2$ and $\theta_2 = \theta - \phi$
phases.

In terms of $\theta$ and $\phi$, the energy density is given by
\begin{equation}
\label{Ethetaphi}
{\cal E} = \frac{1}{2}K_\theta(\nabla\theta)^2
+ \frac{1}{2}K_\phi(\nabla\phi)^2
- K_{\theta\phi}\nabla\theta\cdot\nabla\phi
- J\cos(2\phi),
\end{equation}
with
\begin{eqnarray}
\label{Kthetaphi}
K_\theta&=&\frac{1}{4}K_1 + K_2
\\
K_\phi&=&\frac{1}{4}K_1 + K_2 + 4K_{12}
\\
K_{\theta\phi}&=&-\frac{1}{4}K_1 + K_2.
\end{eqnarray}
The corresponding saddle-point equations $\delta\E/\delta\theta =
\delta \E/\delta\phi=0$ are given by
\begin{eqnarray}
-K_\theta\nabla^2 \theta + K_{\theta\phi} \nabla^2 \phi &=& 0,
\label{ELtheta} \\
-K_\phi\nabla^2\phi + K_{\theta\phi}\nabla^2\theta
+ 2J\sin2\phi &=& 0.
\label{ELphi}
\end{eqnarray}
Eliminating $\theta$ via (\ref{ELtheta}) reduces (\ref{ELphi}) to
the well-studied sine-Gordon equation for $\phi$,
\begin{equation}
-K \nabla^2\phi + 2J \sin2\phi=0,
\label{sineGordon_phi}
\end{equation}
where
\begin{equation}
K = K_\phi - K_{\theta\phi}^2/K_\theta.
\label{K}
\end{equation}
For a straight domain wall oriented along $x$, defined by the
boundary conditions $\phi_\mathrm{dw}(y \to -\infty) = 0$ and
$\phi_\mathrm{dw}(y \to +\infty) = \pi$, the solution is given by
\begin{equation}
\phi_\mathrm{dw}(y) = 2\arctan\left(e^{2y/\xi}\right),
\label{phi_dw}
\end{equation}
illustrated in Fig.\ \ref{domainWalldetail}. The domain wall width
is given by
\begin{eqnarray}
\xi &=& \sqrt{K/J}.
\label{xi_dw}
\end{eqnarray}
The associated $\theta_{dw}(y)$ and corresponding
$\theta_\s^{dw}(y)$ solutions can now be easily obtained from
(\ref{ELtheta}), (\ref{theta}), and (\ref{phi}):
\begin{eqnarray}
\theta_\mathrm{dw}(y) &=& \frac{K_{\theta\phi}}{K_\theta}
\phi_{dw}(y)
\label{thetadw} \\
\theta_1^{dw} &=& \frac{1}{2}
\left(\frac{K_{\theta\phi}}{K_\theta} + 1\right)
\phi_{dw}(y),
\label{theta1dw} \\
\theta_2^\mathrm{dw} &=& \left(\frac{K_{\theta\phi}}{K_\theta}-1\right)
\phi_{dw}(y).
\label{theta2dw}
\end{eqnarray}

Using the above expressions inside the energy density $\E$, Eq.\
(\ref{Ethetaphi}), and taking advantage of the Euler-Lagrange
equations (\ref{ELtheta}), (\ref{ELphi}) one finds that the domain
wall energy is given by
\begin{eqnarray}
E_\mathrm{dw} &=& \int dx dy\ K(\nabla\phi_{dw})^2,
\label{Edw1} \\
&=&\int dx dy\ 2J[1 - \cos(2\phi_{dw})],
\label{Edw2}\\
&=& \sigma_\mathrm{dw} L_x,
\label{Edw3}
\end{eqnarray}
with domain wall line tension (energy per unit of length),
\begin{equation}
\sigma_\mathrm{dw} = 4\sqrt{J K}.
\label{sigma_dw}
\end{equation}

\subsubsection{Point-to-``dumbbell'' atomic vortex transition}

As is clear from the discussion in Sec.\ \ref{vorticesASF}, in the ASF
a $2\pi$ ($p_1 = 1$) atomic elementary vortex is driven by the
Feshbach resonance (internal Josephson) coupling $J$ to be accompanied
by a $4\pi$ ($p_2 = 2$) molecular double vortex. In the limit where $J
\gg K$, corresponding to a broad Feshbach resonance, and deep in the
ASF state, the two superfluids are strongly coupled, and will behave
as a single-component conventional atomic superfluid. Thus the $2\pi$
atomic and $4\pi$ molecular vortices must spatially coincide. This
leads to an isotropic topological defect with energy (measured
relative to the background energy $-J$ of the uniform state) given in
2d by
\begin{equation}
E^{(\mathrm{point})}_v = E_c^{(\mathrm{point})}
+ \pi K_1 \ln(L/\xi_1) + 4 \pi K_2 \ln(L/\xi_2),
\label{Epoint}
\end{equation}
where
\begin{equation}
E_c^{(\mathrm{point})} = E_{1c}^{(2\pi)} + E_{2c}^{(4\pi)}
\label{eq:Ecpoint}
\end{equation}
consists of the atomic ($2\pi$) and molecular ($4\pi$) vortex core
energies.

Such a concentric, isotropic vortex configuration minimizes the
Feshbach resonance energy. However, because it involves a $4\pi$
($p_2 = 2$) molecular vortex that is double the elementary charge,
it raises the large-scale part of the kinetic energy by $2\pi K_2
\ln(L/\xi_2)$ over the energy of a topologically equivalent vortex
configuration consisting of two elementary $2\pi$ ($p_2 = 1$)
molecular vortices---see (\ref{Evortex}). As will be shown below,
this can drive the {\em splitting} of the $4\pi$ double molecular
vortex into its elementary $2\pi$ constituents. This is driven by
the fact that, in the absence of the atomic component, two
elementary vortices repel via a potential $V_2^{(2\pi-2\pi)}(R) = -
2\pi K_2\ln(R/\xi_2)$, where $R$ is the separation. On the other
hand, as shown in Sec.\ \ref{vorticesASF}, in the ASF phase an
elementary $2\pi$ molecular vortex is driven by the internal
Josephson coupling $J$ to be accompanied by a ($p_1 = 1/2$
fractional) atomic $\pi$ vortex and a domain-wall string defect.
Thus, in the ASF state the logarithmic repulsion of two $2\pi$
molecular vortices is arrested by the confining (Josephson coupling)
domain-wall energy that, according to (\ref{Edw3}), grows linearly
with separation $R$. Hence, the energy $E^{(point)}_v$ of a point
vortex (consisting of coincident $2\pi$ atomic and $4\pi$ molecular
vortices) must be compared to a topologically equivalent
``dumbbell'' configuration split by a separation $R$ into two units,
each consisting of coinciding $\pi$ atomic and $2\pi$ molecular
vortices---see Fig.\ \ref{vortex_pi}.\cite{KV04}

The energy of the dumbbell configuration is estimated as
\begin{eqnarray}
E^{(\mathrm{dmbl})}_v &\approx& E_c^{(\mathrm{dmbl})}
+ \pi K_1 \ln L/R + 4\pi K_2 \ln L/R
\nn \\
&&-\ \pi K_1\ln(R/\xi_1) - 4\pi K_2 \ln(R/\xi_2)
+ \sigma_\mathrm{dw} R
\nn \\
\label{Edumbbell1} \\
&\approx& E_c^{(\mathrm{dmbl})} + \pi K_1 \ln(L/\xi_1)
+ 4\pi K_2\ln(L/\xi_2)
\nn \\
&&-\ 2\pi K_1 \ln(R/\xi_1) - 8\pi K_2 \ln(R/\xi_2)
+ \sigma_\mathrm{dw} R,
\nn \\
\label{Edumbbell2}
\end{eqnarray}
where the core energy is
\begin{equation}
E_c^{(\mathrm{dmbl})}
= 2\left[E_{1c}^{(\pi)} + E_{2c}^{(2\pi)} \right].
\label{Ecpoint}
\end{equation}
The first two logarithmic terms in (\ref{Edumbbell1}) estimate the
energy associated with the total $2\pi$ atomic and total $4\pi$
molecular topological charges \emph{outside} the dumbbell of size
$R$. The second two logarithmic terms in (\ref{Edumbbell1}) give the
energy of two $\pi$ atomic vortices and two $2\pi$ molecular
vortices, including their repulsive interaction at separation $R$.
Finally, the last term accounts for the energy of the domain wall in
the atomic superfluid.  Accuracy of the estimate requires that the
system size be much larger than the dumbbell size, which in turn
must be larger than the microscopic scale: $L \gg R \gg
\xi_1,\xi_2$.

To estimate the optimum size $R_0$ of the dumbbell vortex
configuration, one minimizes $E^{(\mathrm{dmbl})}_v(R)$ over $R$,
yielding
\begin{eqnarray}
R_0 &\approx& \frac{\pi}{2} \frac{K_1 + 4 K_2}{\sqrt{J K}},
\label{R0a}\\
&\approx& \frac{\pi\hbar}{\sqrt{m\alpha n_{20}^{1/2}}}
\sqrt{1+2\frac{n_{20}}{n_{10}}},
\label{R0b}
\end{eqnarray}
where (\ref{sigma_dw}) has been used for $\sigma_\mathrm{dw}$, and
the estimate in the last line is made by ignoring depletion effects,
valid in the dilute, weakly interacting limit.

It is clear that there is a range of parameters $K_\s$ and $J$ such
that the energy $E^{(dumbbell)}_v(R_0)$, Eq.\ (\ref{Edumbbell2}), of
the dumbbell vortex is lower than that, Eq.\ (\ref{Epoint}), of the
point vortex configuration. In this case, a 2d system will undergo a
transition to a state in which the $2\pi$ atomic vortices are in a
dumbbell $\pi-\pi$ configuration (locked to a $2\pi-2\pi$ molecular
vortex pair). Because there is no symmetry change associated with a
transition into a state where vortex dumbbells are randomly
oriented, one expects this transition to generically be first
order.\cite{nematicVortex} Since the energy balance between the two
competing states involves core energies, a more detailed microscopic
analysis (that is not pursued here) is necessary to pinpoint the
location of the transition. However, it is clear from the structure
of $E^{(dumbbell)}_v(R_0)$ and $R_0$, that such transition takes place for
a sufficiently large domain-wall width $\sqrt{K/J} \gg \xi_\s$, where
the line tension is small and same-sign vortex repulsion is large.
One thus expects such a transition in narrow (small $\alpha$)
Feshbach resonance systems.

\subsection{ASF-MSF as a confinement-deconfinement transition}

Observe that the atomic condensate density $n_{10}$ vanishes as the
detuning $\nu$ decreases towards a critical value $\nu_c$. Thus the
$2\pi$ dumbbell length
\begin{equation}
R_0 \approx \sqrt{\frac{K_2}{\alpha}}
\frac{2\pi}{n_{10}^{1/2} n_{20}^{1/4}}
\rightarrow \infty,
\;\;\mbox{for}\;\nu \rightarrow \nu_c,
\end{equation}
diverges along with the associated domain-wall width, $\xi$.
Therefore, the ASF--MSF transition in $d=2$ can be complementarily
described as a $2\pi$ molecular vortex deconfinement transition.
While $2\pi$ molecular vortices are confined by a {\em linear}
potential inside the ASF state, in the MSF state this confining
potential (in 2d) is replaced by a much weaker {\em logarithmic}
potential, that binds each $2\pi$ molecular vortex to its oppositely
charged partner.\cite{foot:Lee}

\subsection{Molecular superfluid}

Since molecular $2\pi$ vortices only appear as neutral dipoles in
the MSF phase, the state is characterized by long-range order in the
molecular order parameter, $\Psi_{20} \sim \langle e^{i\theta_2}
\rangle$ (however, as usual in 2d, at finite $T$, $\Psi_{20}$ itself
vanishes, while the molecular helicity modulus, or superfluid
density, $n_{s2}$ remains finite). On the other hand, a deconfined
domain wall (across which the atomic phase $\theta_1$ jumps by
$\pi$) leads to a vanishing of the atomic order parameter, $\Psi_1
\sim \langle e^{i\theta_1} \rangle = 0$.\cite{foot:Lee} The MSF
state exhibits ordinary molecular $2\pi$ point vortices, along with
the atomic and molecular Bogoliubov quasiparticles discussed in
Sec.\ \ref{sec:excite}. It is easy to see that, analogous to the
conventional BCS superconductor, here too an atomic Bogoliubov
quasiparticle (that is gapped) acquires a phase of $\pi$ upon
encircling a molecular $2\pi$ vortex. Thus, these two excitations
interact strongly with each other, with statistical-like
interactions that can be captured by a Chern-Simons field
theory.\cite{fracBalents}

\section{Summary and Conclusions}
\label{sec:summary}

In this paper the thermodynamics of a resonant atomic Bose gas has
been studied. Working within a two-channel model, formulated in
terms of bosonic atoms and their diatomic molecules, the complete
phase diagram of the system has been worked out as a function of
temperature and detuning, the properties of the phases, and the
nature of quantum and classical phase transitions between them,
studied. This analysis was supplemented by a variational calculation
on a one-channel model, whose salient results appear in Sec.\
\ref{sec:results}.

A most notable feature is the appearance of two distinct superfluid
phases, ASF and MSF, separated by an Ising type transition. These
are distinguished by the respective presence and absence of atomic
off-diagonal long-range order, atomic (gapped and gapless)
Bogoliubov spectra, and the nature of topological excitations. In
addition to a distinction based on the atomic momentum occupation
number, these phases can be distinguished through the domain wall
excitations in the ASF (which separate regions in which the atomic
phase aligns with the molecular phase in the two possible different
ways), characteristic of the broken discrete Ising symmetry.

\acknowledgments

L.R. thanks Victor Gurarie and Subir Sachdev for discussions, and the
Kavli Institute for Theoretical Physics for its hospitality during the
``Strongly Correlated Phases in Condensed Matter and Degenerate
Atomic'' workshop, supported in part by the NSF under
grant No.\ PHY05-51164. The authors acknowledge financial support by the
NSF under grant No.\ DMR-0321848 (LR, JP), NIST under the NRC
fellowship (JP), and NASA under contract No.\ NNC04CB23C
(PBW).

\appendix

\section{Details of connection to polar-nematic ordering}
\label{mapNematic}

In this Appendix a connection, stated in Sec.\ref{sec:symmetries},
between the two complex scalar order parameters, characterizing
phases of a resonant atomic Bose gas, to those of a vector-tensor
model of polar, nematic liquid crystals is elaborated.

In a thermodynamic description, the order parameters are derived from
the full free energy density $f_{AM}$ via derivatives with respect to
their conjugate fields:
\begin{equation}
\Psi_{\s 0}({\rv}) = -2\left(\frac{\partial f_{AM}} {\partial
h_\s^*}\right)_{h_\s = 0} = \langle \hat \psi_\s({\rv}) \rangle.
\label{3.18}
\end{equation}
This prescription  is unambiguous for the two-species model
(\ref{H2channel}), but in the one-species model (\ref{Ha}) or
(\ref{HaVar}), $\Psi_{20}$ remains to be defined.  One expects a
molecular composite operator of the form
\bea
\hat{\psi}^\dag_m(\xv) = \int d{\bf r} \phi_0({\bf r})
\hat \psi_1^\dag(\xv+{\bf r}/2)
\hat \psi_1^\dag(\xv-{\bf r}/2).
\label{3.18a}
\eea
to play the role of $\hat \psi_2$, where $\phi_0({\bf r})$ is the
molecular wavefunction. Therefore, to investigate molecular
superfluid ordering, one is motivated to look at the anomalous
correlation function \cite{foot:molop}
\begin{equation}
\Phi_{20}({\xv},{\xv}') = \langle \hat \psi_1({\xv})
\hat \psi_1({\xv}') \rangle.
\label{3.19}
\end{equation}
Since the fundamental object is a two-point function, one generally
lacks a unique definition of the one-point quantity
$\Psi_{20}({\rv})$. However, if the molecular size is much smaller
than their separation, in the spirit of the coarse graining picture of
the molecular operator $\hat \psi_2$, one could define
\begin{equation}
\Psi_{20}({\rv}) = \int d{\rv}' \phi_0({\rv}-{\rv}')
\Phi_{20}({\rv},{\rv}')
\label{3.21}
\end{equation}
in which the two-particle molecular wavefunction $\phi_0$ is used to
weigh the local volume average.

A complex-scalar atomic superfluid order parameter is clearly
isomorphic to a 2d vector order parameter ${\bf M}_0 = \sqrt{2}
(\mathrm{Re} \Psi_{10}, \mathrm{Im} \Psi_{10})$ whose components are
the averages
\begin{equation}
M_{0,1}(\xv) = \langle \hat Q(\xv) \rangle,\
M_{0,2}(\xv) = \langle \hat P(\xv) \rangle
\label{3.22}
\end{equation}
of the corresponding conjugate Hermitian operators,
\bea
\hat Q(\xv) &=& \frac{1}{\sqrt{2}} \left[\hat\psi_1(\xv)
+ \hat\psi_1^\dagger(\xv) \right]
\nn \\
\hat P(\xv) &=& \frac{1}{i\sqrt{2}} \left[\hat\psi_1(\xv)
- \hat\psi_1^\dagger(\xv) \right]
\label{3.23}
\eea
obeying the commutation relation $[\hat Q(\xv), \hat P(\xv')] =
i\delta({\xv}-{\xv}')$.  Phase symmetric combinations of $\Psi_{10}$
(i.e., products of its complex conjugate with itself) correspond to
rotation invariant combinations (i.e., dot products) of ${\bf M}_0$.
Similarly, one defines the 2d conjugate field vector ${\bf H} =
\frac{1}{\sqrt{2}}(\mathrm{Re} h_1, \mathrm{Im} h_1)$.

Since the molecular order parameter is fundamentally a two-point
correlation function, one is motivated to examine the tensor
\bea
{\cal Q}(\xv,\xv')
&=& \left[\begin{array}{cc}
\langle \hat Q(\xv) \hat Q(\xv') \rangle &
\langle \hat Q(\xv) \hat P(\xv') \rangle
\\
\langle \hat P(\xv) \hat Q(\xv') \rangle &
\langle \hat P(\xv) \hat P(\xv') \rangle
\end{array}\right]
\label{3.24} \\
&\equiv& {\cal Q}_s(\xv,\xv') + {\cal Q}_a(\xv,\xv')
+ \frac{1}{2} \Tr[{\cal Q}(\xv,\xv')] \openone
\nn
\eea
in which the symmetric traceless part is given by
\begin{eqnarray}
{\cal Q}_s(\xv,\xv')
&=& \frac{1}{2}\left\{{\cal Q}(\xv,\xv') + {\cal Q}^T(\xv,\xv')
- \Tr[{\cal Q}(\xv,\xv')] \right\}
\nn \\
&=& \left[\begin{array}{cc}
\textrm{Re} \Phi_{20}(\xv,\xv')& \textrm{Im} \Phi_{20}(\xv,\xv') \\
\textrm{Im} \Phi_{20}(\xv,\xv')& -\textrm{Re} \Phi_{20}(\xv,\xv')
\end{array} \right].
\label{3.25}
\end{eqnarray}
and is therefore seen to be precisely equivalent to the anomalous
correlation function $\Phi_{20}(\xv,\xv')$. The antisymmetric part
\begin{eqnarray}
{\cal Q}_a(\xv,\xv')
&=& \frac{1}{2} \left[{\cal Q}(\xv,\xv')
- {\cal Q}^T(\xv,\xv') \right]
\nn \\
&=& \left[\begin{array}{cc}
0 & -\textrm{Im} G(\xv,\xv') \\
\textrm{Im} G(\xv,\xv') & 0
\end{array} \right].
\label{3.26}
\end{eqnarray}
and the trace
\begin{equation}
\Tr[{\cal Q}(\xv,\xv')] = \textrm{Re} G(\xv,\xv')
\label{3.27}
\end{equation}
are given by the real and imaginary parts of the usual
(non-anomalous) two point correlation function
\begin{equation}
G(\xv,\xv') = \langle \hat \psi_1(\xv)
\hat \psi_1^\dagger(\xv') \rangle,
\label{3.28}
\end{equation}
and are therefore phase invariant scalars.  The information about
the molecular order parameter therefore lies entirely in ${\cal
Q}_s$.  One similarly defines the conjugate field tensor
\begin{equation}
{\cal H}_0 = \frac{1}{2} \left(\begin{array}{cc}
\mathrm{Re} h_2 & \mathrm{Im} h_2 \\
\mathrm{Im} h_2 & -\mathrm{Re} h_2
\end{array} \right).
\label{fieldtensor}
\end{equation}

By substituting ${\cal Q}_s(\xv,\xv')$ into (\ref{3.21}) one obtains
a definition of the local tensor order parameter ${\cal Q}_0(\xv)$.
The following identities now follow:
\begin{eqnarray}
\frac{1}{2} M_0^2 &=& |\Psi_{10}|^2
\nn \\
{\bf H}_0 \cdot {\bf M}_0 &=& \mathrm{Re} [h_1^* \Psi_{10}]
\nn \\
\frac{1}{2} \mathrm{Tr}\left({\cal Q}_0^2\right)
&=& -\det \left[{\cal Q}_0 \right] = |\Psi_{20}|^2,
\nn \\
\mathrm{Tr}({\cal H}_0 {\cal Q}_0)
&=& \mathrm{Re} [h_2^* \Psi_{20}]
\nn \\
\frac{1}{2} {\bf M}_0^T {\cal Q}_0 {\bf M}_0
&=& \mathrm{Re} \left[ \Psi_{20}^* \Psi_{10}^2 \right].
\label{3.29}
\end{eqnarray}
This demonstrates that the phase invariant combinations in the
Ginzburg-Landau free energy correspond to 2d rotation invariant
combinations in the vector-tensor representation.

A symmetric traceless tensor order parameter is familiar from a theory
of nematic liquid crystals, encoding the headless-arrow nature of the
nematic state of anisotropic molecules.\cite{deGennes,ChaikinLubensky}
Using the ``dictionary'', Eqs.(\ref{3.29}), in the vector-tensor
representation, ${\bf M}_0$, ${\cal Q}_0$, the mean field Hamiltonian
(\ref{HmuMFT}) takes the form
\begin{eqnarray}
{\cal H}_{\rm mf} &=& -\frac{1}{2} \mu_1 M_0^2
+ \frac{g_1}{8} M_0^4 - {\bf H}_0 \cdot {\bf M}_0
\nn \\
&&-\ \frac{1}{2} \mu_2 \mathrm{Tr}\left({\cal Q}_0^2 \right)
+ \frac{1}{8} g_2 \left[\mathrm{Tr}\left({\cal Q}_0^2 \right)\right]^2
- \mathrm{Tr}({\cal H}_0 {\cal Q}_0)
\nn \\
&&+\ \frac{1}{4} g_{12} M_0^2
\mathrm{Tr}\left({\cal Q}_0^2 \right)
- \frac{\alpha}{2} {\bf M}_0^T {\cal Q}_0 {\bf M}_0,
\label{HmuMFT2}
\end{eqnarray}
representing the theory of an interacting vector and nematic order
parameters.  In the current scalar superfluid context both have two
components, but (\ref{HmuMFT2}) is clearly not limited to this case.
The conjugate field ${\bf H}_0$ is the analogue of a magnetic field,
while ${\cal H}_0$ is the analogue of a nematic liquid crystal
polarization field.

The ``double-headedness'' of the nematic order parameter is
exhibited via the eigenvector decomposition
\begin{equation}
{\cal Q}_0(\xv) = q_0 [{\bf \hat n}(\xv) {\bf \hat n}(\xv)
- {\bf \hat m}(\xv) {\bf \hat m}(\xv)],
\label{3.31}
\end{equation}
in which $q_0 \ge 0$ is the order parameter and ${\bf \hat m},{\bf
\hat n}$ are orthonormal unit eigenvectors characterizing the
nematic order of the MSF. Clearly ${\cal Q}_0$ is invariant under
sign reversal of the unit vectors. Thus, although $q_0$ and ${\bf
\hat n}$ completely define ${\cal Q}_0$, both ${\bf \hat n}$ and
$-{\bf \hat n}$ characterize the same state.

To make contact with the complex molecular order parameter, let
${\bf \hat n} = [\cos(\theta_n),\sin(\theta_n)]$, ${\bf \hat m} =
[-\sin(\theta_n),\cos(\theta_n)]$. From (\ref{3.25}) one obtains
\bea
\Psi_{20} &=& q_0 [(\hat n_1^2 - \hat m_1^2)
+ i (\hat n_1 \hat n_2 - \hat m_1 \hat m_2)]
\nn \\
&=& q_0 e^{2i\theta_n}.
\label{3.32}
\eea
Thus the eigenvalue $q_0 = |\Psi_{20}|$ is the MSF order parameter
magnitude.  Although $\theta_n$ and $\theta_n + \pi$ are equivalent,
$\theta_2 \equiv 2\theta_n$ is uniquely defined.

Inserting (\ref{3.31}) into (\ref{HmuMFT2}), the Feshbach resonance
phase coupling ($\alpha$) term takes the form
\bea
&&-\alpha q_0 \left[({\bf M}_0 \cdot {\bf \hat n})^2
- \frac{1}{2} M_0^2 \right]
\nn \\
&&\ \ \ \ \ \ \ \ \ \ =\ - \alpha q_0 M_0^2
\left[\cos^2(\theta_n - \theta_1) - \frac{1}{2} \right]
\nn \\
&&\ \ \ \ \ \ \ \ \ \ =\ -\frac{1}{2} \alpha q_0 M_0^2
\cos(\theta_2 - 2\theta_1),
\label{3.33}
\eea
where the representation ${\bf M}_0 = M_0 [\cos(\theta_1),
\sin(\theta_1)]$ has been used. The coupling is again clearly
invariant under $\theta_n \to \theta_n + \pi$, and represents an
alignment between the nematic and polar order parameters, familiar
from the theory of polar nematic liquid crystals.
\cite{deGennes,ChaikinLubensky}

\section{Detailed analysis of mean field phase diagram}
\label{mftDetails}

In this Appendix a more detailed analysis of the mean field phase
diagram, summarized in Sec.\ \ref{sec:alphaneq0}, especially Figs.\
\ref{phase_diagramAlphaMuA} and \ref{phase_diagramAlphaMuB}, is
presented. To this end, it is convenient to introduce the following
scaled quantities:
\bea
r_1 &=& -\frac{g_{12}}{\alpha^2} \mu_1,\
r_2 = -\frac{g_1}{\alpha^2} \mu_2,\
\gamma = \frac{g_1 g_2}{g_{12}^2}
\label{scaledVars} \\
\bar \psi_1 &=& \frac{\sqrt{g_1 g_{12}}}{\alpha} \Psi_{10},\
\bar \psi_2 = \frac{g_{12}}{\alpha} \Psi_{20},\
\bar{\cal H} = \frac{g_1 g_{12}^2}{\alpha^4}{\cal H}
\nn
\eea
Substituting these into the mean field Hamiltonian (\ref{HmuMFT}),
one obtains the dimensionless form
\bea
\bar\cH &=& r_1 |\bar \psi_1|^2 + \frac{1}{2} |\bar \psi_1|^4
+ r_2 |\bar\psi_2|^2 + \frac{1}{2} \gamma |\bar \psi_2|^4
\nn \\
&&+\ |\bar \psi_1|^2 |\bar \psi_2|^2
- {\rm Re} [\bar \psi_2^* \bar \psi_1^2]
\label{scaledH}
\eea
The scaling has succeeded in reducing the problem to its essentials,
removing $\alpha$ entirely, and subsuming all interaction strengths
into the single parameter $\gamma$.  The only free parameters are
the two scaled chemical potentials, $r_\sigma$.  One may always
choose the phase of $\bar \psi_1$ so that it is real and
non-negative.  It is then clear from the last term in
(\ref{scaledH}) that $\bar\cH$ is minimized by taking $\bar \psi_2$
to also be real and non-negative, consistent with (\ref{thetaLock}).
With this input the scaled extremum equations take the form
\bea
0 &=& \bar \psi_1 \left(r_1 - \bar \psi_2
+ \bar \psi_1^2 +  \bar \psi_2^2 \right)
\label{SPscaledA} \\
0 &=& -\frac{1}{2}\bar \psi_1^2
+ \bar \psi_2 \left(r_2 + \gamma \bar \psi_2^2
+ \psi_1^2 \right),
\label{SPscaledB}
\eea
and lead to the phase diagrams for $\gamma > 1$ and $\gamma < 1$
shown in Figs.\ \ref{fig:mfpd1} and \ref{fig:mfpd2}, respectively
(and reproduced in Figs.\ \ref{phase_diagramAlphaMuA} and
\ref{phase_diagramAlphaMuB}). These results will now be derived in
detail.

\subsection{ASF free energy in terms of $\bar \psi_2$:}
\label{subsec:fpsi2}

One begins by expressing the energy density in terms of $\bar
\psi_2$ alone by using (\ref{SPscaledA}) and (\ref{SPscaledB}) to
eliminate $\bar \psi_1$. In the ASF phase (the transition to which,
from the N and MSF phases, is the main focus), where $\bar \psi_1
\neq 0$, (\ref{SPscaledA}) gives
\begin{equation}
\bar \psi_1^2 = \bar \psi_2 - \bar \psi_2^2 - r_1,
\label{psi12relate}
\end{equation}
which clearly requires that the right hand side be positive.  The
resulting condition,
\begin{equation}
r_1 \leq \bar \psi_2 - \bar \psi_2^2,
\label{r1r2condition}
\end{equation}
will be important in what follows.

Substituting (\ref{SPscaledA}) into the first line of
(\ref{SPscaledB}), a cubic saddle-point equation is obtained purely
in terms of $\bar\psi_2$
\begin{equation}
0 =\frac{\partial\bar\cH_{\rm ASF}}{\partial\bar\psi_2}\equiv
r_1+ 2 t_2 \bar \psi_2 + 3 \bar \psi_2^2 + 2(\gamma - 1)\bar\psi_2^3,
\label{SPscaledPsi2}
\end{equation}
and the corresponding energy density
\begin{equation}
\bar\cH_{\rm ASF} =
- \frac{1}{2} r_1^2 + r_1 \bar\psi_2 + t_2 \bar \psi_2^2
+ \bar\psi_2^3 + \frac{1}{2}(\gamma-1) \bar \psi_2^4,
\label{HmftPsi2}
\end{equation}
in which the parameter
\begin{equation}
t_2 = r_2 - r_1 - \textstyle{\frac{1}{2}}
\label{tDefine}
\end{equation}
has been defined.

The apparent instability of $\bar\cH_{\rm ASF}$ at large $|\bar
\psi_2|$ for $\gamma < 1$ is in fact illusory: The condition
(\ref{r1r2condition}) restricts $\bar \psi_2$ to a finite interval
$\max\{0,\frac{1}{2}(1-\sqrt{1-4r_1})\} \leq \bar \psi_2 \leq
\frac{1}{2}(1+\sqrt{1-4r_1})$ for any given $r_1$.  Similarly, even
for $\gamma > 1$, (\ref{HmftPsi2}) may have its absolute minimum at
negative $\bar \psi_2$ (e.g., if $r_1 \geq 0$), but only minima at
non-negative $\bar \psi_2$ are of physical interest.

\subsection{Continuous transitions:}
\label{subsec:cts}

It follows from (\ref{SPscaledB}) that if $\bar \psi_1 = 0$, then
either $\bar \psi_2 = 0$ (normal phase) or $\bar \psi_2^2 =
-r_2/\gamma$ (MSF phase, existing only for $r_2 < 0$).  This is the
standard result arising from $\bar\cH = r_2 \bar \psi_2^2 +
\frac{1}{2} \gamma \bar \psi_2^4$, showing that the N--MSF
transition must be continuous, and take place at $r_2 = 0$.

Substituting $\bar \psi_2 = 0$ into (\ref{psi12relate}), one
concludes that any \emph{continuous} N--ASF transition must take
place along the line $r_1 = 0$ (with $r_2 \geq 0$).  For $r_1 < 0$,
it is easy to check that $\bar \cH$ is locally unstable to nonzero
$\bar \psi_1$. Hence, if the transition is first order, representing
some global instability, it must occur for $r_1 > 0$.

Similarly, substituting $\psi_2^2 = -r_2/\gamma$ into
(\ref{psi12relate}), one finds that any \emph{continuous} ASF--MSF
transition must take place along the curve
\begin{equation}
r_1 = r_{1c}(r_2)
\equiv \sqrt{|r_2|/\gamma} + r_2/\gamma.
\label{ASF-MSFtransition}
\end{equation}
For $r_1 < r_{1c}$ it is easy to check that $\bar \cH$ is locally
unstable to nonzero $\bar \psi_1$.  Hence, any first order
transition must occur for $r_1 > r_{1c}$.

\begin{figure}

\includegraphics[width=0.45\textwidth]{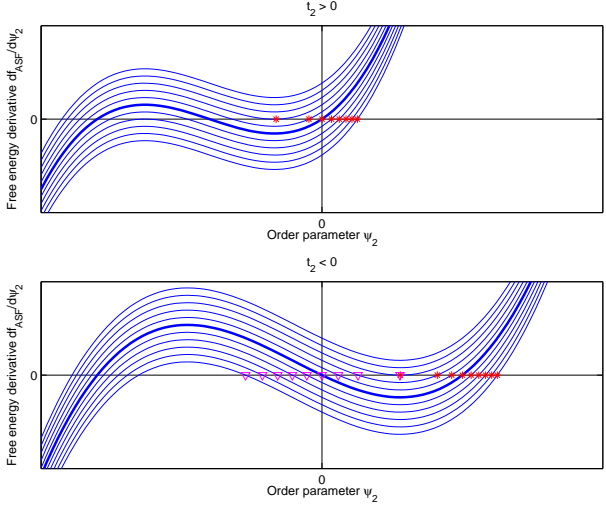}

\caption{Graphical illustration of the solution to
(\ref{SPscaledPsi2}) for $\gamma > 1$ and the cases $t_2 > 0$ (upper
panel) and $t_2 < 0$ (lower panel). The thicker lines correspond to
$r_1 = 0$. Curves above (below) these correspond to $r_1 > 0$ ($r_1
< 0$). For $t_2 > 0$ and $r_1 < 0$ there is a unique positive root
(stars) that approaches the origin as $r_1 \to 0^-$, and then
crosses to unphysical negative values.  As discussed in the text,
this signifies a second order N--ASF transition at $r_1 = 0$. For
$t_2 < 0$ there is a range of negative $r_1$ for which there are two
positive roots.  However, the one (triangles) that vanishes at $r_1
= 0$ is a local maximum of the free energy (\ref{HmftPsi2}), and
therefore unstable. The physical root (stars) is the larger one,
which never reaches the origin, but annihilates with the unstable
one at the positive value at which the cubic discriminant $\Delta$,
Eq.\ (\ref{3.x11}), vanishes. The transition must therefore be first
order, taking place at some intermediate point where the free energy
itself vanishes, but $\Delta$ is still positive.  For $\gamma \leq
1$ the result is qualitatively identical.  For $\gamma < 1$ the far
left root moves to the far right, but remains unphysical since it
corresponds to a local maximum in the free energy. For $\gamma = 1$
it disappears entirely. In either case, the evolution of the other
two roots remains as described.}

\label{fig:roots}
\end{figure}

\subsection{First order N--ASF transition:}
\label{subsec:nasf1}

Consider now the possible existence of a first order N--ASF
transition. For $t_2 \geq 0$ all coefficients in (\ref{HmftPsi2}),
aside from the constant term $r_1$, are positive. A positive $\bar
\psi_2$ root therefore exists only for $r_1 < 0$ (vanishing as $r_1
\to 0^-$), and is clearly unique---see the upper panel in Fig.\
\ref{fig:roots}.  The continuous transition at $r_1 = 0$ therefore
takes place for $t_2 > 0$, i.e., $r_2 > \frac{1}{2}$.

On the other hand, as illustrated in the lower panel of Fig.\
\ref{fig:roots}, for $t_2 < 0$, $\bar\cH_{\rm ASF}$ has a local
minimum at some positive value of $\bar \psi_2$. Thus, for a range
of $r_1 \geq 0$, there will be two non-negative roots, the smaller
of which (vanishing at $r_1 = 0$) corresponds to a maximum of
$\bar\cH_{\rm ASF}$. It is the larger root, existing for a range of
positive $r_1$, that corresponds to the minimum of $\bar\cH_{\rm
ASF}$ associated with the ASF state. For this range of positive
$r_1$ the transition from the normal state to this ASF minimum must
therefore be first order. These considerations hold for both $\gamma
> 1$ and $\gamma \leq 1$.

\begin{figure*}

\includegraphics[width=0.45\textwidth,clip]{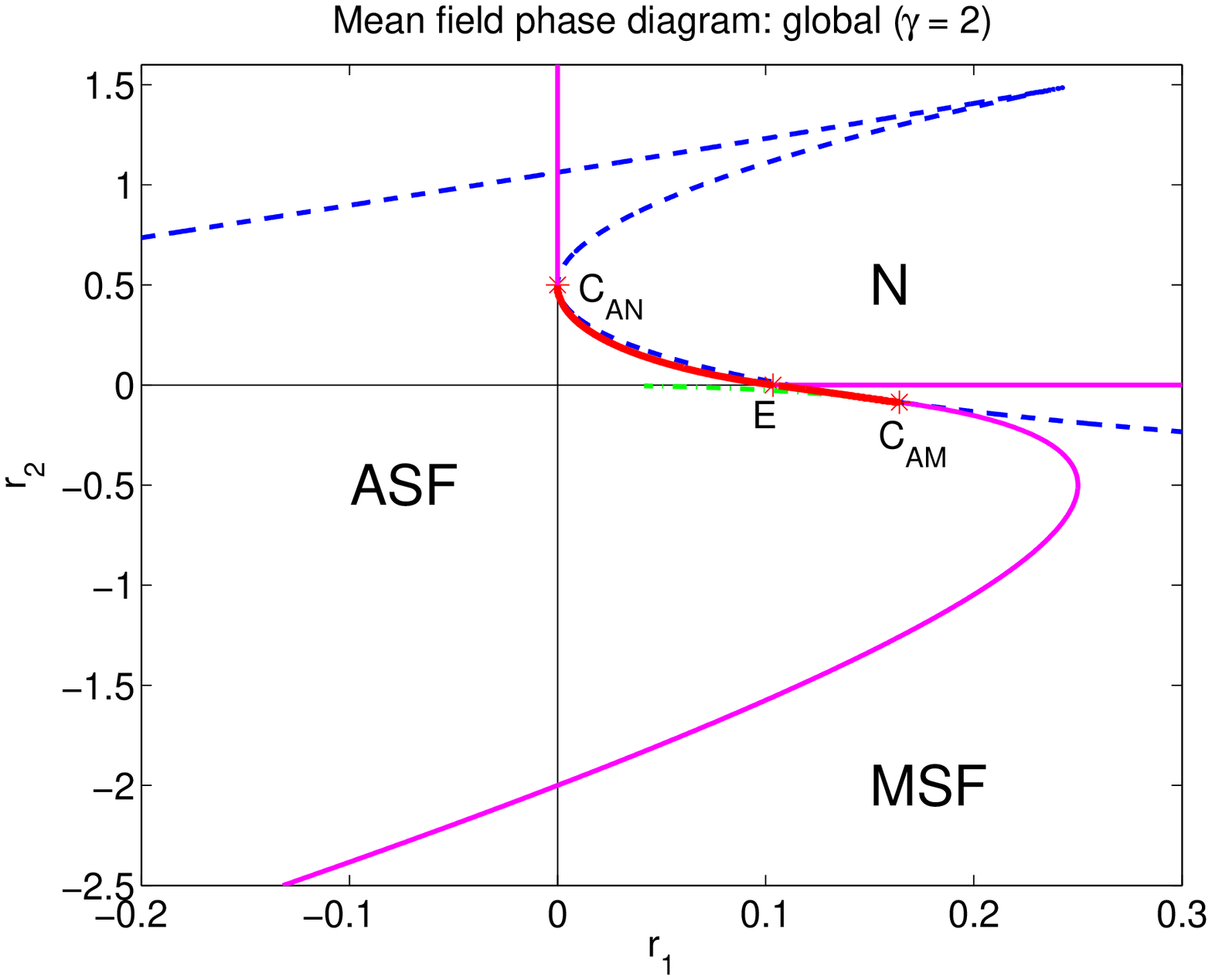}
\qquad
\includegraphics[width=0.45\textwidth,clip]{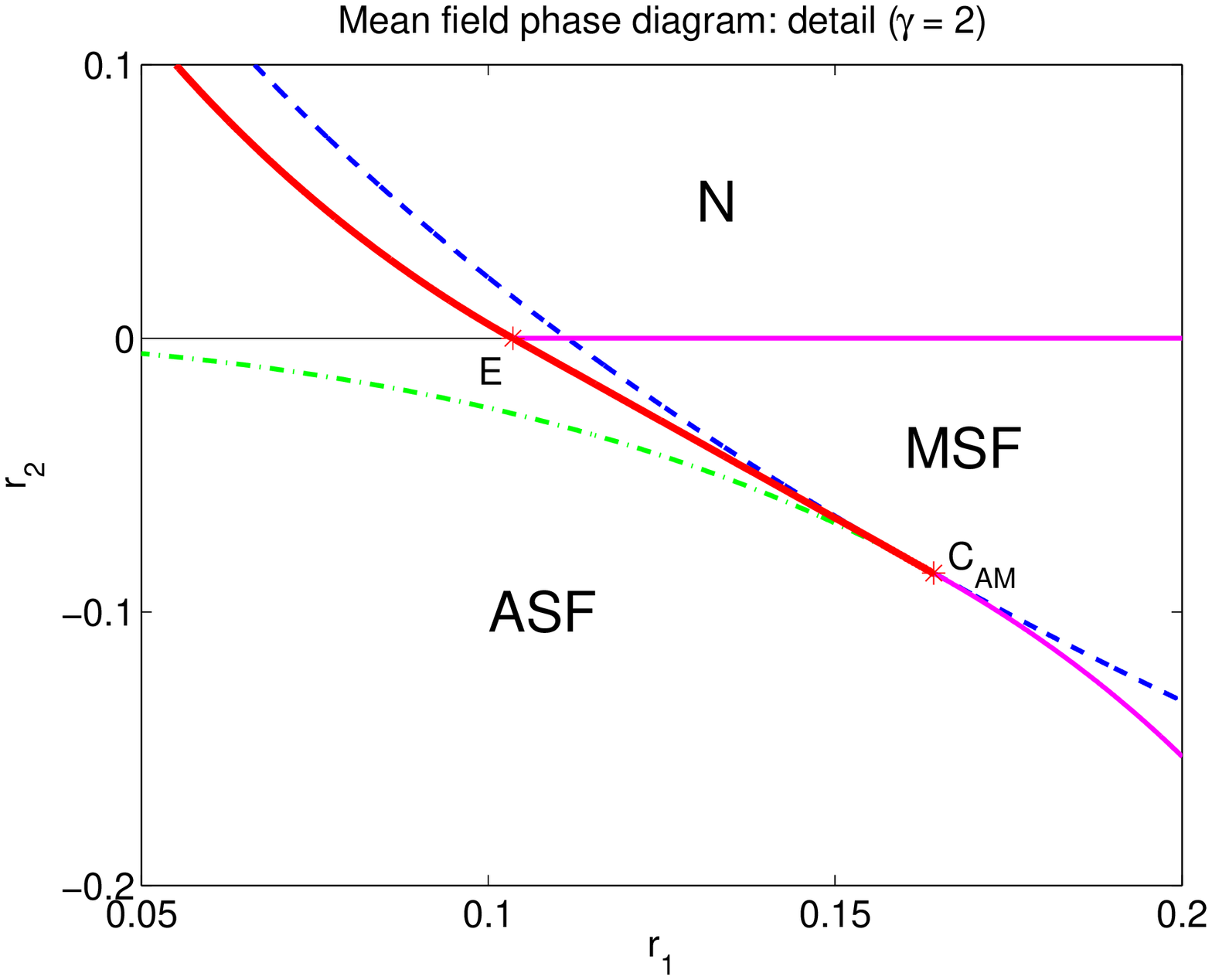}

\caption{Mean-field phase diagram for $\gamma \geq 1$, in the scaled
coordinates defined by (\ref{scaledVars}), with second order phase
boundaries given by the thinner solid lines, and first order phase
boundaries the thicker solid line. The right plot shows an expanded
detail near the labeled points $E$ and $C_{AM}$. The value $\gamma =
2$ is used in the numerical computation here, but the basic
structure remains unchanged for other values. The second order
N--ASF line along the positive $r_2$ axis encounters a tricritical
point $C_{AN}$ at $r_2 = 1/2$, below which the transition turns
first order and the line bends into positive $r_1$.  The second
order N--MSF transition line along the positive $r_1$ axis
terminates on this first order line at a critical endpoint $E$. The
first order line continues below $E$, now separating the ASF and MSF
phases, but turns second order at another tricritical point
$C_{AM}$, and follows the line defined by (\ref{ASF-MSFtransition}).
For large $r_1,r_2$ this line asymptotes to $r_2 = \gamma r_1$,
which agrees with (\ref{ASF-AMSF}) in unscaled units. Its unphysical
continuation toward the origin is shown by the dash-dotted line. The
dashed line is the curve $\Delta = 0$, and therefore provides a
bound on the ASF phase boundary. All three lines osculate at the
point $C_{AM}$. The N--ASF and dashed lines also osculate at
$C_{AN}$. For $\gamma = 0$ the cusp together with the left hand
branch of the $\Delta = 0$ line is pushed off to infinity, but
$C_{AN}$, $E$ and $C_{AM}$ remain finite and well defined.}

\label{fig:mfpd1}
\end{figure*}

\begin{figure*}

\includegraphics[width=0.45\textwidth,clip]{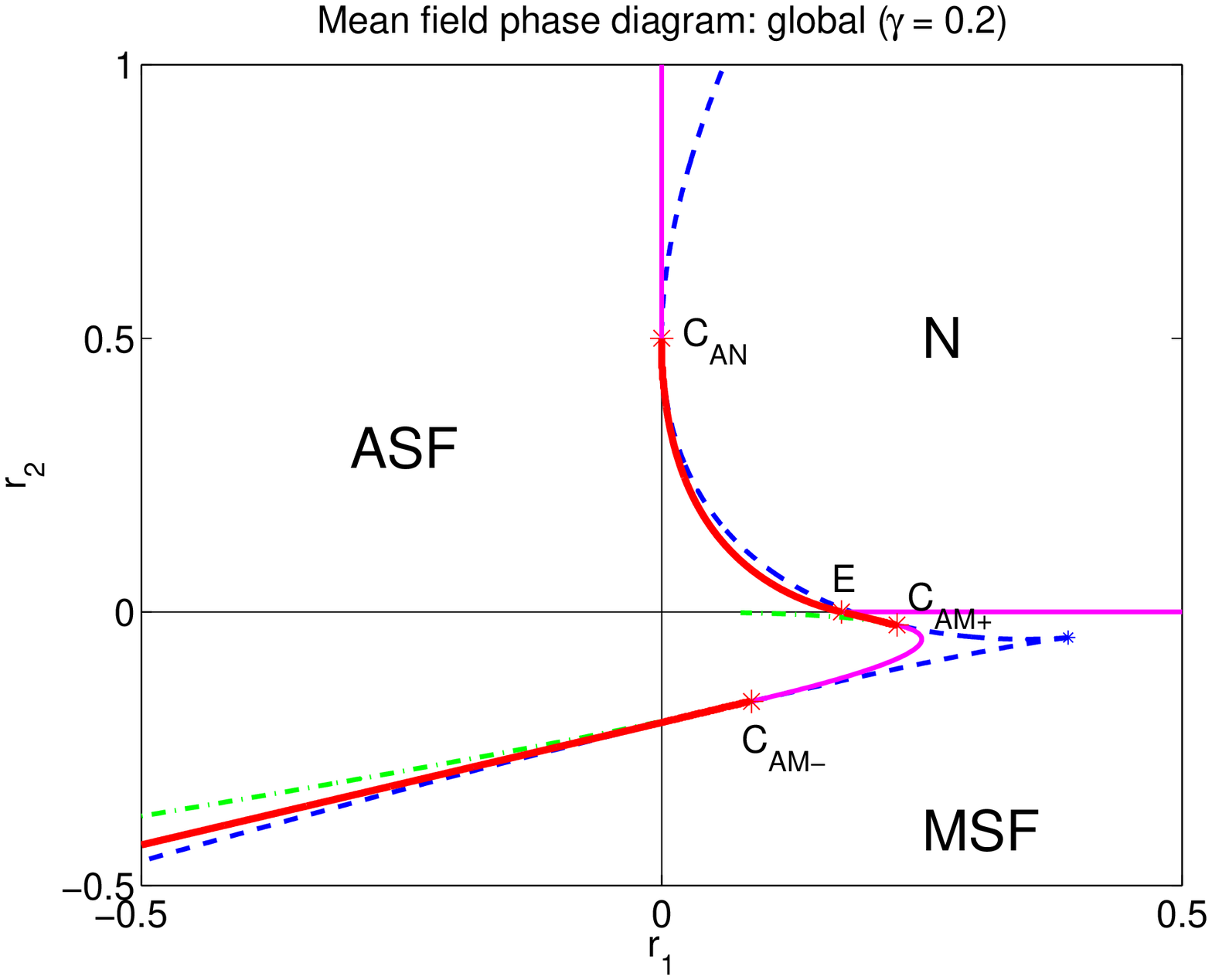}
\qquad
\includegraphics[width=0.45\textwidth,clip]{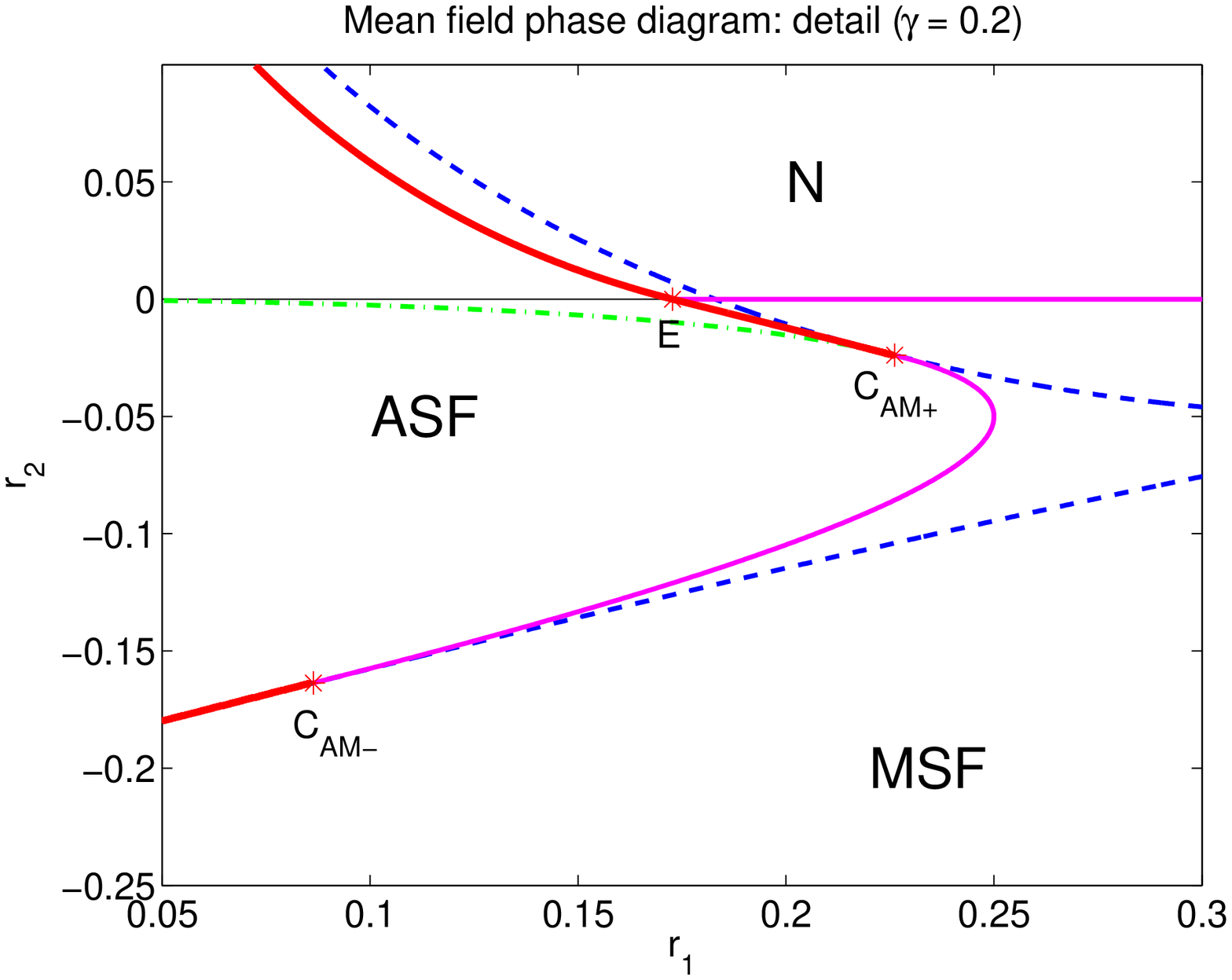}

\caption{Mean field phase diagram for $\gamma < 1$, in the scaled
coordinates defined by (\ref{scaledVars}), with second order phase
boundaries given by the thinner solid lines, and first order phase
boundaries the thicker solid line. The right plot shows an expanded
detail near the labeled points $E$, $C_{AM+}$ and $C_{AM-}$. The
value $\gamma = 0.2$ is used in the numerical computation here, but
the basic structure remains unchanged for other values.  The upper
part of the phase diagram is very similar to that for $\gamma > 1$.
The second order N--ASF line along the positive $r_2$ axis
encounters a tricritical point $C_{AN}$ at $r_2 = 1/2$, below which
the transition turns first order and the line bends into positive
$r_1$. The second order N--MSF transition line along the positive
$r_1$ axis terminates on this first order line at a critical
endpoint $E$. The first order line continues below $E$, now
separating the ASF and MSF phases, but turns second order at another
tricritical point, now labeled $C_{AM+}$, and follows the line
defined by (\ref{ASF-MSFtransition}). However, now the second order
ASF--MSF line turns first order again at a new tricritical point
$C_{AM-}$. The unphysical continuation of this line above $C_{AM+}$
and below $C_{AM-}$ is shown by the dash-dotted line. When $\gamma
\to 1^-$, the latter is pushed out to infinity, while the former
remains finite.  For large $r_1,r_2$ the first order line asymptotes
to $r_2 = \sqrt{\gamma} r_1$, which agrees with (\ref{3.14}) in
unscaled units. The dashed line is the curve $\Delta = 0$, and
therefore provides a bound on the ASF phase boundary. All three
lines (though two different branches of the $\Delta = 0$ line)
osculate at the points $C_{AM+}$ and $C_{AM-}$. The N--ASF and
dashed lines also osculate at $C_{AN}$.}

\label{fig:mfpd2}
\end{figure*}

To quantify this, define the cubic discriminant
\begin{equation}
\Delta = (q/2)^2 + (p/3)^3
\label{3.x11}
\end{equation}
where
\bea
p &=& \frac{1}{\gamma-1}
\left[t_2 - \frac{3}{4(\gamma-1)} \right]
\nn \\
q &=& \frac{1}{2(\gamma-1)}
\left[r_1 - \frac{t_2}{\gamma-1}
+ \frac{1}{2(\gamma-1)^2} \right].
\label{3.x12}
\eea
For $\Delta > 0$, $\bar\cH'_{\rm ASF}$ has only one real root, while
for $\Delta < 0$ it has three real roots.\cite{foot:cubic}  As
$\Delta \to 0^-$ two of the roots merge, and subsequently annihilate
as $\Delta$ changes sign. There are two branches to the $\Delta = 0$
curve (shown as dashed lines in Figs. \ref{fig:mfpd1} and
\ref{fig:mfpd2}) given by
\begin{equation}
r_{1\pm}(t_2) = \frac{\pm \left[1-\frac{4}{3}(\gamma-1) t_2
\right]^{3/2} - 1 + 2(\gamma-1) t_2}{2(\gamma-1)^2},
\label{3.x13}
\end{equation}
of which $r_{1+}$ corresponds to the merging and subsequent
disappearance of the two roots of interest.  The two branches meet
and terminate at a cusp (denoted by a star in Figs.
\ref{fig:mfpd1} and \ref{fig:mfpd2}) at the point
\begin{equation}
t_{2\Delta} = \frac{3}{4(\gamma-1)},\
r_{1\Delta} = \frac{1}{4(\gamma-1)^2}\ \Rightarrow\
r_{2\Delta} = \frac{\gamma (2\gamma-1)}{(\gamma-1)^2},
\label{3.x14}
\end{equation}
where all three roots coincide. The first order transition must
therefore take place at a point $r_{1c}(t_2)$ in the interval $0 <
r_{1c} < r_{1+}$ where the energy density (\ref{HmftPsi2}) itself
vanishes.  For small $t_2$ one finds $r_{1+} = \frac{1}{3} t_2^2\{1
+ O[(\gamma-1)t_2]\}$ (the leading quadratic form being exact to all
orders for $\gamma = 1$), and the very small deviation $r_{1+} -
r_{1c} = - \frac{1}{3} (\frac{2}{3} t_2)^6 [1 + O(t_2)]$.

The point $r_1 = 0$, $r_2 = \frac{1}{2}$, at which the transition
line turns from second order to first is a tricritical point,
labeled $C_{AN}$ in Figs. \ref{fig:mfpd1} and \ref{fig:mfpd2}.
Numerical results for $r_{1c}$ are shown using $\gamma = 2$ and
$\gamma = 0.2$, respectively.

\subsection{First order MSF--ASF transition:}
\label{subsec:msfasf1}

Along the putative second order line (\ref{ASF-MSFtransition}), the
extremum condition (\ref{SPscaledPsi2}) may be factored in the form
\begin{equation}
\bar\cH_{\rm ASF}' = 2(\gamma-1) (\bar \psi_2 - \bar \psi_{20})
(\bar \psi - \bar \psi_{2+})(\bar \psi - \bar \psi_{2-})
\label{3.x15}
\end{equation}
in which $\bar \psi_{20} = \sqrt{|r_2|/\gamma}$ is the MSF state
value, and the other two roots are given by
\bea
\bar \psi_{2\pm} &=& \frac{1}{4(\gamma-1)}
\bigg\{-3-2(\gamma-1) \bar \psi_{20}
\label{3.x16} \\
&&\pm\ \sqrt{[2(\gamma-1) \bar \psi_{20} + 1]^2
+ 8\gamma} \bigg\}.
\nn
\eea
The argument of the square root is positive, so all three roots
are real, and a picture similar to the lower panel of Fig.
\ref{fig:roots} obtains.

\paragraph{The case $\gamma \geq 1$:}
\label{par:gge1}

In order that $\bar \psi_{20}$ be the physical root, the condition
$\bar \psi_{20} \geq \bar \psi_{2+}$ is required, leading to the
inequality
\begin{equation}
\bar\psi_{20} \geq \frac{1}{2(\sqrt{\gamma} + 1)},
\label{3.x17}
\end{equation}
which remains well defined for $\gamma = 1$.  Thus, for
\begin{equation}
r_2 < r_{2T} \equiv -\frac{\gamma}{4(1 + \sqrt{\gamma})^2}
\label{3.x18}
\end{equation}
the transition is indeed second order. For $r_2 > r_{2T}$, $\bar
\psi_{2+}$ becomes the physical root: the transition turns first
order, and takes place when the two energy densities match:
$\bar\cH_{\rm ASF} = \bar\cH_{\rm MSF} = -r_2^2/2\gamma$. The point
$r_2 = r_{2T}$, $r_1 = r_{1T} \equiv (2\sqrt{\gamma} +
1)/4(\sqrt{\gamma} + 1)^2$ is a second tricritical point,\cite{BM05}
labeled $C_{AM}$ in Fig.\ \ref{fig:mfpd1}.  At $r_{2T}$,
(\ref{3.x15}) has a coincident pair of roots, and the line $r_{1+}$
and the transition line must therefore osculate---see Fig.\
\ref{fig:mfpd1}.

\paragraph{The case $\gamma < 1$:}
\label{par:glt1}

For $\gamma < 1$, $\bar \psi_{20}$ must be the intermediate root:
$\psi_{2-} \leq \bar \psi_{20} \leq \bar \psi_{2+}$, leading to the
inequality
\begin{equation}
\frac{1}{2(1 + \sqrt{\gamma})} \leq \bar\psi_{20}
\leq \frac{1}{2(1 - \sqrt{\gamma})}.
\label{3.x19}
\end{equation}
Thus, for
\begin{equation}
r_{2T-} \equiv -\frac{\gamma}{4(1 - \sqrt{\gamma})^2}
< r_2 < r_{2T+} \equiv -\frac{\gamma}{4(1 + \sqrt{\gamma})^2}
\label{3.x20}
\end{equation}
the transition is indeed second order. For $r_2 > r_{2T+}$, $\bar
\psi_{2+}$ becomes the physical root, while for $r_2 < r_{2T-}$,
$\bar \psi_{2-}$ becomes the physical root: in either case the
transition turns first order, and takes place when the ASF and MSF
energies match. The point $r_2 = r_{2T\pm}$, $r_1 = r_{1T\pm} \equiv
[2(\sqrt{\gamma} \pm 1)-1]/4(\sqrt{\gamma} \pm 1)^2$ are both
tricritical points, labeled $C_{AM\pm}$ in Fig.\ \ref{fig:mfpd2}. At
both points (\ref{3.x15}) has a coincident pair of roots, and at
both points the (two different branches of the) $\Delta = 0$ curve
and the transition curve must therefore osculate, as illustrated in
Fig.\ \ref{fig:mfpd2}.  Notice that for $\gamma \to 1-$, $C_{AM+}$
remains well defined, and coincides with $C_{AM}$, while $C_{AM-}$
is pushed to infinity.  The semi-infinite second order line is
therefore recovered in this limit, consistent with the $\gamma \geq
1$ result.

\subsection{Critical endpoint for the N--MSF transition:}
\label{subsec:cep}

Since $\bar\cH_{\rm MSF} \to 0$ as $r_2 \to 0$, the N--ASF and
MSF--ASF first order transition curves must meet at $r_2=0$. The
second order N--MSF curve also terminates at this point, which
therefore represents a critical endpoint, labeled $E$ in Figs.\
\ref{fig:mfpd1} and \ref{fig:mfpd2}.

Figures \ref{fig:mfpd1} and \ref{fig:mfpd2} show complete phase
diagrams, computed for the cases $\gamma = 2$ and $\gamma = 0.2$,
respectively, and correspond to the two phase diagram topologies
illustrated in Figs.\ \ref{phase_diagramAlphaMuA} and
\ref{phase_diagramAlphaMuB} of the main body of the paper.  Here the
first order transition curves are computed numerically.  The
osculation with the $\Delta = 0$ line at the various tricritical
points is also shown.

\section{The two-body molecular binding problem}
\label{app:twobody}

The center of mass Schr\"odinger equation for two particles of mass
$m_A$ interacting via an attractive potential $v({\bf r})$ with
microscopic range $d_0$ is given by
\begin{equation}
-\frac{\hbar^2}{m_A} \nabla^2 \psi({\bf r})
+ v({\bf r}) \psi({\bf r}) = E \psi({\bf r}).
\label{B1}
\end{equation}
Of interest is the regime where the particle is very weakly bound,
with binding energy $E < 0$ obeying $|E| \ll |v|$.  The wavefunction
$\psi$ of such a weakly bound state will extend a distance much
greater than $d_0$ outside the potential. A good approximation is
then to treat $\psi \approx \psi(0)$ as essentially constant over
the potential region $r < d_0$ (this notion may be made rigorous
using effective s-wave scattering parameters). A Fourier analysis of
(\ref{B1}) then yields
\begin{equation}
\hat \psi({\bf k}) = \frac{m_A |v_0|}{\hbar^2}
\frac{\tilde v({\bf k})}{k^2 + \kappa^2} \psi(0)
\label{B2}
\end{equation}
where $\hat v({\bf k}) \equiv v_0 \tilde v({\bf k})$ is the
Fourier transform of the potential, $v_0 = \hat v(0) < 0$ is the
area under the potential, and $\kappa^2 = m_A|E|/\hbar^2$. The
self-consistency condition determining $E$ is therefore
\begin{equation}
\frac{\hbar^2}{m_A |v_0|} = \int \frac{d^dk}{(2\pi)^d}
\frac{\tilde v({\bf k})}{k^2 + \kappa^2}.
\label{B3}
\end{equation}
For $d \leq 2$ the right hand side diverges at $E = 0$.  Therefore
a bound state solution exists for arbitrarily weak potential.  For
$d > 2$ there is a critical potential strength $|v_{0,c}|$, below
which the potential fails to bind, given by
\begin{equation}
\frac{\hbar^2}{m_A |v_{0,c}|} = \int \frac{d^dk}{(2\pi)^d}
\frac{\tilde v({\bf k})}{k^2},
\label{B4}
\end{equation}
and one may write
\begin{equation}
\frac{\hbar^2}{m_A}
\left(\frac{1}{|v_{0,c}|} - \frac{1}{|v_0|} \right)
= \kappa^2 \int \frac{d^d k}{(2\pi)^d}
\frac{\tilde v({\bf k})}{k^2(k^2 + \kappa^2)}.
\label{B5}
\end{equation}
Close to the critical point the integral is dominated by the small
$k$ region, $k \ll \pi/a$.  One may then ignore the $k$-dependence
of $\tilde v$ (the integral remains convergent for $2 < d < 4$), to
obtain
\begin{equation}
\kappa^2 = \left[\frac{\hbar^2}{m_A C_d}
\left(\frac{1}{|v_{0,c}|} - \frac{1}{|v_0|} \right)
\right]^{\frac{2}{d-2}},
\label{B6}
\end{equation}
where
\begin{equation}
C_d = \int \frac{d^du}{(2\pi)^d} \frac{1}{u^2(u^2 + 1)},
\label{B7}
\end{equation}
in particular $C_3 = 1/4\pi$. One therefore obtains the power law
relationship $|E| \sim (|v_0| - |v_{0,c}|)^{2/(d-2)}$ describing the
vanishing of the binding energy upon approach to the critical point.

\section{Thermal and Bose condensate density profiles in a harmonic
trap}
\label{app:density}

The most direct probe of a trapped degenerate atomic gas is through
its spatial density profile, obtained from a freely expanding
cloud. In this Appendix, the details of the density profile
calculations for a trapped Bose gas are presented. For the weakly
interacting case, the profile is well approximated by the
noninteracting expression
\begin{equation}
n(\xv) = \sum_{\nv={\bf 0}}^\infty|\phi_\nv(\xv)|^2 n_\nv,
\end{equation}
with the occupation number $n_\nv$, for a Bose gas given by the
Bose-Einstein distribution
\begin{equation}
n_\nv = \langle\hat a^\dagger_\nv \hat a_\nv\rangle
= \frac{1}{e^{\beta(\eps_\nv - \mu)}-1}.
\end{equation}
The single-particle spectrum $\eps_\nv$ and normalized wavefunctions
$\phi_\nv(\xv)$ are solutions of the single particle Schr\"odinger
equation $\hat h \phi_\nv = [-(\hbar^2/2m)\nabla^2 + V(\xv))\phi_\nv
= \eps_\nv \phi_\nv$ appropriate to a trapping potential $V(\xv)$.
For a 3d harmonic potential $V(\xv) = \frac{1}{2} m \wt^2
r^2-\frac{3}{2}\hbar\wt$ (that for simplicity is taken to be
isotropic)
\begin{eqnarray}
\hspace{-2cm}
\eps_\nv &=& \hbar\wt(n_x + n_y + n_z)\equiv\hbar\wt n,
\\
\phi_\nv(\xv) &=& \prod_{i=x,y,z}
\left(\frac{1}{\sqrt{\pi}\ \rt 2^{n_i} n_i!}\right)^{1/2}
H_{n_i}(\hat r_i)e^{-\hat r_i^2/2},\ \ \ \ \ \
\end{eqnarray}
with $H_{n}(x)$ the $n$th Hermite polynomial, a function of the
normalized coordinates $\hat r_i = r_i/\rt$ ($i=x,y,z$), expressed in
units of the quantum oscillator length $\rt = \sqrt{\hbar/m\wt}$. The
chemical potential, $\mu < 0$ (throughout) is determined by the total
atom constraint
\begin{equation}
N = \int d\xv n(\xv) = \sum_{\nv={\bf 0}}^\infty
\frac{1}{e^{\beta(\eps_\nv - \mu)}-1}.
\end{equation}

In this noninteracting limit, at $T=0$ all atoms go into the lowest
single-particle state $\phi_{\bf 0}$, forming a Bose-Einstein
condensate, with
\begin{equation}
n^{T=0}(r)= \frac{N}{{\pi}^{3/2}\ \rt^3}\  e^{-r^2/\rt^2}.
\label{app:nT=0}
\end{equation}

At finite $T$, a fraction of atoms is thermally excited to higher
single-particle states, and
\begin{eqnarray}
n(r) &=& \langle\xv|\frac{1}{e^{\beta(\hat h-\mu)}-1}|\xv\rangle
\nn \\
&=&\sum_{p=1}^\infty e^{\beta\mu p}\langle\rv|e^{-\beta p \hat h}|\rv\rangle
\nn \\
&=&\sum_{p=1}^\infty e^{\beta\mu p}\rho_{\rm osc}(\rv,\rv; p \beta)
\label{app:nr}
\end{eqnarray}
is expressible purely in terms of diagonal elements of the single-particle
density matrix for a harmonic oscillator
\begin{eqnarray}
\label{rho_osc}
\rho_{\rm osc}(\rv,\rv',\beta)&=&\langle\rv|e^{-\beta\hat h}|\rv'\rangle,
\end{eqnarray}
governed by a single-particle Hamiltonian $\hat h$.  The density
matrix can be found by solving a diffusion equation in a harmonic
potential (or equivalently obtained from analytic continuation of
the harmonic oscillator evolution operator) with the ``initial''
condition of $\rho_{\rm osc}(\rv,\rv';\beta=0) = \delta(\rv-\rv')$
obvious from (\ref{rho_osc}).\cite{FeynmanStatMech} For $d=3$ one
has
\begin{equation}
\rho_{\rm osc}(\rv,\rv;\beta)
=\left[\frac{m\wt e^{\beta\hbar\wt}}
{2\pi\hbar\sinh(\beta\hbar\wt)}\right]^{3/2}
e^{-r^2/\rt^2(\beta)},
\end{equation}
where
\begin{eqnarray}
\rt^2(\beta) &=& \frac{\hbar}{m\wt}\coth\left(\beta\hbar\wt/2\right)
\nn \\
&\approx&
\begin{cases}
\frac{\hbar}{m\wt}, & \hbar\wt/k_B T\gg 1
\\
\frac{k_B T}{\frac{1}{2}m\wt^2}, & \hbar\wt/k_B T \ll 1,
\end{cases}
\label{r0T}
\end{eqnarray}
is the finite-temperature ``oscillator length'' that reduces to the
quantum one $r_0 = \sqrt{\hbar/m\wt}$ at low $T$ and the classical
(thermal) one, $r_T = \sqrt{2 k_B T/m \wt^2}$ (defined by $\frac{1}{2}
m\omega_0^2 r_T^2 = k_B T$) at high $T$.

The $p$ sum in $n(r)$, Eq.\ (\ref{app:nr}), can be evaluated
analytically in various limits.  For $\hbar\omega_0/k_B T\ll 1$
[valid for all but extremely low $T$, where (\ref{app:nT=0}) holds]
the $p$ sum naturally breaks up into two parts with $n(r) = n_T(r) +
n_0(r)$.  The two contributions correspond, respectively, to ranges
$1 \le p < p_c={k_B T/\hbar\omega_0}$ (``thermal'') and $p_c\le p <
\infty$ (``quantum'') with $p_c$ determined by $p_c \beta \hbar
\omega_0 = 1$. The thermal range is characterized by $p$ such that
$p \beta \hbar \omega_0 < 1$, within which the finite temperature
auxiliary oscillator length $r_0(\beta)$ and the corresponding
density matrix $\rho_{\rm osc}(\rv,\rv;p\beta)$ can be approximated
by their thermal classical forms, giving
\begin{eqnarray}
n_T(r) &\approx& \left(\frac{k_B T}{2\pi \hbar\omega_0}\right)^{3/2}
\frac{1}{r_{0}^3} \sum_{p=1}^{p_c-1}
\frac{1}{p^{3/2}} e^{-p(r^2/r_{T}^2 + |\mu|/k_B T)}
\nn \\
&\equiv& \left(\frac{k_B T}{2 \pi\hbar\omega_0}\right)^{3/2}
\frac{1}{r_0^3} \tilde{g}_{3/2} \left(e^{-r^2/r_T^2 - |\mu|/k_B T},
\frac{k_B T}{\hbar\omega_0}\right).
\nn \\
\label{app:n_Tgeneral}
\end{eqnarray}
where a ``cutoff'' extended zeta function,
\begin{eqnarray}
\tilde{g}_\alpha(x,p_c)&=&\sum_{p=1}^{p_c-1} \frac{x^p}{p^\alpha},
\end{eqnarray}
has been defined.

The quantum density contribution $n_0(r)$ is characterized by values
of $p$ such that $p \beta \hbar\omega_0 > 1$, and thus by a
zero-temperature oscillator length $r_0(p\beta) \approx r_0$ and the
density matrix is given by $\rho_{\rm osc}(\rv,\rv;p \beta \hbar
\omega_0) \approx \pi^{-3/2} r_0^{-3} e^{-r^2/r_0^2}$. The resulting
sum is then easily computed, yielding
\begin{equation}
n_{0}(r)\approx \frac{N_0(T)}{\pi^{3/2}r_{0}^{3}}e^{-r^2/r_{0}^2},
\label{app:n0}
\end{equation}
with amplitude factor
\begin{equation}
N_{0}(T) \approx \sum_{p=p_c}^\infty e^{-p|\mu|/k_B T}
= \frac{e^{-(p_c-1)|\mu|/k_B T}}{e^{|\mu|/k_B T}-1}
\label{app:N0T}
\end{equation}
being the number of bosons occupying the single-particle ground
state, reduced by the factor $e^{-(p_c-1)|\mu|/k_B T} \approx
e^{-|\mu|/\hbar\omega_0}$.

Eliminating $p_c = k_B T/\hbar\omega_0$, it is clear that $n_T(r)$
and $n_0(r)$ depend strongly on the ratio of the chemical potential
$\mu$ to the trap level spacing $\hbar\omega_0$, with the former
determined by the temperature through the total atom number
constraint. For $|\mu|/\hbar\omega_0 \gg 1$ (which corresponds to $T
> T_c$), the sum in the expression for $n_T(r)$, Eq.\
(\ref{app:n_Tgeneral}), can be extended to infinity, introducing
only exponentially small error $O(e^{-\mu/\hbar\omega_0}) \ll 1$.
The resulting thermal density $n_T(r)$ is then given by the extended
zeta function
\begin{eqnarray}
n_T(r) &\approx& \left(\frac{k_B T}{2\pi\hbar\omega_0} \right)^{3/2}
\frac{1}{r_{0}^3} g_{3/2}
\left(e^{-r^2/r_T^2 - |\mu|/k_B T}\right),
\nn \\
&&\hspace{3.8cm} \text{for}\ T > T_c,
\label{app:n_T}
\end{eqnarray}
As expected for $T > T_c$, the condensate spatial distribution is
still given by the Gaussian expression (\ref{app:n0}), but with an
exponentially small condensate
\begin{eqnarray}
N_0(T) &\approx& e^{-{|\mu|/\hbar\omega_0}} \approx 0,\
\ \text{for}\ T > T_c,
\end{eqnarray}

At high $T \gg T_{c} = \hbar \omega_0 (N/\zeta(3))^{1/3}$ (where the
gas is nondegenerate), such that $0 < |\mu| \approx -k_B
T\ln[\left(\frac{\hbar\omega_0}{k_B T}\right)^3 N] \approx 3 k_B
T\ln (T/T_{c})\gg k_B T$, the thermal (and therefore total) density
(\ref{app:n_T}) reduces to a pure Gaussian with thermal width $r_T$,
reflecting the high-temperature Boltzmann statistics
\begin{eqnarray}
n(r) &\approx& \left(\frac{k_B T}{2\pi \hbar\omega_0}\right)^{3/2}
\frac{1}{r_0^3} e^{-r^2/r_T^2 - |\mu|/k_B T},
\nn \\
&=& \frac{N}{\pi^{3/2}r_{T}^3} e^{-r^2/r_{T}^2},\ \
\mbox{for} \ \ T \gg T_c
\label{app:nHigh_trap}
\end{eqnarray}
in which the relation $N = (k_B T/\hbar\omega_0)^3 e^{|\mu|/k_B T}$
has been used to satisfy the particle number constraint.

As $T$ is lowered, approaching $T_c$ from above, the magnitude of
the chemical potential drops below $T$ (remaining negative), and
(while the condensate fraction remains vanishingly small) the boson
density profile (\ref{app:n_T}) develops a non-Boltzmann peak
structure even above $T_c$:
\begin{widetext}
\begin{equation}
n(r) \approx \left(\frac{k_B T}{2\pi\hbar\omega_0}\right)^{3/2}
\frac{1}{r_{0}^3}
\begin{cases}
e^{-r^2/r_T^2 - |\mu|/k_B T},\ \ & r \gg r_{T} \\
\zeta(3/2)-2\pi^{1/2}\left(\frac{r^2}{r_{T}^2}
+ \frac{|\mu|}{k_B T}\right)^{1/2},\ \ & r \ll r_{T},
\end{cases}
\label{nMed_trap}
\end{equation}
\end{widetext}
retaining a Gaussian falloff at large $r/r_T$.  The small-$r$ cusp
in $n(r)$, that develops as $T_c$ is approached from above, is
rounded on the cutoff length $r_c(T)$, which for $T > T_c$ is given
by $r_c(T) \approx r_\mu \equiv r_0\sqrt{2|\mu|/\hbar\omega_0}$,
with $r_0 \ll r_c(T) \ll r_T$.

From the prefactor in $n(r)$ it is clear that this thermal form
cannot persist to low temperatures, as the volume under $n(r)$ drops
with $T$. Thus, for even lower temperature, $T <
T_{c}\approx\hbar\omega N^{1/3}$, to accommodate all $N$ particles
$|\mu|$ is forced to drop below the level spacing, $|\mu| \lesssim
\hbar \omega_0$. In this region, the high $p$ terms in
(\ref{app:n_Tgeneral}), estimated to be $O(e^{-p_c|\mu|/k_B T}) =
O(e^{-|\mu|/\hbar\omega_0})$, are no longer exponentially small and
the sum can no longer, in general, be extended to infinity. However,
for $p_c \gg 1$, on length scales longer than the cutoff length
$r_c(T)$, the thermal density $n_T(r)$, Eq.\ (\ref{app:n_Tgeneral}),
is still well approximated by the extended zeta function in Eq.\
(\ref{app:n_T}), but with $\mu/k_B T \approx 0$ and its small $r$
cusp smoothed out on a length scale $r_c(T < T_c) \approx
r_T/\sqrt{p_c} = 2 r_0$ [following from the condition $p_c (r/r_T)^2
= O(1)$].

Correspondingly, the condensate contribution $n_0(r)$ begins to grow
for $T < T_c$ through the growth of the coefficient $N_0(T)$, Eq.\
\ref{app:N0T}.  In fact for $T < T_c$, $|\mu|/k_B T \ll 1$ and
$|\mu|/\hbar\omega_0 \ll 1$, one has $N_0(T < T_c) \approx k_B
T/|\mu|\gg 1$, and the finite fraction of the bosons condensed into
the lowest single-particle oscillator Gaussian state (with a narrow
width $r_0$) are precisely what is required to make up for those
that cannot fit into the thermal distribution $n_T(r)$. Thus for $T
< T_c$ the total atom density profile $n(r)$ changes dramatically,
developing an easily identifiable bimodal distribution $n(r) =
n_{T}(r) + n_{0}(r)$, illustrated in Fig.\ \ref{densityProfiles}.

As discussed in Sec.\ \ref{transitionsTrap}, this analysis easily
generalizes to the two-component Bose gas (bosonic atoms and
molecules) that is the focus of this paper.

\section{Details of Bose-BCS MSF--ASF criticality}
\label{app:bcscrit}

In this appendix, the critical behavior in the vicinity of the ASF
transition line is derived as a function of dimension $d$ via an
analysis of the integral equations (\ref{5.40}) for small deviations
$\rho,\tau,f$.

For $3 < d < 4$ one obtains finite values for the derivatives
\bea
J_1 &\equiv& -\partial_\tau J(0,0)
= \int \frac{d^du}{(2\pi)^d}
\frac{u^2+1}{[u^2(u^2 + 2)]^{3/2}}
\nonumber \\
J_2 &\equiv& \partial_f J(0,0) = \frac{1}{2}
\int \frac{d^du}{(2\pi)^d}
\frac{1}{[u^2(u^2 + 2)]^{3/2}}
\nonumber \\
R_1 &\equiv& -\partial_\tau R(0,0) = J_2/I_d
\nonumber \\
R_2 &\equiv& \partial_f R(0,0) = J_1/4 I_d.
\label{C1}
\eea
Higher order derivatives, however, lead to divergent integrals at
small $u$.  The subtracted integral has leading behavior
\bea
\delta J(\tau,f) &\equiv& J(\tau,f) + J_1 \tau
- J_2 f \approx j_c(s)
\nonumber \\
j_c(s) &\equiv& \frac{1}{\sqrt{2}}
\int \frac{d^d u}{(2\pi)^d}
\left(\frac{1}{u} - \frac{s}{2u^3}
- \frac{1}{\sqrt{u^2 + s}} \right)
\nonumber \\
&=& j_c(1) s^{\frac{d-1}{2}},
\label{C2}
\eea
where
\begin{equation}
s = \frac{(1-\tau)^2-(1-f)}{2(1-\tau)}
\approx \frac{f-2\tau}{2} > 0.
\label{C3}
\end{equation}
With the subtractions, the integral $j_c(1)$ converges at both
large and small $u$. The vanishing of $J$ requires now
\begin{equation}
0 = J_2 f - J_1 \tau + j_c(1) s^{\frac{d-1}{2}}
+ O(f^2,\tau^2),
\label{C4}
\end{equation}
and leads to
\begin{equation}
f = \frac{J_1}{J_2} \tau
- \frac{j_c(1)}{J_2} \left(\frac{J_1}{2J_2} - 1
\right)^{\frac{d-1}{2}} \tau^{\frac{d-1}{2}}
+ O(\tau^2).
\label{C5}
\end{equation}
It is easy to see from (\ref{C1}) that $J_1 > 2J_2$ and that
this solution is consistent with the requirement that $s > 0$.
Similarly, for the density deviation one obtains
\bea
\rho &=& R_2 f - R_1 \tau + \frac{j_c(1)}{2I_d}
s^{\frac{d-1}{2}} + O(f^2,\tau^2)
\nonumber \\
&=& \frac{J_2}{I_d}\left(\frac{J_1^2}{4J_2^2} - 1\right) \tau
- \frac{j_c(1)}{2I_d} \left(\frac{J_1}{2J_2} - 1
\right)^{\frac{d+1}{2}} \tau^{\frac{d-1}{2}}
\nonumber \\
&&+\ O(\tau^2).
\label{C6}
\eea

For $d < 3$ integrals in (\ref{C1}) diverge at small $u$ indicating
singular dependence on $\tau$ and $f$. This infrared singularity has
been isolated by writing
\bea
J(\tau,f) &=& J_c(\tau,f) + \delta J(\tau,f)
\nonumber \\
R(\tau,f) &=& R_c(\tau,f) + \delta R(\tau,f)
\label{C7}
\eea
where
\bea
J_c(\tau,f) &=& \frac{1}{\sqrt{1-\tau}} \tilde j_c(s)
\nonumber \\
R_c(\tau,f) &=& \frac{\sqrt{1-\tau}}{2 I_d} \tilde j_c(s)
\nonumber \\
\tilde j_c(s) &\equiv& \frac{1}{\sqrt{2}}
\int \frac{d^du}{(2\pi)^d}
\left(\frac{1}{u} - \frac{1}{\sqrt{u^2 + s}} \right)
\label{C8}
\eea
where $\tilde j_c(s)$ now requires only the single subtraction for
convergence at large $u$ in this lower dimension. It can be
checked that for $1 < d < 3$ the derivatives of the subtracted
integrals $\delta J$ and $\delta R$ are finite at the critical
point, and so have leading linear dependence on $\tau$ and $f$.
The corresponding coefficients are
\bea
\tilde J_1 &=& \int \frac{d^du}{(2\pi)^d}
\left\{\frac{u^2+1}{[u^2(u^2 + 2)]^{3/2}}
- \frac{1}{2\sqrt{2} u^3} \right\}
\nonumber \\
\tilde J_2 &=& \frac{1}{2}
\int \frac{d^du}{(2\pi)^d}
\left\{\frac{1}{[u^2(u^2 + 2)]^{3/2}}
- \frac{1}{2\sqrt{2} u^3} \right\}
\nonumber \\
\tilde R_1 &=& \tilde J_2/I_d,\
\tilde R_2 = \tilde J_1/4 I_d.
\label{C9}
\eea
One obtains again
\begin{equation}
\tilde j_c(s) = \tilde j_c(1) s^{\frac{d-1}{2}},
\label{C10}
\end{equation}
in which $\tilde j_c(1)$ is finite for $1 < d < 3$ (note that
$d=1$ is the lower critical dimension, below which the phase
transition ceases to exist, and so one expects special behavior
here).  This dominates the linear behavior, and the constraint
$J(\tau,f) = 0$ becomes
\begin{equation}
0 = \tilde j_c(1) s^{\frac{d-1}{2}}
- \tilde J_1 \tau + \tilde J_2 f + O(\tau^2,f^2),
\label{C11}
\end{equation}
which, to leading order, has the solution $s=0$, and hence
\begin{equation}
f = 2\tau + 2\left[\frac{\tilde J_1 - 2\tilde J_2}
{\tilde j_c(1)} \right]^{\frac{2}{d-1}}
\tau^{\frac{2}{d-1}} + O(\tau^2).
\label{C12}
\end{equation}
From (\ref{C9}) it is easy to see that $\tilde J_1 > 2 \tilde
J_2$, so that (\ref{C12}) is again consistent with $s > 0$. Using
(\ref{C7}) and (\ref{C8}), the density deviation is
\bea
\rho &=& \tilde R_2 f - \tilde R_1 \tau
+ \frac{\tilde j_c(1)}{2 I_d}
s^{\frac{d-1}{2}} + O(\tau^2,f^2)
\label{C13} \\
&=& \frac{\tilde J_1 - 2 \tilde J_2}{2 I_d}
\left\{2\tau + \left[\frac{\tilde J_1 - 2\tilde J_2}
{\tilde j_c(1)} \right]^{\frac{2}{d-1}}
\tau^{\frac{2}{d-1}} \right\}+ O(\tau^2).
\nonumber
\eea
The singular corrections to the linear behavior may be identified with
the \emph{energy exponent} $1-\alpha$, where the Gaussian specific
heat exponent is $\alpha = \frac{3-d}{d-1}$ (not to be confused with
the Feshbach resonance coupling).  In a full theory this form would be
replaced by the exact Ising exponent.

At the upper critical dimension $d=3$ itself there will be
logarithmic corrections.  Since $d=3$ is the physically important
dimension, it is worth presenting the results for this case as well.
One may now express the results in terms of elliptic integrals (see
Ref.\ \onlinecite{GR}, pp.\ 232, 235, 905):
\begin{eqnarray}
&&J(\tau,f) = \frac{\sqrt{1 - \tau + \sqrt{1 - f}}}{2\pi^2} E(q) -
\frac{1}{\sqrt{2} \pi^2}
\nonumber \\
&&R(\tau,f) = \frac{\sqrt{1 - \tau + \sqrt{1 - f}}}{12\pi^2 I_3}
\big[-(1 - \tau) E(q)
\nonumber \\
&&\ \ \ +\ (1 - \tau - \sqrt{1 - f}) K(q) \big]
+ \frac{1}{6 \sqrt{2} \pi^2 I_3}
\label{C14}
\end{eqnarray}
where
\begin{equation}
q = \frac{2\sqrt{1-f}}{1-\tau+\sqrt{1-f}}.
\label{C15}
\end{equation}
At small $\tau,f$, hence $q \to 1$, using the asymptotics of the
elliptic integrals (see Ref.\ \onlinecite{GR}, p.\ 906), the $J = 0$
constraint takes the form
\bea
0 &=& -8\tau + (f-2\tau)
\ln\left(\frac{64 e^{-3}}{f-2\tau} \right)
\nonumber \\
&&+\ O[\tau^2 \ln(f-2\tau),f^2 \ln(f-2\tau)],
\label{C16}
\eea
with solution
\begin{equation}
f = 2\tau \left\{1 + \frac{4}{\ln(8/e^3\tau)}
+ O\left[\frac{\ln \ln(8/e^3 \tau)}
{\ln^2(8/e^3\tau)} \right] \right\}.
\label{C17}
\end{equation}
The density deviation takes the form
\bea
\rho &=& \frac{1}{96 \sqrt{2} \pi^2 I_3}
\left[12f + 3(f - 2\tau)
\ln\left(\frac{64 e^{-3}}{2\tau - f} \right) \right]
\nonumber \\
&&+\ O[\tau^2 \ln(2\tau-f),f^2 \ln(2\tau-f)]
\nonumber \\
&=& \frac{\tau}{2\sqrt{2}\pi^2 I_3}
\left\{1 + \frac{2}{\ln(8/e^2\tau)}
+ O\left[\frac{\ln \ln(8/e^2 \tau)}
{\ln^2(8/e^3\tau)} \right] \right\}.
\nn \\
\label{C18}
\eea


\begin{thebibliography}{7}
\bibitem{vibration} In the experiments the bound state actually
exists in a high vibrational state.  Observation of a
quasi-equilibrium molecular condensate requires that the lifetime of
this state be much larger than collisional equilibration time of the
vapor. Only then can the lower vibrational states to which this
state must eventually decay be neglected.
\bibitem{decoupledBEC} In the noninteracting BEC limit ($\alpha=0$)
the ASF will not be accompanied by a molecular condensate as it is
for a finite Feshbach resonance coupling. As discussed in Sec.\
\ref{sec:alphaneq0}, for a finite $\alpha$ a molecular condensate
will necessarily appear in the ASF state, but will be small in the
weakly interacting limit.
\bibitem{commentFirstOrder} It will be seen in Sec.\ \ref{sec:mft}
that this transition may also be driven first order in certain
regions of the phase diagram (see Figs.\ref{phasediagramNuT},
\ref{phase_diagramAlphaMuA},\ref{phase_diagramAlphaMuB}). This
possibility has been explored previously within a renormalization
group framework,\cite{FB97,LL04}, where runaway flows generated by
the coupling between the incipient atomic order and fluctuations in
the molecular order parameter, are interpreted in this fashion. Near
the point (critical endpoint) where all three phases meet, this is
also found unambiguously even at the mean field level.
\bibitem{nuclearspin} In the presence of a hyperfine interaction,
open and closed channels are coupled, and the inter-channel
transition is accompanied by (an angular momentum compensating)
nuclear spin flip. The coupling of the latter to the external
magnetic field is much smaller than that of the electrons (by the
ratio of the nuclear to the Bohr magneton) and therefore plays a
negligible role in the Feshbach resonance parameters.
\bibitem{nonmonopotential} At low energies the same scattering
amplitude structure arises for two atoms interacting through a
potential with a large potential barrier separating a short-range
attractive minimum from the rapidly decaying repulsive part at
infinity.  For a sufficiently deep minimum, a (negative energy)
bound state (molecule) exists. For a sufficiently shallow minimum, a
(positive energy) quasi-bound state exists with a finite lifetime
(resonance).  However, there is also an intermediate range in which
there is no bound state, but the real part of the energy pole is
negative, and the metastable molecule interpretation fails.
\bibitem{RGpart} From a renormalization group point of view, one
may also think of the two species model as a partially renormalized
Hamiltonian in which all fluctuations below the scale of the
molecular diameter have been integrated out. It will be seen later
that the dissociation amplitude $\alpha$ will ``dress'' the
$\psi_2^\dagger$ particles with a virtual cloud of $\psi_1^\dagger$
pairs that diverges in size as resonance is approached.  Thus, such
a model captures all further renormalization, including critical
fluctuations.  In particular, the multi-body parameters in $\hat H$,
Eq.\ \ref{H2channel} should all be treated as constants throughout
the neighborhood of the Feshbach resonance.
\bibitem{foot:trap} In the presence of a trapping potential, the
spectrum should be observably discrete for excitation wavelengths
comparable to the system size, and will depend on the trap shape.
\bibitem{foot:roton} In future work, it might be interesting to
revisit the usual Feynman argument that predicts a downturn (roton
minimum) in the energy spectrum in the neighborhood of the peak in
the structure factor $S(k)$. For the dense fluid there is a strong
peak at $k \sim n^{1/3}$, the inverse of the mean interparticle
spacing, reflecting the oscillation of the density-density
correlation function on this scale. For the dilute, weakly
interacting fluid this ``incipient crystal'' structure disappears.
However, since the gas parameter $n a_{\rm eff}^3$ can be large near
resonance, interesting structure in $S(k)$ could return, and in turn
generate interesting structure in $E_{\kv \s}$ at higher momenta.
\bibitem{foot:d4} For $d \geq 4$ a similar analysis may be
performed, but one now needs to impose an ultraviolet (large $k$)
cutoff, which physically arises from the vanishing of the scattering
coefficients at large $k$.
\bibitem{commentZ2} In the ``hard-spin'' $\psi_\s = e^{i\theta_\s}$
description, the Feshbach interconversion term reduces to $\alpha
\cos(2\theta_1-\theta_2)$, locking the atomic phase to the phase of
the molecular condensate up to $\pi$, corresponding to the unbroken
${\cal Z}_2$ symmetry of the MSF. In terms of a vector order
parameter, the MSF is characterized by a finite quadrupole moment
($l=2$ angular harmonic, akin to the nematic, ``headless arrow''
order in liquid crystals), and ASF by a vector order parameter
($l=1$ angular harmonic). In this language, in the ASF both the
``nematic director'' (headless arrow, $\psi_2$) and the vector
($\psi_1$) are aligned, with their directions locked together.  In
contrast, in the MSF, while the nematic director is ordered, the
vector order is disordered by virtue of only being locked to the
nematic director up to an angle $\pi$.  Clearly this unlocking
corresponds to a restoration of the ${\cal Z}_2$ Ising symmetry of
the MSF. Further discussion of this, and related, issues can be
found in Appendix \ref{mapNematic}.
\bibitem{commentIrrelevant} The ``diamagnetic'' nonlinearity,
$(\nabla\theta_2)^2|\psi_1|^2$ in Eq.\ (\ref{S_1}) is clearly
subdominant to $|\nabla\theta_2|^2$. To see that the
``paramagnetic'' term, $\nabla\theta_2 \cdot \psi^*_1 \nabla\psi_1$
is irrelevant it is easiest to integrate out $\nabla \theta_2$,
obtaining a local current-current interaction, that (because of the
two extra powers of the gradient) at long wavelengths is subdominant
to the local quartic coupling.
\bibitem{commentConnection} The connection to a compressible Ising
model can be made even closer by a duality transformation. See for
example Ref.\ \onlinecite{FB97}.
\bibitem{gapcomment} The single particle energy spectrum should
indeed be gapped in the MSF phase. The fact that it is gapped even
in the ASF phase is an artifact of the variational
calculation,\cite{Nozieres} and can be cured by more sophisticated
treatments.\cite{W88,Popov} In particular, the usual Bogoliubov
approximation for $g > 0$ yields gapless excitations, but the
variational approach, which includes additional higher order terms
in the Hamiltonian coefficients (\ref{5.17}) to account for the
possibility of molecular order, reopens the gap. Reclosing it
requires careful inclusion of further higher order terms.
\bibitem{nematicVortex} The existence of highly anisotropic dumbbell
$2\pi$ atomic vortices allows for the possibility of liquid
crystal\cite{deGennes,ChaikinLubensky} normal phases. These can, in
principle, appear when the anisotropic $2\pi$ vortices proliferate
but exhibit some form of (e.g., smectic or nematic) spatial order,
that distinguishes the resulting liquid-crystal normal state from a
fully disordered normal state, where vortices form a homogeneous,
isotropic liquid.
\bibitem{foot:cubic} The parameters $p$ and $q$ defined in
(\ref{3.x11}) and (\ref{3.x12}) are obtained by expressing the cubic
equation $f'_{\rm ASF} = 0$ in the form $y^3 + py + q = 0$ via the
shift $y = \bar \psi_2 + 1/2(\gamma-1)$. Let $u_\pm = (-q/2 \pm
\sqrt{\Delta})^{1/3}$, both defined to be real for $\Delta \geq 0$,
and complex conjugates for $\Delta < 0$. Then the exact roots are
given by $y_l = e^{i2\pi l/3} u_+ + e^{-i2\pi l/3} u_-$, $l=0,1,2$.
For $\Delta = 0$ one obtains $y_0 = -2(q/2)^{1/3}$, $y_1 = y_2 =
(q/2)^{1/3}$.
\end{thebibliography}
\end{document}